\documentclass[linenumbers,superscriptaddress,longbibliography,nobibnotes,10pt,prx,twocolumn]{revtex4-1}

\usepackage{array}
\usepackage{graphicx}
\usepackage{verbatim}
\usepackage{color} 
\usepackage{amsfonts}
\usepackage{amsmath}
\usepackage{fancyhdr}
\usepackage{bbold}
\usepackage{subfigure}
\usepackage{leftidx}
\usepackage{textcomp} 
\usepackage{ulem} 
\usepackage{lineno}

\newcommand{\ket}[1]{\left\vert{#1}\right\rangle}

 \newcommand {\beq}{\begin{equation}}
\newcommand {\eeq}{\end{equation}}

\usepackage[dvipsnames]{xcolor}

\usepackage{textcomp}

\usepackage[colorlinks=true]{hyperref}
\hypersetup{
    linkcolor=blue,
    citecolor=blue,
    urlcolor=blue
}

\begin{document}
\title{Silicon edge-dot architecture for quantum computing with global control and integrated trimming}

\author{M.~A.~Fogarty}
\email{michael@quantummotion.tech}
\affiliation{London Centre for Nanotechnology, UCL, 17-19 Gordon St, London WC1H 0AH, United Kingdom}
\affiliation{Quantum Motion, 9 Sterling Way, London N7 9HJ, United Kingdom}

\nolinenumbers

\begin{abstract}
A scalable quantum information processing architecture based on silicon metal-oxide-semiconductor technology is presented, combining quantum hardware elements from planar and 3D silicon-on-insulator technologies. This architecture is expressed in the ``unit cell'' approach, where tiling cells in two dimensions and allowing inter-cellular nearest-neighbour interactions makes the architecture compatible with the surface code for fault tolerant quantum computation. The architecture utilises global control methods, substantially reducing processor complexity with scale: Single-qubit control is achieved using globally applied spin-resonance techniques and two-qubit interactions are mediated by large quantum dots. Further, a solution to device variation is proposed through integration of electronics for individual trimming of quantum dot voltage references. Such a combined set of solutions addresses several major barriers to scaling quantum machines within completely silicon based architectures.
\end{abstract}
\date{\today}
\nolinenumbers
\maketitle
 
Universal quantum computers with fully scalable architectures are necessary to solve meaningful problems that are intractable on digital computers. Through quantum error detection schemes applied to qubit lattices~\cite{terhal2015quantum}, these computations can be made fault-tolerant. This process involves encoding partitions of many physical qubits into separate logical qubits, and works by invoking a trade-off between the number of physical qubits and their error rates. This approach is expected to result in significant qubit overheads required to achieve meaningful computational capabilities. For example, it is predicted that $\mathcal{O}(10^{8})$ qubits operating with error rates at or below $10^{-3}$ are required for the non-trivial execution of Shor's factoring algorithm~\cite{ogorman2017quantum}. This requirement makes scaled architectures in silicon particularly attractive; the low form-factor of a silicon quantum dot produced by 300 mm wafer technologies in contemporary foundries~\cite{hutin2019soi,li2020flexible,zwerver2021qubits} results in achievable qubit densities as high as $\mathcal{O}(10^{9})$cm$^{-2}$, while the compatibility with a highly developed and globally accessible silicon device fabrication industry presents several advantages, ranging from on-chip integration with metal-oxide-semiconductor (MOS) digital hardware~\cite{vandersypen2017interfacing,gonzalez2020scaling,pauka2021cryogenic} to high-volume device production and dissemination. 

Hardware architecture approaches for fault-tolerant machines within the silicon MOS materials platform have largely focused on devices formed at, or near, a planar silicon / silicon oxide interface~\cite{veldhorst2017silicon,li2018crossbar,hollenberg2006two,pica2016surface,tosi2017silicon} with the lithographically defined metallic gates (or gate stacks) used to confine and/or define qubits. In direct contrast to these planar device approaches, the concept of an ``edge-dot'' is introduced to the reader here. Quantum dots of this type are similarly produced in 3D silicon-on-insulator (SOI) technology, where carrier confinement is naturally produced by the electric field concentration in the cross-sectional corner of an etched silicon nanowire, applied by an overlapping gate electrode~\cite{voisin2014few}. The length of the gate in the overlapped region acts to confine the quantum dot in the remaining dimension. These systems have been extended to produce bilinear arrays of quantum dots through the development of ``split-gate'' devices~\cite{dupont2013coherent}, and can extend into an arbitrary arrays of 2$\times$N quantum dots~\cite{hutin2019gate,ansaloni2020single,chanrion2020charge} simply based on the number of electrodes patterned. Recently, the concept of utilising floating gate electrodes~\cite{trifunovic2012long} as a method to sense or couple charges in parallel running nanowires has been achieved~\cite{duan2019remote,gilbert2020single}, opening this nanowire approach to scalable architectures beyond the dimensions of 2$\times$N. 

Here, this approach is conceptually extended by asserting that these quantum dot structures do not need to be produced in the corners of a thin nanowire; but instead these dots can form a 1$\times$N array along a single edge of some silicon ``plateau'', which is of arbitrary width in the dimension perpendicular to the edge defining the dot array. This use of topography within the silicon layer has the potential to combine advantages from two prevalent approaches of forming quantum structures in semiconductors including planar~\cite{veldhorst2014addressable,veldhorst2015two} and 3D SOI~\cite{hutin2019soi} which have, to date, remained as separately developed host technologies for quantum devices. This work illustrates how this hybridised approach to formation of silicon quantum devices has advantages when approaching the challenge of integrating elements of a quantum-classical interfacing layer~\cite{reilly2015engineering,reilly2019challenges,franke2019rent,gonzalez2020scaling}, particularly focusing on the formation of 2D qubit lattices for the execution of error correction codes.

In this paper a ``unit cell'' hardware sub-structure is presented which directly reflects the tile-like nature of error detection algorithms when applied to the physical qubit layer. Here, the surface code~\cite{fowler2012surface} is studied as it has advantages of requiring a 2D qubit lattice with nearest neighbour interactions and can tolerate compound errors as high as 1\%. This unit cell approach has additional advantages towards device scalability through the definition of a single, reproduced structure which forms the foundations of the quantum machine. Further, through shared control of all cells~\cite{li2018crossbar}, the input signal overheads required can be drastically reduced~\cite{vandersypen2017interfacing,gonzalez2020scaling,franke2019rent} to scale with the unit cell size, rather than the total number of physical qubits within the quantum machine.

The remainder of this work is structured as follows: the physical hardware unit-cell is first presented in \S~\ref{SSec:EdgeDot}, followed by a discussion in context with the current state-of-the-art on the operations required to execute universal quantum computing with this cell in \S~\ref{SSec:QubitSignals}. This physical architecture is mirrored by an algorithmic unit-cell protocol for execution of the surface code, as outlined in \S~\ref{SSec:SC_Unit}. This section highlights how the error detection scheme can be implemented while applying completely global electron spin resonance pulses to all spins (data and ancilla) for the execution of the necessary single qubit gates, with the feasibility and control methods discussed in \S~\ref{SSec:Cavity}. Finally, it is acknowledged that slight variations in the hardware can easily result in one location of any two different unit cells exhibiting very different behaviour under the same globally applied control conditions. Therefore, for the unit cell approach to be a success, contingencies intrinsic to the hardware cell must be present to ensure enough uniformity can be realised artificially. A method to address this challenge is proposed in \S~\ref{SSec:Trim}, through circuitry designed to ``trim'' the DC voltage offsets supplied to each physical qubit via the integration of non-volatile analogue memories as a silicon MOS hardware overhead. The example presented here integrates floating gate transistors into the control circuitry for each quantum dot.

\begin{figure*}[t]
 \centering
  \begin{center}
  \includegraphics[width=18cm]{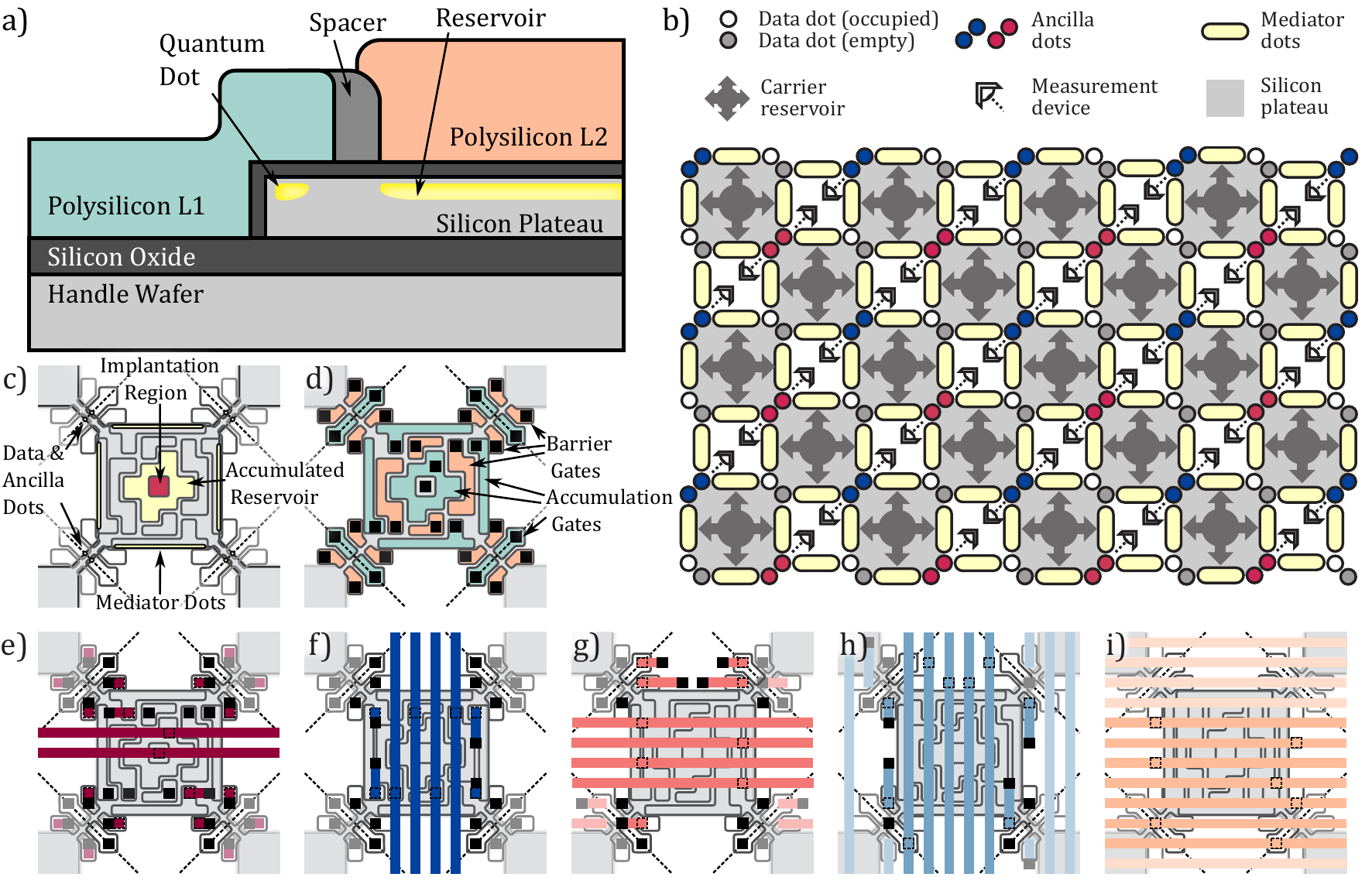}
  \end{center}
  \vspace*{-6mm}
  \caption{\textbf{$\vert$ Edge-topology architecture based on silicon plateaus.} \textbf{a)} Quantum structures are confined naturally due to electric field enhancement at the edge of the silicon plateau through overlapping conductive electrodes. Additional electrodes patterned wholly on the plateau region can result in either planar quantum or classical structures, with an accumulated carrier reservoir shown in the image. \textbf{b)} Array-based architectures are developed with this technology by integrating the silicon plateau as a core element within the fundamental unit-cell structure, patterned repeatedly under translational symmetry to construct a larger quantum processor. \textbf{c)} Front End Of Line (FEOL) of a unit cell based around a single silicon plateau (light grey). This hosts an implantation site, accumulated charge reservoir and locations on edge-defined quantum dots based on potentials applied to the \textbf{d)} gate electrodes shown. Black squares indicate contact vias that connect features to the above metallic layers (out of page), with corresponding dashed squares showing these locations in the subsequent layer. \textbf{e-i)} Shows the sequence of Back End Of Line (BEOL) routing layers which interconnect the features between unit cells. Faded elements are from adjacent unit cells and are shown for illustration of cell tessellation. Not present are the control elements (and required BEOL routing) some of which can exist on the FEOL within the hardware unit cell boundary illustrated by diagonal black-dashed lines.}\label{fig:FTQEC_Architecture}
\end{figure*} \vspace{-4mm}

\section{Hardware unit cell for an edge-dot architecture}\label{SSec:EdgeDot}
\vspace{-4mm}
Each hardware unit cell is constructed around a single silicon plateau which facilitates the formation of quantum dot structures along the plateau boundary while also hosting standard techniques seen in planar device formation in locations away from the boundary as shown in Fig.~\ref{fig:FTQEC_Architecture}a). These include the formation of implantation regions and carrier reservoirs, tunnel barriers and quantum dots. The quantum dots formed at the plateau boundary take advantage of the electric field concentration in the cross-sectional corner due to the overlapping gate electrode, resulting in strong carrier confinement~\cite{voisin2014few,corna2018electrically}. The length of the gate in the overlapped region acts to confine the quantum dot in the remaining dimension.

A recent scaled architecture proposal~\cite{cai2019silicon} illustrated several advantages of how the integration of a spinless mediator quantum dot~\cite{srinivasa2015tunable,malinowski2019fast} connected to a charge reservoir could facilitate robustness against certain types of leakage errors in the form of physical charge movement, which would otherwise be highly detrimental to a schemes based on repeated execution of stabilizer cycles. A blueprint for a scaled quantum machine utilising the edge-dot approach, producing a 2D array of qubits interconnected via mediator dots, is illustrated in Fig.~\ref{fig:FTQEC_Architecture}b). 

A more detailed picture of the primitive hardware unit cell for this scaled quantum machine is shown in Fig.~\ref{fig:FTQEC_Architecture}c-i), including the Front End of Line (FEOL) consisting of the physical qubit layer and supporting hardware defined by the silicon plateau, and Back End of Line (BEOL) consisting of routing metal layers for control. Each corner of the plateau is connected to a neighbouring plateau (the next adjacent hardware cell) via a shared nanowire with patterned split-gate elements acting to form a double quantum dot site. These quantum dots can be populated with charge carriers from the reservoir via the mediator quantum dots~\cite{cai2019silicon}, and act to house the physical qubits upon which the computations and error correction codes are performed. The charge reservoir consists of an ohmic implantation site illustrated in Fig.~\ref{fig:FTQEC_Architecture}c) as an exposed central region of the plateau which is directly contacted by a metallic via (black squares in Fig.\ref{fig:FTQEC_Architecture}d-h). The planar surface of this silicon plateau is further utilised to distribute charge from this central reservoir to each edge through a metallic accumulation gate patterned in polycrystalline silicon (poly-silicon) Layer 1 in Fig.~\ref{fig:FTQEC_Architecture}d). Tunnel barriers between the accumulated reservoirs and the edge defined quantum dots can be formed either though engineered gaps~\cite{rochette2019quantum,hutin2019soi} or dedicated barrier gates~\cite{yang2013spin,zajac2018resonantly} seen patterned in poly-silicon Layer 2 in Fig.~\ref{fig:FTQEC_Architecture}d). The plateau edges are overlapped with gate electrodes in poly-silicon Layer 1 which act to confine elongated quantum dots through the concentration of electric fields in the gate-edge overlap region. The plateaus are interconnected at the corners by a narrow silicon channel, commonly referred to as a ``nanowire'', where split-gate technology~\cite{dupont2013coherent} is used to form a double quantum dot via features in poly-silicon Layer 1 and tunable tunnel barriers via electrodes in poly-silicon Layer 2. The tunnel barriers are utilised to interface quantum structures formed in the nanowires with those formed at the edges of connected plateaus. 
For the long plateau edges, a mediator quantum dot is formed~\cite{malinowski2019fast}, which serves to transfer qubit information between the quantum dots at each endpoint~\cite{srinivasa2015tunable,cai2019silicon}. 
The mediator dots have the additional advantage of introducing a means through which quantum information processing hardware elements can be physically separated to allow for efficient signal routing and integration of control or sensing peripherals required at the intra-hardware-cell level. In the example illustrated, the mediator dot is required to be $\sim5\times$ the pitch of the BEOL metal routing, allowing for inter-cell connective routing of the FEOL elements as well as routing between metal layers. Space in the FEOL for integrated peripheral hardware can be seen as the (empty) regions between plateaus (grey), amounting to an area approximately equivalent to the plateaus. A diagram illustrating interconnected plateaus is provided in Supplementary Fig.~\ref{fig:UCFab_Int}.
\begin{figure*}[t]
 \centering
  \begin{center}
  \includegraphics[width=18cm]{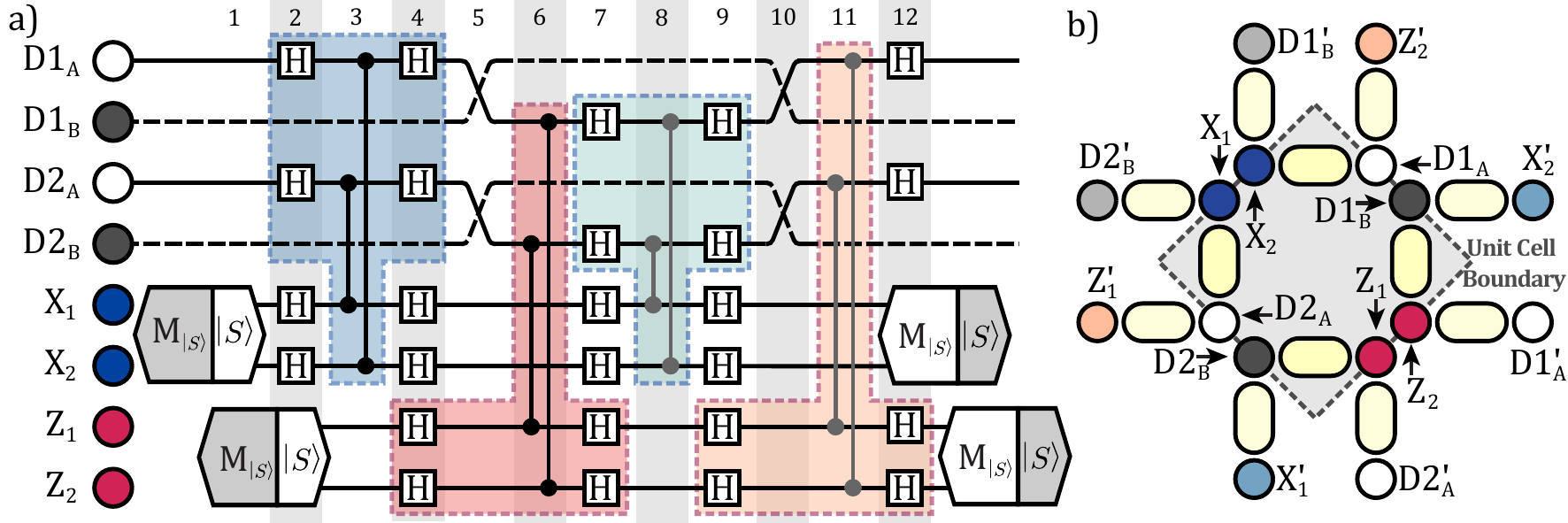}
  \end{center}
  \vspace*{-6mm}
  \caption{\textbf{$\vert$ Surface code unit cell with global ESR control} \textbf{a)} The algorithmic unit cell for the surface code represented by the interleaved application of the XXXX and ZZZZ stabilisers upon the two data qubits within the hardware unit cell. Single spin data qubits shuttle between two physical dot locations, with empty locations indicated by dashed traces. CNOT operations are compositions of selectively applied CZ and globally applied Hadamard gates utilising the `Tick-Tock' approach~\cite{jones2018logical} with the time step numbering indicated across the top of the circuit. Grey operations indicate a CZ mediated between elements in adjacent hardware unit cells, as indicated in \textbf{b)} by a primed (e.g. $X_1'$) notation. Elements encoded with quantum information are shared at the boundary of the hardware unit cell, with the code cycle interfacing multiple adjacent hardware cells as indicated by the elements outside the cell boundary. Ancilla element labelling ${1,2}$ alternates for adjacent cells, and can alternate species in ${X,Z}$ upon consecutive stabiliser cycles due to unpaired data Hadamard gates at step 12 (optionally mitigated through Hadamards applied at step 1 of the next cycle - not shown).} \label{fig:SC_Unit}
\end{figure*}


\section{Silicon Qubit Initialisation, Readout and Control}\label{SSec:QubitSignals}
\vspace{-4mm}
For the execution of the surface code cycle, as detailed in \S~\ref{SSec:SC_Unit}, two varieties of qubit are utilised simultaneously; single electron spin qubits for data qubits and singlet-triplet qubits for ancillas ($X$ and $Z$ syndromes). 
For the control of single electron spins, electron spin resonance (ESR) methods have recently shown control fidelities of up to $99.96\%$ utilising optimised pulse schemes~\cite{yang2019silicon} operating over $8$~\textmu s timescales. As shown in later sections, pulse optimisation schemes such as these transfer well to globally applied spin manipulation.  

For readout, the effect of Pauli spin blockade on interdot tunnelling~\cite{ono2002current,petta2005coherent} enables in-situ double-dot readout for the syndrome qubit state, with high fidelity single-shot Pauli spin blockade detected in several implementations of silicon MOS quantum devices~\cite{harvey2018high,urdampilleta2019gate,zhao2019single}. This readout method, combined with gate-based reflectometry measurement techniques~\cite{urdampilleta2019gate,west2019gate,ibberson2021large} results in a scalable approach to rapidly reading qubit states without the need for integrating additional charge-sensing hardware infrastructure within the FEOL. Some examples in silicon devices have shown readout fidelity of $99.7\%$ in 300 ns~\cite{ibberson2021large}, with others showing extrapolations to $99.9\%$ fidelity achievable in comparable timescales to the single spin qubit gates ~\cite{zheng2019rapid}. 

For initialisation, the triplet lifetimes within the spin blockade region is seen to vary based on species, from 200~\textmu s for $\ket{T_0}$ and up to 5~ms for polarised triplets~\cite{west2019gate,seedhouse2021pauli}, however for initialisation of the singlet state, this triplet lifetime can potentially be reduced through the use of relaxation hot-spots~\cite{srinivasa2013simultaneous,huang2019fidelity}, particularly the spin-orbit driven $S/T^\pm_{(1,1)}$ anti-crossings found either side of the primary singlet anti-crossing, or through selective tunneling with nearby reservoirs~\cite{maune2012coherent,jock2021silicon}.

For two-qubit gates, the nearest-neighbour exchange interaction has lead to several realisations~\cite{veldhorst2015two,watson2018programmable,he2019two,zajac2018resonantly,huang2019fidelity,sigillito2019coherent}, resulting in scaling proposals based on densely-packed two-dimensional arrays of quantum dots~\cite{veldhorst2017silicon,li2018crossbar}, or protocols involving the transport of qubits along long chains of dots~\cite{boter2019sparse}. Other solutions such as photon-mediated two qubit interactions are accessible through hybrid material platforms~\cite{borjans2020resonant,clerk2020hybrid}. 

Alternative methods which maintain compatibility with the materials and processes used in the silicon device industry is to mediate a next-nearest-neighbour exchange through empty~\cite{baart2017coherent}, or multi-electron dots~\cite{malinowski2019fast}.
The mediating structures focused upon here take the form of elongated quantum dots which have nearest-neighbour exchange coupling between itself and a quantum dot at each endpoint (see Fig.~\ref{fig:FTQEC_Architecture}b,c). This elongated quantum dot does not contain any quantum information and only acts to mediate a next-nearest-neighbour interaction between the two sites at each endpoint through the Ruderman-Kittel-Kasuya-Yosida exchange interactions~\cite{srinivasa2015tunable}. For a multielectron dot occupied by the first two electrons (or equivalent S=0 ground-state), the exchange energy $J_{\rm DA}$ between a data (D) dot and ancilla (A) dot, through the mediator (M), is approximated by~\cite{srinivasa2015tunable,cai2019silicon}
\begin{equation}
    J_{\rm DA} \simeq \frac{t_{\rm DM}^2t_{\rm AM}^2}{\varepsilon_{\rm DM}\varepsilon_{\rm AM}\delta_{\rm M}},\label{Eq.ExMed}
\end{equation}
where $t_{\rm DM}$ ($t_{\rm AM}$) is the tunnel coupling between the data-mediator (ancilla-mediator) dots, $\delta_{\rm M}$ is the excited state energy on the mediator with $\varepsilon_{\rm DM}$ ($\varepsilon_{\rm AM}$) the energy detuning of the data (ancilla) dot from the mediator excited state.
A final element required for this scheme is coherent shuttling of a single electron spin qubit between double-dot locations. Shuttling has been shown to have negligible effect on spin projection, with a recent study showing spin polarization is maintained when shuttling between two sites with a fidelity of $>99.9\%$, with $>99\%$ for phase coherence~\cite{yoneda2021coherent}. Further, accurate charge shuttling across arrays of up to 9 dots~\cite{mills2019shuttling}, and relative shuttling uncertainties below 50 parts per million~\cite{rossi2014accurate} have also been demonstrated.

\section{Surface Code Unit Cell}\label{SSec:SC_Unit}
\vspace{-4mm}
The algorithic unit cell representing the suface code is illustrated in Figure~\ref{fig:SC_Unit}a), executing an interleaved XXXX and ZZZZ stabilizer. Utilizing strategic timing for state preparation and measurement of syndromes, the code cycle can be made compatible with the ``tick-tock'' method of executing algorithms using globally applied single qubit Hadamard gates and selective CZ gates~\cite{jones2018logical}. The effective CNOT operations are highlighted by coloured grouping of Hadamard and CZ gates in Fig.~\ref{fig:SC_Unit}a). The use of globally applied single qubit operations results in all unit cells within the quantum machine being controlled by the same input ESR signal. The approach requires embedding the quantum machine within a large peripheral 3D cavity, and the feasibility considerations for this solution is discussed in \S~\ref{SSec:Cavity}. The schematic of a single edge-dot hardware cell capable of executing this stabilizer code is illustrated Fig.~\ref{fig:SC_Unit}b). To execute the error correction process across the surface of the quantum machine, the unit cell must be interconnected to adjacent cells. This is achieved by positioning the ancillar and data quantum dots on the unit cell boundary, while the mediator dots (which result in the CZ operations between qubit locations), are enclosed within a cell boundary. The inter-cell operations are shown as the grey CZ connections within the surface code protocol in Fig.~\ref{fig:SC_Unit}a), executed through the mediators external to the hardware cell boundary in Fig.~\ref{fig:SC_Unit}b) where the equivalent dot positions in adjacent cells are indicated by a prime notation (e.g. $X_1', Z_2'$). 

Much like the primitive cell of a crystal lattice structure, where elements are shared when located at a vertex, edge or face, the unit cell here contains on average 4 dots and (for this implementation) 3 spin 1/2 particles attributed to qubits, plus an additional 4 mediator dots. The two data qubits D1 and D2 bounding the cells consist of a single spin 1/2 particle, each contained within one of two possible quantum dots $A$ or $B$. During the execution of the code cycle, the data spin qubit is shuttled between the dots $A$ and $B$ in order to access and complete all necessary operations between ancillas. In Fig.~\ref{fig:SC_Unit}, dot locations D1$_{A,B}$ and D2$_{A,B}$ are illustrated, with the physical spin qubits indicated in Fig.~\ref{fig:SC_Unit}a) as the solid traces while the empty occupancy of dots are denoted by the dashed traces.

A detailed breakdown of the 12 individual time steps within the surface code cycle seen in Fig.~\ref{fig:SC_Unit}a) is presented in Supplementary Note~\ref{SupNote:SC_TimeSteps}. The use of singlet-triplet based syndrome qubits is compatible with the global ESR schemes as the singlet and triplets remain within their spin manifolds under global rotations. This gives rise to additional simplifications to the above protocol, however these have been omitted from the figures for clarity. For example, the initialisation and measurement process for the two ancillas could be spread over all time steps outside the two-qubit CZ operations (i.e inclusive of 9-2 for $X$ and 12-5 for $Z$), which could facilitate additional time in the measurement and initialisation phases without extending the total time required for the code-cycle.

\begin{figure*}[t]
 \centering
  \begin{center}
  \includegraphics[width=18cm]{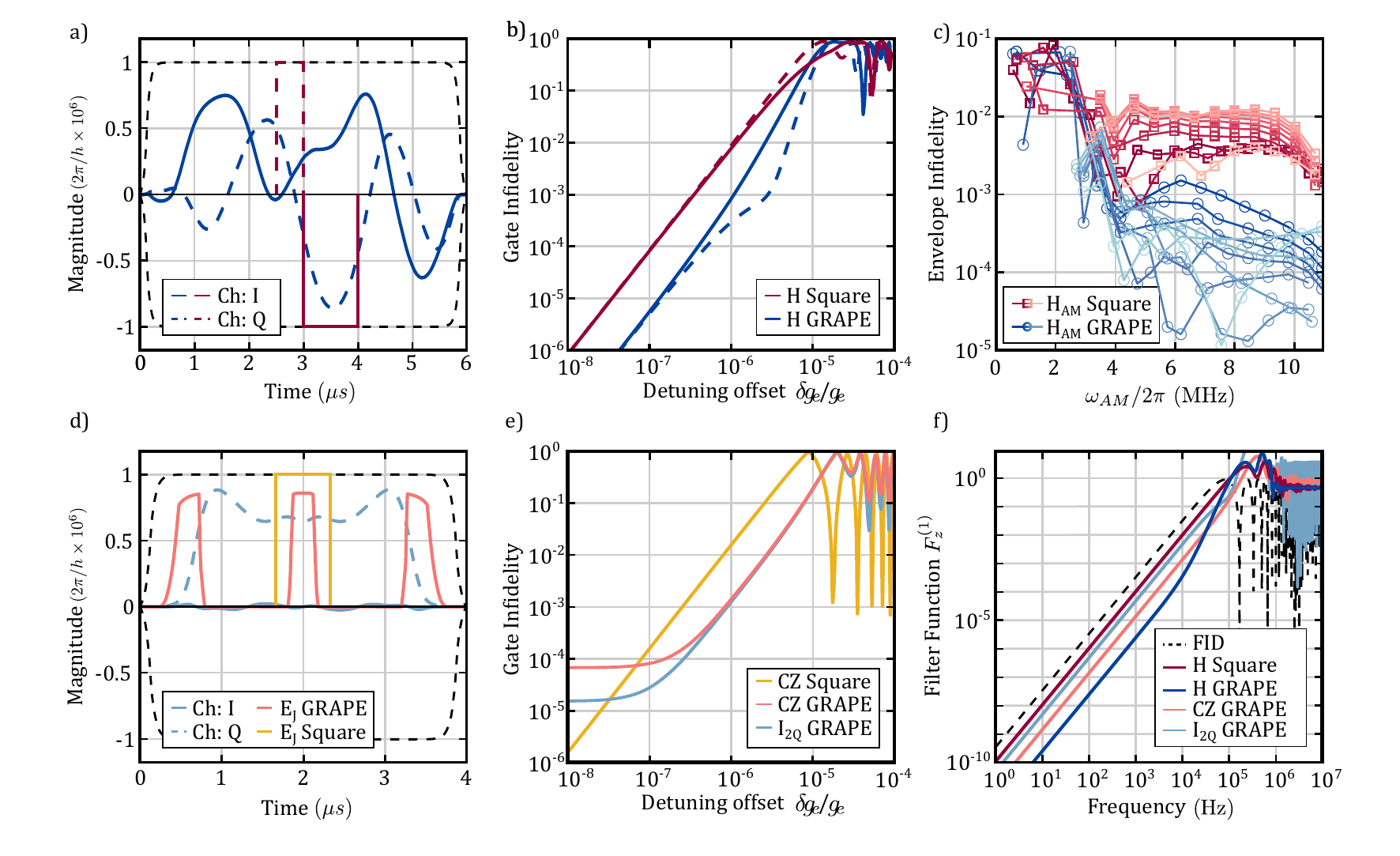}
  \end{center}
  \vspace*{-6mm}
  \caption{\textbf{$\vert$ Global ESR Control Scheme} \textbf{a)} GRAPE pulse envelope executing a Hadamard gate utilising 6$\times$ the duration required for a $\pi$ rotation under a square pulse approach. The black dashed lines indicate a filter function equivalent to a cavity quality factor of Q=250. \textbf{b)} Pulse infidelity as a function of qubit detuning referred to the electron $g$-factor, comparing the square and GRAPE solutions for the Hadamard gate with positive (solid) and negative (dashed) detuning offsets. \textbf{c)} Maximum infidelity envelope for each of the 10 sidebands for both the GRAPE and Square pulse Hadamard gate methods when embedded within the Amplitude Modulation scheme outlined in the main text. Colour darkens for increasing sideband frequency. The rapid decrease in infidelity as a function of increasing side-band separation $\omega_{AM}$ for the GRAPE pulse compared to the Square pulse indicates reduced effects of cross-talk capable for pulse engineering solutions. \textbf{d)} A Controlled-Z (CZ) pulse envelope executed by a square pulse on Exchange energy $E_J$ compared to a GRAPE solution. Globally applied Ch:I, Ch:Q and selectively applied $E_J$ GRAPE signals can be combined to produce a decoupled CZ pulse when $E_J$ GRAPE is combined with Ch:I and Ch:Q signals. A decoupled identity operation on the two qubits is in effect for locations when $E_J$ GRAPE is set to zero. \textbf{e)} Pulse infidelity as a function of $g$-factor referred detuning offset applied to a single qubit. \textbf{f)} First order filter functions for single-qubit and two-qubit control signals described in a) and d). The filter function given by a Free Induction Decay (FID) over 6~\textmu s is indicated for comparison. The CZ Square pulse is omitted due to close similarity to the FID.} \label{fig:CavityControl}
\end{figure*}

\section{Global control of spins using a cavity peripheral}\label{SSec:Cavity}
\vspace{-4mm}
It is asserted that universally applied, or ``global", signals acting as a control input for all qubits across all cells is a desirable property for this approach to scaled quantum systems.  
Several proposals have also highlighted the potential use of globally applied AC fields for scaled systems of spin qubits, controlling either the entire ensemble of qubits~\cite{vandersypen2017interfacing}, or sub-ensembles~\cite{veldhorst2017silicon,li2018crossbar}. One approach to achieving this is to embed the silicon devices within a microwave cavity~\cite{vahapoglu2020single}. Here, the feasibility of achieving sufficiently uniform global control on the ensemble of single qubit spins within the silicon quantum dot array is discussed.

For electron spins in silicon, the spin-orbit interaction is weak when compared to other semiconductors, but still remains appreciable. The surface-roughness of the Si/SiOx interface is predicted to result in an uncontrollable distribution in the Land\'e $g$-factor of up to $\delta g = \mathcal{O}(10^{-2})$~\cite{ferdous2018interface}. It has been shown that this value is tunable based on changes in the electrostatic field environment~\cite{veldhorst2014addressable,hwang2017impact,jock2017probing}, however this degree of the Stark shift seen experimentally remains $\Delta g \simeq \mathcal{O}(10^{-4}) - \mathcal{O}(10^{-3})$~\cite{veldhorst2015two,hwang2017impact}. Thus, for a quantum device operating at appreciable applied magnetic field (B$_0\sim1$T), the Stark shift of the electron $g$-factor can act as a potential source of local tunability, but cannot compensate for the entire $\delta g$ distribution. Therefore, the quantum machine must operate within the limits of broadband microwave pulsing techniques, requiring fields substantially lower than 1T~\cite{vandersypen2017interfacing,li2018crossbar} or otherwise operate with composite pulsing schemes requiring a train of pulses~\cite{vandersypen2005nmr}. For this proposal, high B$_0$ fields are necessary for the implementation of the charge leakage protection schemes~\cite{cai2019silicon}, however pulse trains that are substantially longer than the single qubit $\pi$-rotation time are an undesirable solution as they can absorb significant fractions of the $T_2$ coherence time budget for active qubits. As a compromise, amplitude modulation (AM) techniques can be integrated into the control.

For a quantum gate applied via a resonant microwave AC driving field, it is convenient to define the control signals $I_0(t)$ and $Q_0(t)$ as the in-phase and quadrature components of an envelope function with carrier frequency $\omega_0$~\cite{yang2019silicon}. This carrier is typically set by the Larmor frequency of the individual electron $g$-factor $g_e$, here it is set by the mean of the $\delta g$ distribution. In this rotating frame, the qubit control Hamiltonian appears as
\begin{equation}\label{Eq.Qubit}
    H_Q = I_0(t)\sigma_x + Q_0(t)\sigma_y + \delta \omega \sigma_z,
\end{equation}
where $\sigma_i, i\in\{x,y,z\}$ are the Pauli matrices, and $\delta \omega$ represents the frequency detuning of any single qubit from the mean of the $
\delta g$ distribution. Amplitude modulation can be employed as a method to expand the number of resonant peaks in the frequency domain from a single peak at $\omega_0$, to multiple sets of side-bands that are frequency-shifted copies of the pulse envelope set by $I_0(t)$ and $Q_0(t)$.
An amplitude modulation scheme which can be utilised to increase the number of resonant peaks output by the cavity from a single peak, to $N$ peaks separated by frequency $\omega_{AM}$ and centred at $\omega_0$ is detailed further in Supplementary Note~\ref{SupNote:AM_Scheme}. Thus, the Stark shift can operate such that $\delta \omega = 0$ for the qubit Hamiltonian in Eq.~\eqref{Eq.Qubit} by tuning the resonant frequency of each electron to the nearest side-band within the set $\omega_0\pm n\cdot \omega_{AM}$ for $n\in[0,(N-1)/2]$ (odd $N$) or $\omega_0\pm (n+1/2)\cdot \omega_{AM}$ $n\in[0,N/2]$ (even $N$).
It is expected that $N\simeq 10 - 100$ is required to span the frequency bandwidth created by the range of $g$-factor distribution $\delta g = \mathcal{O}(10^{-2})$, given a stark shift range of $\Delta g \simeq \mathcal{O}(10^{-4}) - \mathcal{O}(10^{-3})$.

Recent studies have shown how Gradient Ascent Pulse Engineering (GRAPE) techniques~\cite{khaneja2005optimal,de2011second} have been utilised to design broadband pusles that can account for local environmental noise, pushing single qubit control fidelity to the limit of incoherence~\cite{yang2019silicon}. For systems in which individual qubit tunability is necessary for the uniformity of global control signals across a large array of qubits, these GRAPE methods can also be viewed as a scheme for mitigating errors in qubit tuning. Figure~\ref{fig:CavityControl}a) shows a GRAPE solution for the $I_0(t)$ and $Q_0(t)$ envelopes required to produce a Hadamard gate, delivering a high fidelity operation over a larger bandwidth for robustness against tuning errors or small drifts in $\delta g$. The offset error in the tuning via the Stark shift $\Delta g$ which the gate can tolerate for given target fidelity can be seen in Fig.~\ref{fig:CavityControl}b). The solution results in a more complex trajectory over the Bloch sphere as seen in Supplementary Figure~\ref{fig:GrapeSpheres}, requiring 6$\times$ the time of a $\pi$ rotation produced by a square pulse with the same amplitude limits (set as a 1~\textmu s square pulse $\pi$ rotation at $B_0 = 1$T). The resulting GRAPE pulse has better performance against low frequency noise coupled via the $\delta \omega$ term in Eq.~\eqref{Eq.Qubit} as illustrated by the first order filter transfer function in detuning $F^{(1)}_z$~\cite{green2013arbitrary} as seen in Fig.~\ref{fig:CavityControl}f). Similar to the methods in Ref.~\cite{yang2019silicon}, the operation is trained against realistic qubit environmental noise~\cite{chan2018assessment}. While GRAPE methods can result in frequency-broadened qubit drive-lines offering a robustness to tuning errors, combining this with amplitude modulation results in the presence of multiple broadened side-bands that can also interfere with each-other. The infidelity of N=10 simultaneously driven peaks as a function of modulation frequency $\omega_{AM}$ is seen in Figure~\ref{fig:CavityControl}c), showing disruptive levels interference between side-bands under small $\omega_{AM}$ peak separations, however a much smaller separation in $\omega_{AM}$ is seen to be required by a solution utilising GRAPE when compared to the square pulse solution. Figure~\ref{fig:Cavity_AM10Env} illustrates the full infidelity-envelope for the amplitude modulation method described in Eq.~\eqref{Eq.AM}, where N=10 peaks are simultaneously produced. A limit of $\omega_{AM} > 4$MHz is observed to gain a region of Hadamard infidelity $1-F_H < 1\times 10^{-3}$ corresponding to a $g$-factor tunability range of $\Delta g \simeq 4.5\times 10^{-4}$ to move between two neighbouring peaks, or equivalently $\Delta V = 22.5$mV utilising $\partial g/\partial V = 0.002/V$ motivated from Ref.~\cite{veldhorst2015two,hwang2017impact}. The requirement on the Stark shift of $\Delta g \simeq 4.5\times 10^{-4}$ falls within the expected range of $\mathcal{O}(10^{-4}) - \mathcal{O}(10^{-3})$, although this threshold could be improved through more advanced optimisation acting to reduce the cross-talk. 

Pulse engineering methods can also be utilised for two-qubit gates~\cite{ball2021software}, where the globally applied ESR field can assist in decoupling action from local qubit noise. The mediated exchange energy in Eq.~\eqref{Eq.ExMed} can be incorporated as a voltage controlled signal $E_J(t) \propto J_{DA}$ in the following two qubit Hamiltonian:
\begin{align*}
    H_{Q2} = & I_0(t)(\sigma_x\otimes\mathcal{I} + \mathcal{I}\otimes\sigma_x) \\
    & + Q_0(t)(\sigma_y\otimes\mathcal{I} + \mathcal{I}\otimes\sigma_y) \\
    & +\delta \omega_1 (\sigma_z\otimes\mathcal{I}) +\delta\omega_2 (\mathcal{I}\otimes\sigma_z) \\
    & + E_J(t)\left( \begin{matrix}
            0 & 0 & 0 & 0 \\
            0 & -1 & 0 &0\\
            0 & 0 & -1 &0\\
            0 & 0 & 0 & 0
            \end{matrix} \right),\tag{3}\label{Eq.Qubit2}
\end{align*}
where $\mathcal{I}$ is the identity matrix in the single qubit subspace and $\otimes$ represents the Kronecker product. The component of the Hamiltonian in Eq.~\eqref{Eq.Qubit2} responsible for interaction between spins is valid for $| E_J(t) |_{\rm max} \ll E_Z$,~\cite{meunier2011efficient} and has been demonstrated to produce effective CZ operations~\cite{veldhorst2015two,watson2018programmable}. The pulse shapes in Fig.~\ref{fig:CavityControl}d have been designed such that when $E_J(t)$ is applied with global $I_0(t)$ and $Q_0(t)$ signals, the resulting operation is a decoupled CZ operation. In contrast, when $E_J(t)$ is not applied, i.e. for $E_J(t) = 0$, the global $I_0(t)$ and $Q_0(t)$ signals result instead in a decoupling identity gate applied to both independent spins. The infidelity of the decoupled CZ and Identity gates as a function of qubit detunings $\delta \omega_1$ or $\delta \omega_2$ (referred to a single electron $\delta g$ offset) is shown in Fig.~\ref{fig:CavityControl}e). The GRAPE solution is contrasted against a simple square pulse in $E_J$, showing improved robustness against tuning errors up to fidelity targets of $~99.99\%$, which is above the threshold for a surface code implementation~\cite{fowler2012surface}. The resulting 1Q and 2Q GRAPE pulses also have improved performance against low frequency noise coupled via the $\delta \omega$ terms in Eq.~\eqref{Eq.Qubit} and Eq.~\eqref{Eq.Qubit2} compared to Square-pulse implementations, as illustrated by the first order filter transfer function in detuning $F^{(1)}_z$~\cite{green2013arbitrary,ball2021software} as seen in Fig.~\ref{fig:CavityControl}f).  

\section{Integrated Qubit trimming circuitry}\label{SSec:Trim}
\vspace{-4mm}

\begin{figure}[t]
 \centering
  \begin{center}
  \includegraphics[width=9cm]{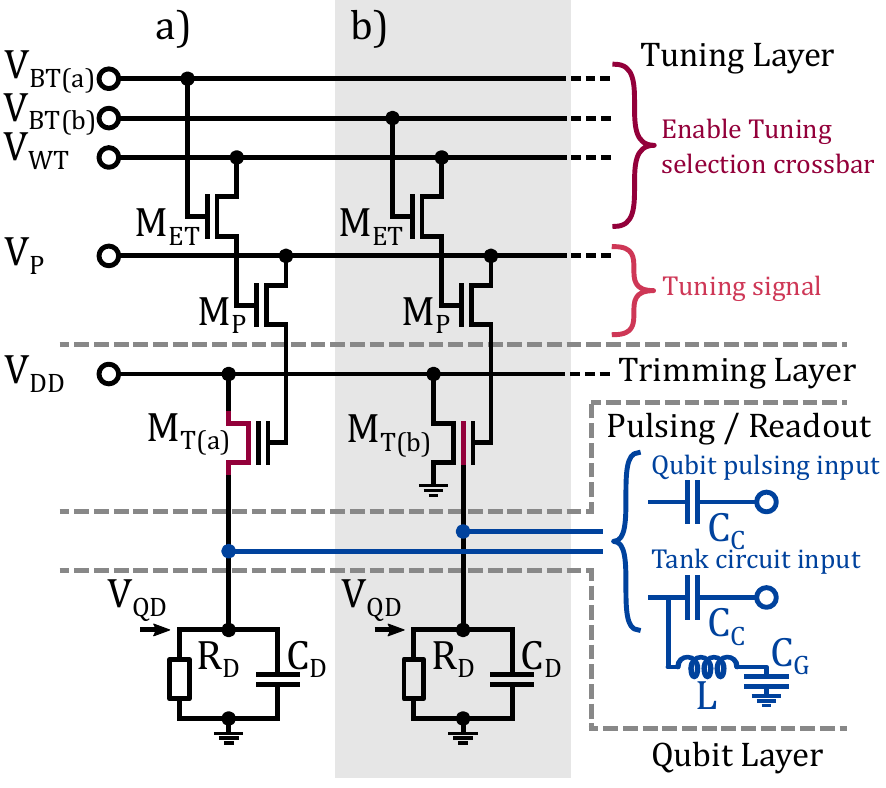}
  \end{center}
  \vspace*{-6mm}
  \caption{\textbf{$\vert$ In-situ non-volatile quantum dot voltage trimming circuit.} Schematic of the embedded circuitry for trimming individual voltages applied to a quantum dot (equivalent circuit shown in the Qubit Layer). A Tuning Layer consisting of a crossbar array allows for the selected enabling of the tuning via FETs $M_{ET}$, through word-line and bit-line addressing. By enabling $M_P$, the tuning signal $V_P$ is passed to the target FET(s) $M_T$ that exist within the Trimming Layer. Suitable $M_T$ behaviour is similar to that of floating-gate MOSFETs, which are non-volatile and can be tuned in a quasi-analogue fashion through control signals applied at the $V_P$ terminal. In \textbf{(a)}, the $M_T$ device is in series with the quantum dot, such that the voltage $V_{QD}$ is reduced from $V_{DD}$ due to the voltage division between channel resistance $M_T$ and total down-stream resistance $R_D$. Conversely, in \textbf{(b)}, the $M_T$ device can be directly integrated with the quantum dot, such that the top-gate electrode of the quantum dot is controlled directly by an element in $M_T$ that which holds an adjustable voltage. Any necessary control signals such as those used for pulsing and readout are AC-coupled to the quantum dots, shown as the optional (blue) connections.}\label{fig:Trimmer}
\end{figure}

The introduction of the Stark shift as a means of tuning individual spins to a nearby resonance peak gains the ability to perform global qubit control. However this approach results in the transformation of the problem from simultaneously controlling an ensemble of qubit frequencies into one of requiring to provide as many tuned voltage references as there are qubits in the quantum machine. In effect, transferring the scaling problem from a ``frequency crowding'' problem into a ``signal bottleneck'' problem~\cite{franke2019rent}. 

In order to achieve complete uniformity across all unit cells, the device variation must be addressable at the intra-cell level, resulting in a need for the integration of circuitry fit to address this variation. Here, the additional integrated circuitry is referred to as a ``trimmer'' circuit and is proposed to consist of a flash-memory-like device. Such a solution can take advantage of desirable characteristics including long-term stability and non-volatility of these integrated circuit elements, to retain voltage set-points over the lifetime of the quantum machine~\cite{hasler2021cryogenic}. Figure~\ref{fig:Trimmer} illustrates a circuit schematic diagram showing two potential methods for the working principles of the trimming circuit; Fig.~\ref{fig:Trimmer}a) which operates as a source follower (buffer circuit) configuration through reduction of the single setpoint voltage $V_{DD}$ through resistive division and Fig.~\ref{fig:Trimmer}b) which operates through supplying the stored potential as a direct reference for the quantum device. The resistive element $R_D$ within the equivalent circuit of a single quantum dot, is produced by the cumulative gate leakage to ground present in CMOS processes~\cite{schaal2018conditional}. The circuit elements $M_{\rm T(a,b)}$ shown in Figure~\ref{fig:Trimmer} are central to performing the trimming, possessing salient characteristics similar to that of a floating-gate MOSFET including a tunable threshold voltage and non-volatility.

By trimming the threshold voltage of the device, the channel resistance in deep sub-threshold operation of the device $M_{\rm T(a)}$ in Fig.~\ref{fig:Trimmer}a) can become comparable to $R_D$, resulting in an active voltage division which reduces the value of $V_{DD}$ down to some desired value (which is presumed in this instance to target a qubit resonance line though the Stark shift discussed in \S~\ref{SSec:Cavity}). A more detailed feasibility study of this resistive trimmer configuration is presented in Supplementary Note~\ref{SupNote:Trimmer_resistive}. The threshold voltage is tuned through charge storage within the device, converted to a voltage through the capacitance relationship. For $M_{\rm T(b)}$ in Fig.~\ref{fig:Trimmer}b), the stored charge within the device can be more directly converted into a voltage reference in a similar fashion to a previous proposal utilising dynamic random access memory (DRAM)~\cite{veldhorst2017silicon}. Here, the voltage reference is delivered directly to the quantum device through a direct connection between the gate electrode defining the quantum dot and the charge storage element within the memory component $M_{\rm T(b)}$.
For the solution shown in Figure~\ref{fig:Trimmer}, three additional MOS devices must be integrated for each trimmed object within the unit cell. A cross-bar addressing scheme is utilised for individual trimming devices $M_{\rm T(a,b)}$ across the unit cell, which becomes active during a pre-computation tuning phase for the quantum machine. The threshold voltage is selectively tuned through signal input $V_P$ which is connected, through activated $M_P$, to the $M_T$ trimming devices. Word-line and bit-line voltage $V_{WT}$ and $V_{BT}$ combine through tuning-enable transistors $M_{ET}$, activating a selected $M_P$. Based on the hardware unit cell architecture discussed in \S~\ref{SSec:EdgeDot}, the elongated mediator dots which interconnect data and ancilla qubits allow for a certain amount of physical space~\cite{cai2019silicon} in-between the silicon plateaus where these devices can be laid out in the FEOL layer, avoiding vertical circuit integration in the form of stacking control transistor layers above the the quantum FEOL~\cite{veldhorst2017silicon}. The characteristic of non-volatility assumes operation within a consistent thermal environment and is deliverable by standard MOS memory hardware~\cite{hasler2021cryogenic}. This is essential for the device $M_{\rm T(a,b)}$ as, after tuning each device to the desired set-point, the selector crossbar-architecture as shown in Fig.~\ref{fig:Trimmer} becomes idle, allowing the circuitry to be powered down more completely, reducing latent power consumption or heat load. Stability characteristics of the device must be such that the set memory state does not drift appreciably over time, as this would contribute to errors in $\Delta g$. In this instance the stability requirements are defined by the high-fidelity region of the broadband pulses discussed in \S~\ref{SSec:Cavity}.  

\section*{Discussion}

Here, a solution is proposed to manage resources at the quantum-classical interface within a scaled processor through integration of MOS structures at the FEOL quantum hardware level. The additional hardware is required to combat device-to-device variation which is a principle challenge when scaling quantum machines~\cite{laucht2021roadmap}. The different aspect ratio between the qubit layer and the data processing layers can also complicate this quantum-classical interface~\cite{veldhorst2017silicon}, which can be avoided through the use of elongated mediator dots~\cite{cai2019silicon} or shutting qubits through 1D dot chains~\cite{boter2019sparse}, as well as de-embedding single qubit control to be executed globally via a 3D microwave cavity~\cite{vahapoglu2020single}. This scaled quantum machine can be achieved through the hybridisation of SOI nanowire technologies~\cite{hutin2019soi} with planar quantum dot structures~\cite{yang2013spin}, producing an ``edge dot'' platform where the quantum dots are defined at the geometric boundary of a raised silicon plateau. 

The hardware cell presented here is also extremely flexible, with the capacity to be re-configured to operate the surface code for different control schemes. For example, forgoing globally applied ESR of electron spins in the pursuit of an all-electrical control scheme can still utilise this same hardware architecture, with the addition of integrated micromagnet arrays~\cite{singh2020quantum} into the FEOL. This approach can potentially utilise the space outside the plateau in each alternating cell to incorporate diagonally aligned micromagnets between the data qubit locations. This produces an engineered magnetic field gradient across each data-qubit double-dot, which can facilitate single qubit rotations~\cite{pioro2008electrically}. In combination with the capability of tuning the stark shift via the trimming circuit, the amplitude modulation scheme presented here is also directly transferable to a globally applied EDSR control signal for this implementation. 

The type of qubits used in this hardware is also flexibly defined, based on the configuration and number of charges within each double-dot site. It is shown here that this architecture can facilitate a co-existing combination of single-electron spin qubits and two-electron Singlet-T$_0$ Triplet qubits. A similar micromagnet configuration as the one discussed above could also be utilised for an all-electrical control implementation with both data and syndrome qubits defined in the singlet-triplet basis~\cite{wu2014two}. However, it is noted that approaches involving micromagnets can constitute a deviation from the materials used in the CMOS industry. Without the integrated micromagnet array, all-electrical control can still be achieved in silicon via several implementations. Single-hole spins with EDSR control~\cite{maurand2016cmos} leverage higher spin-orbit couplings compared to electrons and singlet-triplet qubits can rely on the naturally present spin-orbit coupling for single qubit rotations~\cite{jock2017probing,jock2021silicon}. Other all-electrical qubit species involve a (2,1) electron occupancy for the hybrid qubit implementation~\cite{shi2012fast} and (1,0) occupancy for qubits defined in the charge basis~\cite{hayashi2003coherent}. Note that for each of these qubit implementations listed above, tailored control schemes implementing the achievable gate-sets for these qubit varieties must be devised and are considered out of the current scope of this study.

A unit cell approach to constructing scalable quantum information processors benefits from a drasitic reduction in input overheads due to high levels of parallelisation between each cell. These hardware-based unit cells can also strongly compliment the tile-like nature of many error correction codes applied to 2D lattices of qubits. In order to successfully carry out this goal, it is necessary to tailor the design and execution towards the use of global signal control strategies such as noise-robust pulses and parallelisation schemes. From the hardware perspective, this also requires the integration of robustness against inter-cell device variations.
For any scaled qubit implementation, the variation in qubit control parameters must be overcome. In the case of single electron spin qubits, utilising a 3D cavity as a control peripheral~\cite{vahapoglu2020single} in which the silicon chip is embedded allows for many qubit to be addressed across a large spatial range. However, contemporary results in planar MOS devices show the expected distribution in the electron $g$-factor~\cite{ferdous2018interface} far exceeds the range available through Stark-shift tunability~\cite{veldhorst2015two,hwang2017impact,huang2019fidelity}. Thus, the concept of tuning via stark shifts must be augmented when operating at appreciable magnetic fields for global operations to be applied to the spin ensemble. Here, the solution presented involves the use of amplitude modulation for the production of discrete side-bands, with the Stark-shift providing individual $g$-factor tuning towards the nearest band. This approach is subject to a trade-off between the separation between the side-bands given a certain tunable range in $g$-factor, and the cross-talk between side-bands observed at small separations. As shown here, the same engineered pulses which increase robustness against small tuning deviations in the $g$-factor can also result in reduced cross-talk between side-bands compared to equivalent square-pulse implementations. With the added design element of intentionally reducing cross-talk between side-bands within the optimisation process, this trade-off between cross-talk and Stark-shift could be further improved.  
The signals executing the CZ operation can also be made compatible with global control operations through the electrical tuning of the exchange energy between features in the unit cell. Tuning the Stark shift will result in alterations to the $\varepsilon_{\{D,A\}M}$ terms in Eq.~\eqref{Eq.ExMed}, however by tuning of the potential on the barrier gates situated between the Dots and the Mediators in Fig.~\ref{fig:SC_Unit}b) can compensate via directly tuning the $t^2_{\{D,A\}M} / \varepsilon_{\{D,A\}M}$ ratio, ensuring uniformity in $J_{DA}$ across all sites. The magnitude of $J_{DA}$ can then be modulated through signal $V_M$ applied to the Mediator accumulation gate, which has a linear relationship to both $\varepsilon_{DM}$ and $\varepsilon_{AM}$, resulting in $J_{DA}\propto V_M^{-2}$

For the global control solutions presented here, the surface code cycle is executed within $\sim 46$~\textmu s ($5\times\tau_H+4\times\tau_{CZ}$), assuming near-negligible electron shuttling times~\cite{yoneda2021coherent}. This is well below the state-of-the-art single-spin coherence time of $T_2^{RB} = 9.4$~ms~\cite{yang2019silicon} as derived from a Randomized Benchmarking experiment utilising GRAPE pulses. The collective ancilla qubit measurement and initialisation time for this protocol is 12-28~\textmu s depending on the implementation. Current measurements in silicon nanowires have lead to the determination of a PSB signal with $>$99\% fidelity within 5.6~\textmu s~\cite{oakes2022fast}, leaving approximately a 6-22~\textmu s budget for $\ket{S}$-state initialisation via mediator/reservoirs in the first instance. This time budget can also be extended over multiple code-cycles if necessary, through adopting plaquette sequencing protocols discussed in Ref.~\cite{cai2019silicon}.

Focusing on the mature CMOS industry for the development of quantum processors has the core benefit of being able to draw upon many different advancements and techniques for information processing and storage~\cite{gonzalez2020scaling}. A similar approach for addressing device variation includes embedding a quantum machine into a DRAM-style circuit~\cite{veldhorst2017silicon}, where the 6T3C/dot cell stores a pre-tuned voltage supplied from a variable source on a capacitor near each quantum dot, and refreshed over a cyclic period. For this approach, the hold-capacitor must be large enough for a sufficiently stable voltage (and thus resonant frequency via the Stark shift in the $g$-factor) for high fidelity qubit operations. Conversely, this capacitive element must also be small enough not to dominate the integrated dispersive readout signals~\cite{schaal2019cmos}. In contrast, the method presented here proposes the specific integration of non-volatile memory elements for the storage of these pre-tuned voltage settings. These voltage references are continually applied by the non-volatile elements and are therefore not subject to the same limitations set by capacitive decay constants or cycle-to-cycle variations, but will still be limited by Johnson noise generated by the voltage reference elements in a similar way. Both approaches, however, require a potentially high set-up cost represented by a pre-computation trimming phase which identifies the correct operating conditions for each qubit element, however an advantage for the implementation presented here is that this tuning circuitry has the option to become dormant during computation phases due to the non-volatility of the integrated memory elements, potentially reducing the latent power consumption of the device. 

With the addition of trimming internal to the hardware unit cell, the number of inputs required to set the state of the quantum machine can be drastically reduced through the interconnection of cells. This results in a hardware in which the number of inputs scales with the complexity of the unit cell, rather than the number of unit cells required to produce the quantum machine~\cite{franke2019rent}, ensuring the extensibility of the qubit platform.
The physical layout of the CMOS elements in the FEOL, and the routing between the elements at the quantum-classical interface, is considered to be out of scope for this initial work, however for the solution presented in \S~\ref{SSec:Trim}, 48 elements are required per unit cell to connect 16 structures which require trimming (the reservoir accumulation gate, and barriers between the mediators and reservoir are discounted here, as these do not require precise tuning to function). As the area between silicon plateaus is $\sim (8\times p)^2$ for the direct routing solution presented in Fig.~\ref{fig:SC_Unit}, where $p$ is the BEOL routing pitch, this affords an area budget of $\sim 1.3p^2$ per control element before the size of these elements impacts the length of the mediator dots which are $ \sim5\times p$ in this example. A potential alternative method for spin transfer would be to replace the mediator with a spin-shuttling chain of quantum dots, which has been studied elsewhere~\cite{boter2019sparse}, and has different trade-offs regarding increased numbers of electrodes and control signals.

While the extension of this architecture towards lattice surgery methods is also considered out of the scope for a study of the individual hardware cell, it is noted that a deviation from globally applied electrical control signals towards grouping regions of the surface into distinct areas, perhaps governed by separate DACs, can facilitate lattice splitting and merging required for lattice surgery~\cite{horsman2012surface}. For example, inactivity of CZ mediator signals along a selected row/column results in dormant regions in the surface structure, producing a split between two distinct regions. Further, it is also feasible to have intermittently placed, dedicated control circuitry offsetting entire regions of quantum unit cells in the FEOL, as the surface code has been shown to be robust against both time-resolved, and/or spatial interruptions defects~\cite{strikis2021quantum}. The exact geometric topology and layout of the quantum VS classical regions in the expanded FEOL is considered to be beyond the scope of the unit cell as studied here.     

\section*{Conclusion}
Defining a hardware unit cell which is complimentary to a specific stabiliser code cycle can lead to a highly parallelised approach to the execution of large scale quantum information processing. Here, a case study is presented which utilises a hybrid between two prevalent silicon MOS technologies, combining the advantages of current state-of-the-art solid-state quantum hardware approaches with memory storage techniques. The choice of design to include integrated mediator quantum dots makes the  stored quantum information additionally robust against leakage error types which cannot be protected against through standard quantum error detection protocols. The adapted code-cycle presented here allows for the implementation of globally applied single-qubit rotations across the entire ensemble of qubit resonant frequencies, as well as selectively applied two-qubit CZ operations from a global signal source. The inclusion of non-volatile memory elements within the hardware unit cell also reduces the signal overheads to expand with the unit cell size rather than with the number of cells. The result is a complete unit cell approach to constructing a robust quantum information processing machine of arbitrary scale in silicon.

\section*{Acknowledgments}
\vspace*{-5mm}
The author would like to thank J.~J.~L.~Morton, M.~F.~Gonzalez-Zalba, S.~C.~Benjamin and Z.~Cai for valuable discussions and comments on the manuscript\\
\section*{Author Information}
\vspace*{-5mm}
The Author is supported by Quantum Motion, a start-up developing silicon-based quantum computing.
\nolinenumbers

\hyphenpenalty=10000
\bibliography{EdgeDotArchitecture.bib}

\begin{thebibliography}{84}%
\makeatletter
\providecommand \@ifxundefined [1]{%
 \@ifx{#1\undefined}
}%
\providecommand \@ifnum [1]{%
 \ifnum #1\expandafter \@firstoftwo
 \else \expandafter \@secondoftwo
 \fi
}%
\providecommand \@ifx [1]{%
 \ifx #1\expandafter \@firstoftwo
 \else \expandafter \@secondoftwo
 \fi
}%
\providecommand \natexlab [1]{#1}%
\providecommand \enquote  [1]{``#1''}%
\providecommand \bibnamefont  [1]{#1}%
\providecommand \bibfnamefont [1]{#1}%
\providecommand \citenamefont [1]{#1}%
\providecommand \href@noop [0]{\@secondoftwo}%
\providecommand \href [0]{\begingroup \@sanitize@url \@href}%
\providecommand \@href[1]{\@@startlink{#1}\@@href}%
\providecommand \@@href[1]{\endgroup#1\@@endlink}%
\providecommand \@sanitize@url [0]{\catcode `\\12\catcode `\$12\catcode
  `\&12\catcode `\#12\catcode `\^12\catcode `\_12\catcode `\%12\relax}%
\providecommand \@@startlink[1]{}%
\providecommand \@@endlink[0]{}%
\providecommand \url  [0]{\begingroup\@sanitize@url \@url }%
\providecommand \@url [1]{\endgroup\@href {#1}{\urlprefix }}%
\providecommand \urlprefix  [0]{URL }%
\providecommand \Eprint [0]{\href }%
\providecommand \doibase [0]{http://dx.doi.org/}%
\providecommand \selectlanguage [0]{\@gobble}%
\providecommand \bibinfo  [0]{\@secondoftwo}%
\providecommand \bibfield  [0]{\@secondoftwo}%
\providecommand \translation [1]{[#1]}%
\providecommand \BibitemOpen [0]{}%
\providecommand \bibitemStop [0]{}%
\providecommand \bibitemNoStop [0]{.\EOS\space}%
\providecommand \EOS [0]{\spacefactor3000\relax}%
\providecommand \BibitemShut  [1]{\csname bibitem#1\endcsname}%
\let\auto@bib@innerbib\@empty
\bibitem [{\citenamefont {Terhal}(2015)}]{terhal2015quantum}%
  \BibitemOpen
  \bibfield  {author} {\bibinfo {author} {\bibfnamefont {B.~M.}\ \bibnamefont
  {Terhal}},\ }\bibfield  {title} {\enquote {\bibinfo {title} {Quantum error
  correction for quantum memories},}\ }\href@noop {} {\bibfield  {journal}
  {\bibinfo  {journal} {Reviews of Modern Physics}\ }\textbf {\bibinfo {volume}
  {87}},\ \bibinfo {pages} {307} (\bibinfo {year} {2015})}\BibitemShut
  {NoStop}%
\bibitem [{\citenamefont {O'Gorman}\ and\ \citenamefont
  {Campbell}(2017)}]{ogorman2017quantum}%
  \BibitemOpen
  \bibfield  {author} {\bibinfo {author} {\bibfnamefont {J.}~\bibnamefont
  {O'Gorman}}\ and\ \bibinfo {author} {\bibfnamefont {E.~T.}\ \bibnamefont
  {Campbell}},\ }\bibfield  {title} {\enquote {\bibinfo {title} {Quantum
  computation with realistic magic-state factories},}\ }\href@noop {}
  {\bibfield  {journal} {\bibinfo  {journal} {Physical Review A}\ }\textbf
  {\bibinfo {volume} {95}},\ \bibinfo {pages} {032338} (\bibinfo {year}
  {2017})}\BibitemShut {NoStop}%
\bibitem [{\citenamefont {Hutin}\ \emph
  {et~al.}(2019{\natexlab{a}})\citenamefont {Hutin}, \citenamefont {Bertrand},
  \citenamefont {Niquet}, \citenamefont {Hartmann}, \citenamefont {Sanquer},
  \citenamefont {De~Franceschi}, \citenamefont {Meunier},\ and\ \citenamefont
  {Vinet}}]{hutin2019soi}%
  \BibitemOpen
  \bibfield  {author} {\bibinfo {author} {\bibfnamefont {L.}~\bibnamefont
  {Hutin}}, \bibinfo {author} {\bibfnamefont {B.}~\bibnamefont {Bertrand}},
  \bibinfo {author} {\bibfnamefont {Y.-M.}\ \bibnamefont {Niquet}}, \bibinfo
  {author} {\bibfnamefont {J.-M.}\ \bibnamefont {Hartmann}}, \bibinfo {author}
  {\bibfnamefont {M.}~\bibnamefont {Sanquer}}, \bibinfo {author} {\bibfnamefont
  {S.}~\bibnamefont {De~Franceschi}}, \bibinfo {author} {\bibfnamefont
  {T.}~\bibnamefont {Meunier}}, \ and\ \bibinfo {author} {\bibfnamefont
  {M.}~\bibnamefont {Vinet}},\ }\bibfield  {title} {\enquote {\bibinfo {title}
  {Soi mos technology for spin qubits},}\ }\href@noop {} {\bibfield  {journal}
  {\bibinfo  {journal} {ECS Transactions}\ }\textbf {\bibinfo {volume} {93}},\
  \bibinfo {pages} {35} (\bibinfo {year} {2019}{\natexlab{a}})}\BibitemShut
  {NoStop}%
\bibitem [{\citenamefont {Li}\ \emph {et~al.}(2020)\citenamefont {Li},
  \citenamefont {Stuyck}, \citenamefont {Kubicek}, \citenamefont {Jussot},
  \citenamefont {Chan}, \citenamefont {Mohiyaddin}, \citenamefont {Elsayed},
  \citenamefont {Shehata}, \citenamefont {Simion}, \citenamefont {Godfrin},
  \citenamefont {Canvel}, \citenamefont {Ivanov}, \citenamefont {Goux},
  \citenamefont {Govoreanu},\ and\ \citenamefont {Radu}}]{li2020flexible}%
  \BibitemOpen
  \bibfield  {author} {\bibinfo {author} {\bibfnamefont {R.}~\bibnamefont
  {Li}}, \bibinfo {author} {\bibfnamefont {N.~I.~D.}\ \bibnamefont {Stuyck}},
  \bibinfo {author} {\bibfnamefont {S.}~\bibnamefont {Kubicek}}, \bibinfo
  {author} {\bibfnamefont {J.}~\bibnamefont {Jussot}}, \bibinfo {author}
  {\bibfnamefont {B.~T.}\ \bibnamefont {Chan}}, \bibinfo {author}
  {\bibfnamefont {F.~A.}\ \bibnamefont {Mohiyaddin}}, \bibinfo {author}
  {\bibfnamefont {A.}~\bibnamefont {Elsayed}}, \bibinfo {author} {\bibfnamefont
  {M.}~\bibnamefont {Shehata}}, \bibinfo {author} {\bibfnamefont
  {G.}~\bibnamefont {Simion}}, \bibinfo {author} {\bibfnamefont
  {C.}~\bibnamefont {Godfrin}}, \bibinfo {author} {\bibfnamefont
  {Y.}~\bibnamefont {Canvel}}, \bibinfo {author} {\bibfnamefont
  {T.}~\bibnamefont {Ivanov}}, \bibinfo {author} {\bibfnamefont
  {L.}~\bibnamefont {Goux}}, \bibinfo {author} {\bibfnamefont {B.}~\bibnamefont
  {Govoreanu}}, \ and\ \bibinfo {author} {\bibfnamefont {I.~P.}\ \bibnamefont
  {Radu}},\ }\bibfield  {title} {\enquote {\bibinfo {title} {A flexible 300 mm
  integrated si mos platform for electron-and hole-spin qubits exploration},}\
  \ }(\bibinfo {organization} {IEEE},\ \bibinfo {year} {2020})\ pp.\ \bibinfo
  {pages} {38.3.1--38.3.4}\BibitemShut {NoStop}%
\bibitem [{\citenamefont {Zwerver}\ \emph {et~al.}(2022)\citenamefont
  {Zwerver}, \citenamefont {Kr{\"a}henmann}, \citenamefont {Watson},
  \citenamefont {Lampert}, \citenamefont {George}, \citenamefont
  {Pillarisetty}, \citenamefont {Bojarski}, \citenamefont {Amin}, \citenamefont
  {Amitonov}, \citenamefont {Boter}, \citenamefont {Caudillo}, \citenamefont
  {Corras-Serrano}, \citenamefont {Dehollain}, \citenamefont {Droulers},
  \citenamefont {Henry}, \citenamefont {Kotlyar}, \citenamefont {Lodari},
  \citenamefont {Luthi}, \citenamefont {Michalak}, \citenamefont {Mueller},
  \citenamefont {Neyens}, \citenamefont {Roberts}, \citenamefont {Samkharadze},
  \citenamefont {Zheng}, \citenamefont {Zietz}, \citenamefont {Veldhorst},
  \citenamefont {Vandersypen},\ and\ \citenamefont
  {Clarke}}]{zwerver2021qubits}%
  \BibitemOpen
  \bibfield  {author} {\bibinfo {author} {\bibfnamefont {A.~M.~J.}\
  \bibnamefont {Zwerver}}, \bibinfo {author} {\bibfnamefont {T.}~\bibnamefont
  {Kr{\"a}henmann}}, \bibinfo {author} {\bibfnamefont {T.~F.}\ \bibnamefont
  {Watson}}, \bibinfo {author} {\bibfnamefont {L.}~\bibnamefont {Lampert}},
  \bibinfo {author} {\bibfnamefont {H.~C.}\ \bibnamefont {George}}, \bibinfo
  {author} {\bibfnamefont {R.}~\bibnamefont {Pillarisetty}}, \bibinfo {author}
  {\bibfnamefont {S.~A.}\ \bibnamefont {Bojarski}}, \bibinfo {author}
  {\bibfnamefont {P.}~\bibnamefont {Amin}}, \bibinfo {author} {\bibfnamefont
  {S.~V.}\ \bibnamefont {Amitonov}}, \bibinfo {author} {\bibfnamefont {J.~M.}\
  \bibnamefont {Boter}}, \bibinfo {author} {\bibfnamefont {R.}~\bibnamefont
  {Caudillo}}, \bibinfo {author} {\bibfnamefont {D.}~\bibnamefont
  {Corras-Serrano}}, \bibinfo {author} {\bibfnamefont {J.~P.}\ \bibnamefont
  {Dehollain}}, \bibinfo {author} {\bibfnamefont {G.}~\bibnamefont {Droulers}},
  \bibinfo {author} {\bibfnamefont {E.~M.}\ \bibnamefont {Henry}}, \bibinfo
  {author} {\bibfnamefont {R.}~\bibnamefont {Kotlyar}}, \bibinfo {author}
  {\bibfnamefont {M.}~\bibnamefont {Lodari}}, \bibinfo {author} {\bibfnamefont
  {F.}~\bibnamefont {Luthi}}, \bibinfo {author} {\bibfnamefont {D.~J.}\
  \bibnamefont {Michalak}}, \bibinfo {author} {\bibfnamefont {B.~K.}\
  \bibnamefont {Mueller}}, \bibinfo {author} {\bibfnamefont {S.}~\bibnamefont
  {Neyens}}, \bibinfo {author} {\bibfnamefont {J.}~\bibnamefont {Roberts}},
  \bibinfo {author} {\bibfnamefont {N.}~\bibnamefont {Samkharadze}}, \bibinfo
  {author} {\bibfnamefont {G.}~\bibnamefont {Zheng}}, \bibinfo {author}
  {\bibfnamefont {G.}~\bibnamefont {Zietz}, \bibfnamefont {O.~K.~Scappucci}},
  \bibinfo {author} {\bibfnamefont {M.}~\bibnamefont {Veldhorst}}, \bibinfo
  {author} {\bibfnamefont {L.~M.~K.}\ \bibnamefont {Vandersypen}}, \ and\
  \bibinfo {author} {\bibfnamefont {J.~S.}\ \bibnamefont {Clarke}},\ }\bibfield
   {title} {\enquote {\bibinfo {title} {Qubits made by advanced semiconductor
  manufacturing},}\ }\href@noop {} {\bibfield  {journal} {\bibinfo  {journal}
  {Nature Electronics}\ }\textbf {\bibinfo {volume} {5}},\ \bibinfo {pages}
  {184--190} (\bibinfo {year} {2022})}\BibitemShut {NoStop}%
\bibitem [{\citenamefont {Vandersypen}\ \emph {et~al.}(2017)\citenamefont
  {Vandersypen}, \citenamefont {Bluhm}, \citenamefont {Clarke}, \citenamefont
  {Dzurak}, \citenamefont {Ishihara}, \citenamefont {Morello}, \citenamefont
  {Reilly}, \citenamefont {Schreiber},\ and\ \citenamefont
  {Veldhorst}}]{vandersypen2017interfacing}%
  \BibitemOpen
  \bibfield  {author} {\bibinfo {author} {\bibfnamefont {L.~M.~K.}\
  \bibnamefont {Vandersypen}}, \bibinfo {author} {\bibfnamefont
  {H.}~\bibnamefont {Bluhm}}, \bibinfo {author} {\bibfnamefont {J.~S.}\
  \bibnamefont {Clarke}}, \bibinfo {author} {\bibfnamefont {A.~S.}\
  \bibnamefont {Dzurak}}, \bibinfo {author} {\bibfnamefont {R.}~\bibnamefont
  {Ishihara}}, \bibinfo {author} {\bibfnamefont {A.}~\bibnamefont {Morello}},
  \bibinfo {author} {\bibfnamefont {D.~J.}\ \bibnamefont {Reilly}}, \bibinfo
  {author} {\bibfnamefont {L.~R.}\ \bibnamefont {Schreiber}}, \ and\ \bibinfo
  {author} {\bibfnamefont {M.}~\bibnamefont {Veldhorst}},\ }\bibfield  {title}
  {\enquote {\bibinfo {title} {Interfacing spin qubits in quantum dots and
  donors—hot, dense, and coherent},}\ }\href@noop {} {\bibfield  {journal}
  {\bibinfo  {journal} {npj Quantum Information}\ }\textbf {\bibinfo {volume}
  {3}},\ \bibinfo {pages} {1--10} (\bibinfo {year} {2017})}\BibitemShut
  {NoStop}%
\bibitem [{\citenamefont {Gonzalez-Zalba}\ \emph {et~al.}(2021)\citenamefont
  {Gonzalez-Zalba}, \citenamefont {de~Franceschi}, \citenamefont {Charbon},
  \citenamefont {Meunier}, \citenamefont {Vinet},\ and\ \citenamefont
  {Dzurak}}]{gonzalez2020scaling}%
  \BibitemOpen
  \bibfield  {author} {\bibinfo {author} {\bibfnamefont {M.~F.}\ \bibnamefont
  {Gonzalez-Zalba}}, \bibinfo {author} {\bibfnamefont {S.}~\bibnamefont
  {de~Franceschi}}, \bibinfo {author} {\bibfnamefont {E.}~\bibnamefont
  {Charbon}}, \bibinfo {author} {\bibfnamefont {T.}~\bibnamefont {Meunier}},
  \bibinfo {author} {\bibfnamefont {M.}~\bibnamefont {Vinet}}, \ and\ \bibinfo
  {author} {\bibfnamefont {A.~S.}\ \bibnamefont {Dzurak}},\ }\bibfield  {title}
  {\enquote {\bibinfo {title} {Scaling silicon-based quantum computing using
  cmos technology: State-of-the-art, challenges and perspectives},}\
  }\href@noop {} {\bibfield  {journal} {\bibinfo  {journal} {Nature
  Electronics}\ }\textbf {\bibinfo {volume} {4}},\ \bibinfo {pages} {872--884}
  (\bibinfo {year} {2021})}\BibitemShut {NoStop}%
\bibitem [{\citenamefont {Pauka}\ \emph {et~al.}(2021)\citenamefont {Pauka},
  \citenamefont {Das}, \citenamefont {Kalra}, \citenamefont {Moini},
  \citenamefont {Yang}, \citenamefont {Trainer}, \citenamefont {Bousquet},
  \citenamefont {Cantaloube}, \citenamefont {Dick}, \citenamefont {Gardner},
  \citenamefont {Manfra},\ and\ \citenamefont {Reilly}}]{pauka2021cryogenic}%
  \BibitemOpen
  \bibfield  {author} {\bibinfo {author} {\bibfnamefont {S.~J.}\ \bibnamefont
  {Pauka}}, \bibinfo {author} {\bibfnamefont {K.}~\bibnamefont {Das}}, \bibinfo
  {author} {\bibfnamefont {R.}~\bibnamefont {Kalra}}, \bibinfo {author}
  {\bibfnamefont {A.}~\bibnamefont {Moini}}, \bibinfo {author} {\bibfnamefont
  {Y.}~\bibnamefont {Yang}}, \bibinfo {author} {\bibfnamefont {M.}~\bibnamefont
  {Trainer}}, \bibinfo {author} {\bibfnamefont {A.}~\bibnamefont {Bousquet}},
  \bibinfo {author} {\bibfnamefont {C.}~\bibnamefont {Cantaloube}}, \bibinfo
  {author} {\bibfnamefont {N.}~\bibnamefont {Dick}}, \bibinfo {author}
  {\bibfnamefont {G.~C.}\ \bibnamefont {Gardner}}, \bibinfo {author}
  {\bibfnamefont {M.~J.}\ \bibnamefont {Manfra}}, \ and\ \bibinfo {author}
  {\bibfnamefont {D.~J.}\ \bibnamefont {Reilly}},\ }\bibfield  {title}
  {\enquote {\bibinfo {title} {A cryogenic cmos chip for generating control
  signals for multiple qubits},}\ }\href@noop {} {\bibfield  {journal}
  {\bibinfo  {journal} {Nature Electronics}\ }\textbf {\bibinfo {volume} {4}},\
  \bibinfo {pages} {64--70} (\bibinfo {year} {2021})}\BibitemShut {NoStop}%
\bibitem [{\citenamefont {Veldhorst}\ \emph {et~al.}(2017)\citenamefont
  {Veldhorst}, \citenamefont {Eenink}, \citenamefont {Yang},\ and\
  \citenamefont {Dzurak}}]{veldhorst2017silicon}%
  \BibitemOpen
  \bibfield  {author} {\bibinfo {author} {\bibfnamefont {M.}~\bibnamefont
  {Veldhorst}}, \bibinfo {author} {\bibfnamefont {H.~G.~J.}\ \bibnamefont
  {Eenink}}, \bibinfo {author} {\bibfnamefont {C.-H.}\ \bibnamefont {Yang}}, \
  and\ \bibinfo {author} {\bibfnamefont {A.~S.}\ \bibnamefont {Dzurak}},\
  }\bibfield  {title} {\enquote {\bibinfo {title} {Silicon cmos architecture
  for a spin-based quantum computer},}\ }\href@noop {} {\bibfield  {journal}
  {\bibinfo  {journal} {Nature communications}\ }\textbf {\bibinfo {volume}
  {8}},\ \bibinfo {pages} {1--8} (\bibinfo {year} {2017})}\BibitemShut
  {NoStop}%
\bibitem [{\citenamefont {Li}\ \emph {et~al.}(2018)\citenamefont {Li},
  \citenamefont {Petit}, \citenamefont {Franke}, \citenamefont {Dehollain},
  \citenamefont {Helsen}, \citenamefont {Steudtner}, \citenamefont {Thomas},
  \citenamefont {Yoscovits}, \citenamefont {Singh}, \citenamefont {Wehner},
  \citenamefont {Vandersypen}, \citenamefont {Clarke},\ and\ \citenamefont
  {Veldhorst}}]{li2018crossbar}%
  \BibitemOpen
  \bibfield  {author} {\bibinfo {author} {\bibfnamefont {R.}~\bibnamefont
  {Li}}, \bibinfo {author} {\bibfnamefont {L.}~\bibnamefont {Petit}}, \bibinfo
  {author} {\bibfnamefont {D.~P.}\ \bibnamefont {Franke}}, \bibinfo {author}
  {\bibfnamefont {J.~P.}\ \bibnamefont {Dehollain}}, \bibinfo {author}
  {\bibfnamefont {J.}~\bibnamefont {Helsen}}, \bibinfo {author} {\bibfnamefont
  {M.}~\bibnamefont {Steudtner}}, \bibinfo {author} {\bibfnamefont {N.~K.}\
  \bibnamefont {Thomas}}, \bibinfo {author} {\bibfnamefont {Z.~R.}\
  \bibnamefont {Yoscovits}}, \bibinfo {author} {\bibfnamefont {K.~J.}\
  \bibnamefont {Singh}}, \bibinfo {author} {\bibfnamefont {S.}~\bibnamefont
  {Wehner}}, \bibinfo {author} {\bibfnamefont {L.~M.~K.}\ \bibnamefont
  {Vandersypen}}, \bibinfo {author} {\bibfnamefont {J.~S.}\ \bibnamefont
  {Clarke}}, \ and\ \bibinfo {author} {\bibfnamefont {M.}~\bibnamefont
  {Veldhorst}},\ }\bibfield  {title} {\enquote {\bibinfo {title} {A crossbar
  network for silicon quantum dot qubits},}\ }\href@noop {} {\bibfield
  {journal} {\bibinfo  {journal} {Science advances}\ }\textbf {\bibinfo
  {volume} {4}},\ \bibinfo {pages} {eaar3960} (\bibinfo {year}
  {2018})}\BibitemShut {NoStop}%
\bibitem [{\citenamefont {Hollenberg}\ \emph {et~al.}(2006)\citenamefont
  {Hollenberg}, \citenamefont {Greentree}, \citenamefont {Fowler},\ and\
  \citenamefont {Wellard}}]{hollenberg2006two}%
  \BibitemOpen
  \bibfield  {author} {\bibinfo {author} {\bibfnamefont {L.~C.~L.}\
  \bibnamefont {Hollenberg}}, \bibinfo {author} {\bibfnamefont {A.~D.}\
  \bibnamefont {Greentree}}, \bibinfo {author} {\bibfnamefont {A.~G.}\
  \bibnamefont {Fowler}}, \ and\ \bibinfo {author} {\bibfnamefont {C.~J.}\
  \bibnamefont {Wellard}},\ }\bibfield  {title} {\enquote {\bibinfo {title}
  {Two-dimensional architectures for donor-based quantum computing},}\
  }\href@noop {} {\bibfield  {journal} {\bibinfo  {journal} {Physical Review
  B}\ }\textbf {\bibinfo {volume} {74}},\ \bibinfo {pages} {045311} (\bibinfo
  {year} {2006})}\BibitemShut {NoStop}%
\bibitem [{\citenamefont {Pica}\ \emph {et~al.}(2016)\citenamefont {Pica},
  \citenamefont {Lovett}, \citenamefont {Bhatt}, \citenamefont {Schenkel},\
  and\ \citenamefont {Lyon}}]{pica2016surface}%
  \BibitemOpen
  \bibfield  {author} {\bibinfo {author} {\bibfnamefont {G.}~\bibnamefont
  {Pica}}, \bibinfo {author} {\bibfnamefont {B.~W.}\ \bibnamefont {Lovett}},
  \bibinfo {author} {\bibfnamefont {R.~N.}\ \bibnamefont {Bhatt}}, \bibinfo
  {author} {\bibfnamefont {T.}~\bibnamefont {Schenkel}}, \ and\ \bibinfo
  {author} {\bibfnamefont {S.~A.}\ \bibnamefont {Lyon}},\ }\bibfield  {title}
  {\enquote {\bibinfo {title} {Surface code architecture for donors and dots in
  silicon with imprecise and nonuniform qubit couplings},}\ }\href@noop {}
  {\bibfield  {journal} {\bibinfo  {journal} {Physical Review B}\ }\textbf
  {\bibinfo {volume} {93}},\ \bibinfo {pages} {035306} (\bibinfo {year}
  {2016})}\BibitemShut {NoStop}%
\bibitem [{\citenamefont {Tosi}\ \emph {et~al.}(2017)\citenamefont {Tosi},
  \citenamefont {Mohiyaddin}, \citenamefont {Schmitt}, \citenamefont {Tenberg},
  \citenamefont {Rahman}, \citenamefont {Klimeck},\ and\ \citenamefont
  {Morello}}]{tosi2017silicon}%
  \BibitemOpen
  \bibfield  {author} {\bibinfo {author} {\bibfnamefont {G.}~\bibnamefont
  {Tosi}}, \bibinfo {author} {\bibfnamefont {F.~A.}\ \bibnamefont
  {Mohiyaddin}}, \bibinfo {author} {\bibfnamefont {V.}~\bibnamefont {Schmitt}},
  \bibinfo {author} {\bibfnamefont {S.}~\bibnamefont {Tenberg}}, \bibinfo
  {author} {\bibfnamefont {R.}~\bibnamefont {Rahman}}, \bibinfo {author}
  {\bibfnamefont {G.}~\bibnamefont {Klimeck}}, \ and\ \bibinfo {author}
  {\bibfnamefont {A.}~\bibnamefont {Morello}},\ }\bibfield  {title} {\enquote
  {\bibinfo {title} {Silicon quantum processor with robust long-distance qubit
  couplings},}\ }\href@noop {} {\bibfield  {journal} {\bibinfo  {journal}
  {Nature communications}\ }\textbf {\bibinfo {volume} {8}},\ \bibinfo {pages}
  {1--11} (\bibinfo {year} {2017})}\BibitemShut {NoStop}%
\bibitem [{\citenamefont {Voisin}\ \emph {et~al.}(2014)\citenamefont {Voisin},
  \citenamefont {Nguyen}, \citenamefont {Renard}, \citenamefont {Jehl},
  \citenamefont {Barraud}, \citenamefont {Triozon}, \citenamefont {Vinet},
  \citenamefont {Duchemin}, \citenamefont {Niquet}, \citenamefont
  {De~Franceschi},\ and\ \citenamefont {Sanquer}}]{voisin2014few}%
  \BibitemOpen
  \bibfield  {author} {\bibinfo {author} {\bibfnamefont {B.}~\bibnamefont
  {Voisin}}, \bibinfo {author} {\bibfnamefont {V.-H.}\ \bibnamefont {Nguyen}},
  \bibinfo {author} {\bibfnamefont {J.}~\bibnamefont {Renard}}, \bibinfo
  {author} {\bibfnamefont {X.}~\bibnamefont {Jehl}}, \bibinfo {author}
  {\bibfnamefont {S.}~\bibnamefont {Barraud}}, \bibinfo {author} {\bibfnamefont
  {F.}~\bibnamefont {Triozon}}, \bibinfo {author} {\bibfnamefont
  {M.}~\bibnamefont {Vinet}}, \bibinfo {author} {\bibfnamefont
  {I.}~\bibnamefont {Duchemin}}, \bibinfo {author} {\bibfnamefont {Y.-M.}\
  \bibnamefont {Niquet}}, \bibinfo {author} {\bibfnamefont {S.}~\bibnamefont
  {De~Franceschi}}, \ and\ \bibinfo {author} {\bibfnamefont {M.}~\bibnamefont
  {Sanquer}},\ }\bibfield  {title} {\enquote {\bibinfo {title} {Few-electron
  edge-state quantum dots in a silicon nanowire field-effect transistor},}\
  }\href@noop {} {\bibfield  {journal} {\bibinfo  {journal} {Nano letters}\
  }\textbf {\bibinfo {volume} {14}},\ \bibinfo {pages} {2094--2098} (\bibinfo
  {year} {2014})}\BibitemShut {NoStop}%
\bibitem [{\citenamefont {Dupont-Ferrier}\ \emph {et~al.}(2013)\citenamefont
  {Dupont-Ferrier}, \citenamefont {Roche}, \citenamefont {Voisin},
  \citenamefont {Jehl}, \citenamefont {Wacquez}, \citenamefont {Vinet},
  \citenamefont {Sanquer},\ and\ \citenamefont
  {De~Franceschi}}]{dupont2013coherent}%
  \BibitemOpen
  \bibfield  {author} {\bibinfo {author} {\bibfnamefont {E.}~\bibnamefont
  {Dupont-Ferrier}}, \bibinfo {author} {\bibfnamefont {B.}~\bibnamefont
  {Roche}}, \bibinfo {author} {\bibfnamefont {B.}~\bibnamefont {Voisin}},
  \bibinfo {author} {\bibfnamefont {X.}~\bibnamefont {Jehl}}, \bibinfo {author}
  {\bibfnamefont {R.}~\bibnamefont {Wacquez}}, \bibinfo {author} {\bibfnamefont
  {M.}~\bibnamefont {Vinet}}, \bibinfo {author} {\bibfnamefont
  {M.}~\bibnamefont {Sanquer}}, \ and\ \bibinfo {author} {\bibfnamefont
  {S.}~\bibnamefont {De~Franceschi}},\ }\bibfield  {title} {\enquote {\bibinfo
  {title} {Coherent coupling of two dopants in a silicon nanowire probed by
  landau-zener-st{\"u}ckelberg interferometry},}\ }\href@noop {} {\bibfield
  {journal} {\bibinfo  {journal} {Physical review letters}\ }\textbf {\bibinfo
  {volume} {110}},\ \bibinfo {pages} {136802} (\bibinfo {year}
  {2013})}\BibitemShut {NoStop}%
\bibitem [{\citenamefont {Hutin}\ \emph
  {et~al.}(2019{\natexlab{b}})\citenamefont {Hutin}, \citenamefont {Bertrand},
  \citenamefont {Chanrion}, \citenamefont {Bohuslavskyi}, \citenamefont
  {Ansaloni}, \citenamefont {Yang}, \citenamefont {Michniewicz}, \citenamefont
  {Niegemann}, \citenamefont {Spence}, \citenamefont {Lundberg}, \citenamefont
  {Chatterjee}, \citenamefont {Crippa}, \citenamefont {Li}, \citenamefont
  {Maurand}, \citenamefont {Jehl}, \citenamefont {Sanquer}, \citenamefont
  {Gonzalez-Zalba}, \citenamefont {Kuemmeth}, \citenamefont {Niquet},
  \citenamefont {De~Franceschi}, \citenamefont {Urdampilleta}, \citenamefont
  {Meunier},\ and\ \citenamefont {Vinet}}]{hutin2019gate}%
  \BibitemOpen
  \bibfield  {author} {\bibinfo {author} {\bibfnamefont {L.}~\bibnamefont
  {Hutin}}, \bibinfo {author} {\bibfnamefont {B.}~\bibnamefont {Bertrand}},
  \bibinfo {author} {\bibfnamefont {E.}~\bibnamefont {Chanrion}}, \bibinfo
  {author} {\bibfnamefont {H.}~\bibnamefont {Bohuslavskyi}}, \bibinfo {author}
  {\bibfnamefont {F.}~\bibnamefont {Ansaloni}}, \bibinfo {author}
  {\bibfnamefont {T.-Y.}\ \bibnamefont {Yang}}, \bibinfo {author}
  {\bibfnamefont {J.}~\bibnamefont {Michniewicz}}, \bibinfo {author}
  {\bibfnamefont {D.~J.}\ \bibnamefont {Niegemann}}, \bibinfo {author}
  {\bibfnamefont {C.}~\bibnamefont {Spence}}, \bibinfo {author} {\bibfnamefont
  {T.}~\bibnamefont {Lundberg}}, \bibinfo {author} {\bibfnamefont
  {A.}~\bibnamefont {Chatterjee}}, \bibinfo {author} {\bibfnamefont
  {A.}~\bibnamefont {Crippa}}, \bibinfo {author} {\bibfnamefont
  {J.}~\bibnamefont {Li}}, \bibinfo {author} {\bibfnamefont {R.}~\bibnamefont
  {Maurand}}, \bibinfo {author} {\bibfnamefont {X.}~\bibnamefont {Jehl}},
  \bibinfo {author} {\bibfnamefont {M.}~\bibnamefont {Sanquer}}, \bibinfo
  {author} {\bibfnamefont {M.~F.}\ \bibnamefont {Gonzalez-Zalba}}, \bibinfo
  {author} {\bibfnamefont {F.}~\bibnamefont {Kuemmeth}}, \bibinfo {author}
  {\bibfnamefont {Y.-M.}\ \bibnamefont {Niquet}}, \bibinfo {author}
  {\bibfnamefont {S.}~\bibnamefont {De~Franceschi}}, \bibinfo {author}
  {\bibfnamefont {M.}~\bibnamefont {Urdampilleta}}, \bibinfo {author}
  {\bibfnamefont {T.}~\bibnamefont {Meunier}}, \ and\ \bibinfo {author}
  {\bibfnamefont {M.}~\bibnamefont {Vinet}},\ }\bibfield  {title} {\enquote
  {\bibinfo {title} {Gate reflectometry for probing charge and spin states in
  linear si mos split-gate arrays},}\ \ }(\bibinfo {organization} {IEEE},\
  \bibinfo {year} {2019})\BibitemShut {NoStop}%
\bibitem [{\citenamefont {Ansaloni}\ \emph {et~al.}(2020)\citenamefont
  {Ansaloni}, \citenamefont {Chatterjee}, \citenamefont {Bohuslavskyi},
  \citenamefont {Bertrand}, \citenamefont {Hutin}, \citenamefont {Vinet},\ and\
  \citenamefont {Kuemmeth}}]{ansaloni2020single}%
  \BibitemOpen
  \bibfield  {author} {\bibinfo {author} {\bibfnamefont {F.}~\bibnamefont
  {Ansaloni}}, \bibinfo {author} {\bibfnamefont {A.}~\bibnamefont
  {Chatterjee}}, \bibinfo {author} {\bibfnamefont {H.}~\bibnamefont
  {Bohuslavskyi}}, \bibinfo {author} {\bibfnamefont {B.}~\bibnamefont
  {Bertrand}}, \bibinfo {author} {\bibfnamefont {L.}~\bibnamefont {Hutin}},
  \bibinfo {author} {\bibfnamefont {M.}~\bibnamefont {Vinet}}, \ and\ \bibinfo
  {author} {\bibfnamefont {F.}~\bibnamefont {Kuemmeth}},\ }\bibfield  {title}
  {\enquote {\bibinfo {title} {Single-electron operations in a
  foundry-fabricated array of quantum dots},}\ }\href@noop {} {\bibfield
  {journal} {\bibinfo  {journal} {Nature communications}\ }\textbf {\bibinfo
  {volume} {11}},\ \bibinfo {pages} {1--7} (\bibinfo {year}
  {2020})}\BibitemShut {NoStop}%
\bibitem [{\citenamefont {Chanrion}\ \emph {et~al.}(2020)\citenamefont
  {Chanrion}, \citenamefont {Niegemann}, \citenamefont {Bertrand},
  \citenamefont {Spence}, \citenamefont {Jadot}, \citenamefont {Li},
  \citenamefont {Mortemousque}, \citenamefont {Hutin}, \citenamefont {Maurand},
  \citenamefont {Jehl}, \citenamefont {Sanquer}, \citenamefont {De~Franceschi},
  \citenamefont {B\"auerle}, \citenamefont {Balestro}, \citenamefont {Niquet},
  \citenamefont {Vinet}, \citenamefont {Meunier},\ and\ \citenamefont
  {Urdampilleta}}]{chanrion2020charge}%
  \BibitemOpen
  \bibfield  {author} {\bibinfo {author} {\bibfnamefont {E.}~\bibnamefont
  {Chanrion}}, \bibinfo {author} {\bibfnamefont {D.~J.}\ \bibnamefont
  {Niegemann}}, \bibinfo {author} {\bibfnamefont {B.}~\bibnamefont {Bertrand}},
  \bibinfo {author} {\bibfnamefont {C.}~\bibnamefont {Spence}}, \bibinfo
  {author} {\bibfnamefont {B.}~\bibnamefont {Jadot}}, \bibinfo {author}
  {\bibfnamefont {J.}~\bibnamefont {Li}}, \bibinfo {author} {\bibfnamefont
  {P.-A.}\ \bibnamefont {Mortemousque}}, \bibinfo {author} {\bibfnamefont
  {L.}~\bibnamefont {Hutin}}, \bibinfo {author} {\bibfnamefont
  {R.}~\bibnamefont {Maurand}}, \bibinfo {author} {\bibfnamefont
  {X.}~\bibnamefont {Jehl}}, \bibinfo {author} {\bibfnamefont {M.}~\bibnamefont
  {Sanquer}}, \bibinfo {author} {\bibfnamefont {S.}~\bibnamefont
  {De~Franceschi}}, \bibinfo {author} {\bibfnamefont {C.}~\bibnamefont
  {B\"auerle}}, \bibinfo {author} {\bibfnamefont {F.}~\bibnamefont {Balestro}},
  \bibinfo {author} {\bibfnamefont {Y.-M.}\ \bibnamefont {Niquet}}, \bibinfo
  {author} {\bibfnamefont {M.}~\bibnamefont {Vinet}}, \bibinfo {author}
  {\bibfnamefont {T.}~\bibnamefont {Meunier}}, \ and\ \bibinfo {author}
  {\bibfnamefont {M.}~\bibnamefont {Urdampilleta}},\ }\bibfield  {title}
  {\enquote {\bibinfo {title} {Charge detection in an array of cmos quantum
  dots},}\ }\href@noop {} {\bibfield  {journal} {\bibinfo  {journal} {Physical
  Review Applied}\ }\textbf {\bibinfo {volume} {14}},\ \bibinfo {pages}
  {024066} (\bibinfo {year} {2020})}\BibitemShut {NoStop}%
\bibitem [{\citenamefont {Trifunovic}\ \emph {et~al.}(2012)\citenamefont
  {Trifunovic}, \citenamefont {Dial}, \citenamefont {Trif}, \citenamefont
  {Wootton}, \citenamefont {Abebe}, \citenamefont {Yacoby},\ and\ \citenamefont
  {Loss}}]{trifunovic2012long}%
  \BibitemOpen
  \bibfield  {author} {\bibinfo {author} {\bibfnamefont {L.}~\bibnamefont
  {Trifunovic}}, \bibinfo {author} {\bibfnamefont {O.}~\bibnamefont {Dial}},
  \bibinfo {author} {\bibfnamefont {M.}~\bibnamefont {Trif}}, \bibinfo {author}
  {\bibfnamefont {J.~R.}\ \bibnamefont {Wootton}}, \bibinfo {author}
  {\bibfnamefont {R.}~\bibnamefont {Abebe}}, \bibinfo {author} {\bibfnamefont
  {A.}~\bibnamefont {Yacoby}}, \ and\ \bibinfo {author} {\bibfnamefont
  {D.}~\bibnamefont {Loss}},\ }\bibfield  {title} {\enquote {\bibinfo {title}
  {Long-distance spin-spin coupling via floating gates},}\ }\href@noop {}
  {\bibfield  {journal} {\bibinfo  {journal} {Physical review X}\ }\textbf
  {\bibinfo {volume} {2}},\ \bibinfo {pages} {011006} (\bibinfo {year}
  {2012})}\BibitemShut {NoStop}%
\bibitem [{\citenamefont {Duan}\ \emph {et~al.}(2020)\citenamefont {Duan},
  \citenamefont {Fogarty}, \citenamefont {Williams}, \citenamefont {Hutin},
  \citenamefont {Vinet},\ and\ \citenamefont {Morton}}]{duan2019remote}%
  \BibitemOpen
  \bibfield  {author} {\bibinfo {author} {\bibfnamefont {J.}~\bibnamefont
  {Duan}}, \bibinfo {author} {\bibfnamefont {M.~A.}\ \bibnamefont {Fogarty}},
  \bibinfo {author} {\bibfnamefont {J.}~\bibnamefont {Williams}}, \bibinfo
  {author} {\bibfnamefont {L.}~\bibnamefont {Hutin}}, \bibinfo {author}
  {\bibfnamefont {M.}~\bibnamefont {Vinet}}, \ and\ \bibinfo {author}
  {\bibfnamefont {J.~J.~L.}\ \bibnamefont {Morton}},\ }\bibfield  {title}
  {\enquote {\bibinfo {title} {Remote capacitive sensing in two-dimensional
  quantum-dot arrays},}\ }\href@noop {} {\bibfield  {journal} {\bibinfo
  {journal} {Nano Letters}\ }\textbf {\bibinfo {volume} {20}},\ \bibinfo
  {pages} {7123--7128} (\bibinfo {year} {2020})}\BibitemShut {NoStop}%
\bibitem [{\citenamefont {Gilbert}\ \emph {et~al.}(2020)\citenamefont
  {Gilbert}, \citenamefont {Saraiva}, \citenamefont {Lim}, \citenamefont
  {Yang}, \citenamefont {Laucht}, \citenamefont {Bertrand}, \citenamefont
  {Rambal}, \citenamefont {Hutin}, \citenamefont {Escott}, \citenamefont
  {Vinet},\ and\ \citenamefont {Dzurak}}]{gilbert2020single}%
  \BibitemOpen
  \bibfield  {author} {\bibinfo {author} {\bibfnamefont {W.}~\bibnamefont
  {Gilbert}}, \bibinfo {author} {\bibfnamefont {A.}~\bibnamefont {Saraiva}},
  \bibinfo {author} {\bibfnamefont {W.~H.}\ \bibnamefont {Lim}}, \bibinfo
  {author} {\bibfnamefont {C.~H.}\ \bibnamefont {Yang}}, \bibinfo {author}
  {\bibfnamefont {A.}~\bibnamefont {Laucht}}, \bibinfo {author} {\bibfnamefont
  {B.}~\bibnamefont {Bertrand}}, \bibinfo {author} {\bibfnamefont
  {N.}~\bibnamefont {Rambal}}, \bibinfo {author} {\bibfnamefont
  {L.}~\bibnamefont {Hutin}}, \bibinfo {author} {\bibfnamefont {C.~C.}\
  \bibnamefont {Escott}}, \bibinfo {author} {\bibfnamefont {M.}~\bibnamefont
  {Vinet}}, \ and\ \bibinfo {author} {\bibfnamefont {A.~S.}\ \bibnamefont
  {Dzurak}},\ }\bibfield  {title} {\enquote {\bibinfo {title} {Single-electron
  operation of a silicon-cmos 2 × 2 quantum dot array with integrated charge
  sensing},}\ }\href@noop {} {\bibfield  {journal} {\bibinfo  {journal} {Nano
  Letters}\ }\textbf {\bibinfo {volume} {20}},\ \bibinfo {pages} {7882--7888}
  (\bibinfo {year} {2020})}\BibitemShut {NoStop}%
\bibitem [{\citenamefont {Veldhorst}\ \emph {et~al.}(2014)\citenamefont
  {Veldhorst}, \citenamefont {Hwang}, \citenamefont {Yang}, \citenamefont
  {Leenstra}, \citenamefont {de~Ronde}, \citenamefont {Dehollain},
  \citenamefont {Muhonen}, \citenamefont {Hudson}, \citenamefont {Itoh},
  \citenamefont {Morello},\ and\ \citenamefont
  {Dzurak}}]{veldhorst2014addressable}%
  \BibitemOpen
  \bibfield  {author} {\bibinfo {author} {\bibfnamefont {M.}~\bibnamefont
  {Veldhorst}}, \bibinfo {author} {\bibfnamefont {J.~C.~C.}\ \bibnamefont
  {Hwang}}, \bibinfo {author} {\bibfnamefont {C.~H.}\ \bibnamefont {Yang}},
  \bibinfo {author} {\bibfnamefont {A.~W.}\ \bibnamefont {Leenstra}}, \bibinfo
  {author} {\bibfnamefont {B.}~\bibnamefont {de~Ronde}}, \bibinfo {author}
  {\bibfnamefont {J.~P.}\ \bibnamefont {Dehollain}}, \bibinfo {author}
  {\bibfnamefont {J.~T.}\ \bibnamefont {Muhonen}}, \bibinfo {author}
  {\bibfnamefont {F.~E.}\ \bibnamefont {Hudson}}, \bibinfo {author}
  {\bibfnamefont {K.~M.}\ \bibnamefont {Itoh}}, \bibinfo {author}
  {\bibfnamefont {A.}~\bibnamefont {Morello}}, \ and\ \bibinfo {author}
  {\bibfnamefont {A.~S.}\ \bibnamefont {Dzurak}},\ }\bibfield  {title}
  {\enquote {\bibinfo {title} {An addressable quantum dot qubit with
  fault-tolerant control-fidelity},}\ }\href@noop {} {\bibfield  {journal}
  {\bibinfo  {journal} {Nature nanotechnology}\ }\textbf {\bibinfo {volume}
  {9}},\ \bibinfo {pages} {981--985} (\bibinfo {year} {2014})}\BibitemShut
  {NoStop}%
\bibitem [{\citenamefont {Veldhorst}\ \emph {et~al.}(2015)\citenamefont
  {Veldhorst}, \citenamefont {Yang}, \citenamefont {Hwang}, \citenamefont
  {Huang}, \citenamefont {Dehollain}, \citenamefont {Muhonen}, \citenamefont
  {Simmons}, \citenamefont {Laucht}, \citenamefont {Hudson}, \citenamefont
  {Itoh}, \citenamefont {Morello},\ and\ \citenamefont
  {Dzurak}}]{veldhorst2015two}%
  \BibitemOpen
  \bibfield  {author} {\bibinfo {author} {\bibfnamefont {M.}~\bibnamefont
  {Veldhorst}}, \bibinfo {author} {\bibfnamefont {C.~H.}\ \bibnamefont {Yang}},
  \bibinfo {author} {\bibfnamefont {J.~C.~C.}\ \bibnamefont {Hwang}}, \bibinfo
  {author} {\bibfnamefont {W.}~\bibnamefont {Huang}}, \bibinfo {author}
  {\bibfnamefont {J.~P.}\ \bibnamefont {Dehollain}}, \bibinfo {author}
  {\bibfnamefont {J.~T.}\ \bibnamefont {Muhonen}}, \bibinfo {author}
  {\bibfnamefont {S.}~\bibnamefont {Simmons}}, \bibinfo {author} {\bibfnamefont
  {A.}~\bibnamefont {Laucht}}, \bibinfo {author} {\bibfnamefont {F.~E.}\
  \bibnamefont {Hudson}}, \bibinfo {author} {\bibfnamefont {K.~M.}\
  \bibnamefont {Itoh}}, \bibinfo {author} {\bibfnamefont {A.}~\bibnamefont
  {Morello}}, \ and\ \bibinfo {author} {\bibfnamefont {A.~S.}\ \bibnamefont
  {Dzurak}},\ }\bibfield  {title} {\enquote {\bibinfo {title} {A two-qubit
  logic gate in silicon},}\ }\href@noop {} {\bibfield  {journal} {\bibinfo
  {journal} {Nature}\ }\textbf {\bibinfo {volume} {526}},\ \bibinfo {pages}
  {410--414} (\bibinfo {year} {2015})}\BibitemShut {NoStop}%
\bibitem [{\citenamefont {Reilly}(2015)}]{reilly2015engineering}%
  \BibitemOpen
  \bibfield  {author} {\bibinfo {author} {\bibfnamefont {D.~J.}\ \bibnamefont
  {Reilly}},\ }\bibfield  {title} {\enquote {\bibinfo {title} {Engineering the
  quantum-classical interface of solid-state qubits},}\ }\href@noop {}
  {\bibfield  {journal} {\bibinfo  {journal} {npj Quantum Inf.}\ }\textbf
  {\bibinfo {volume} {1}},\ \bibinfo {pages} {15011} (\bibinfo {year}
  {2015})}\BibitemShut {NoStop}%
\bibitem [{\citenamefont {Reilly}(2019)}]{reilly2019challenges}%
  \BibitemOpen
  \bibfield  {author} {\bibinfo {author} {\bibfnamefont {D.~J.}\ \bibnamefont
  {Reilly}},\ }\bibfield  {title} {\enquote {\bibinfo {title} {Challenges in
  scaling-up the control interface of a quantum computer},}\ \ }(\bibinfo
  {organization} {IEEE},\ \bibinfo {year} {2019})\ pp.\ \bibinfo {pages}
  {31.7.1--31.7.6}\BibitemShut {NoStop}%
\bibitem [{\citenamefont {Franke}\ \emph {et~al.}(2019)\citenamefont {Franke},
  \citenamefont {Clarke}, \citenamefont {Vandersypen},\ and\ \citenamefont
  {Veldhorst}}]{franke2019rent}%
  \BibitemOpen
  \bibfield  {author} {\bibinfo {author} {\bibfnamefont {D.~P.}\ \bibnamefont
  {Franke}}, \bibinfo {author} {\bibfnamefont {J.~S.}\ \bibnamefont {Clarke}},
  \bibinfo {author} {\bibfnamefont {L.~M.~K.}\ \bibnamefont {Vandersypen}}, \
  and\ \bibinfo {author} {\bibfnamefont {M.}~\bibnamefont {Veldhorst}},\
  }\bibfield  {title} {\enquote {\bibinfo {title} {Rent’s rule and
  extensibility in quantum computing},}\ }\href@noop {} {\bibfield  {journal}
  {\bibinfo  {journal} {Microprocessors and Microsystems}\ }\textbf {\bibinfo
  {volume} {67}},\ \bibinfo {pages} {1--7} (\bibinfo {year}
  {2019})}\BibitemShut {NoStop}%
\bibitem [{\citenamefont {Fowler}\ \emph {et~al.}(2012)\citenamefont {Fowler},
  \citenamefont {Mariantoni}, \citenamefont {Martinis},\ and\ \citenamefont
  {Cleland}}]{fowler2012surface}%
  \BibitemOpen
  \bibfield  {author} {\bibinfo {author} {\bibfnamefont {A.~G.}\ \bibnamefont
  {Fowler}}, \bibinfo {author} {\bibfnamefont {M.}~\bibnamefont {Mariantoni}},
  \bibinfo {author} {\bibfnamefont {J.~M.}\ \bibnamefont {Martinis}}, \ and\
  \bibinfo {author} {\bibfnamefont {A.~N.}\ \bibnamefont {Cleland}},\
  }\bibfield  {title} {\enquote {\bibinfo {title} {Surface codes: Towards
  practical large-scale quantum computation},}\ }\href@noop {} {\bibfield
  {journal} {\bibinfo  {journal} {Physical Review A}\ }\textbf {\bibinfo
  {volume} {86}},\ \bibinfo {pages} {032324} (\bibinfo {year}
  {2012})}\BibitemShut {NoStop}%
\bibitem [{\citenamefont {Corna}\ \emph {et~al.}(2018)\citenamefont {Corna},
  \citenamefont {Bourdet}, \citenamefont {Maurand}, \citenamefont {Crippa},
  \citenamefont {Kotekar-Patil}, \citenamefont {Bohuslavskyi}, \citenamefont
  {Lavi{\'e}ville}, \citenamefont {Hutin}, \citenamefont {Barraud},
  \citenamefont {Jehl}, \citenamefont {Vinet}, \citenamefont {De~Franceschi},
  \citenamefont {Niquet},\ and\ \citenamefont
  {Sanquer}}]{corna2018electrically}%
  \BibitemOpen
  \bibfield  {author} {\bibinfo {author} {\bibfnamefont {A.}~\bibnamefont
  {Corna}}, \bibinfo {author} {\bibfnamefont {L.}~\bibnamefont {Bourdet}},
  \bibinfo {author} {\bibfnamefont {R.}~\bibnamefont {Maurand}}, \bibinfo
  {author} {\bibfnamefont {A.}~\bibnamefont {Crippa}}, \bibinfo {author}
  {\bibfnamefont {D.}~\bibnamefont {Kotekar-Patil}}, \bibinfo {author}
  {\bibfnamefont {H.}~\bibnamefont {Bohuslavskyi}}, \bibinfo {author}
  {\bibfnamefont {R.}~\bibnamefont {Lavi{\'e}ville}}, \bibinfo {author}
  {\bibfnamefont {L.}~\bibnamefont {Hutin}}, \bibinfo {author} {\bibfnamefont
  {S.}~\bibnamefont {Barraud}}, \bibinfo {author} {\bibfnamefont
  {X.}~\bibnamefont {Jehl}}, \bibinfo {author} {\bibfnamefont {M.}~\bibnamefont
  {Vinet}}, \bibinfo {author} {\bibfnamefont {S.}~\bibnamefont
  {De~Franceschi}}, \bibinfo {author} {\bibfnamefont {Y.-M.}\ \bibnamefont
  {Niquet}}, \ and\ \bibinfo {author} {\bibfnamefont {M.}~\bibnamefont
  {Sanquer}},\ }\bibfield  {title} {\enquote {\bibinfo {title} {Electrically
  driven electron spin resonance mediated by spin--valley--orbit coupling in a
  silicon quantum dot},}\ }\href@noop {} {\bibfield  {journal} {\bibinfo
  {journal} {npj quantum information}\ }\textbf {\bibinfo {volume} {4}},\
  \bibinfo {pages} {1--7} (\bibinfo {year} {2018})}\BibitemShut {NoStop}%
\bibitem [{\citenamefont {Cai}\ \emph {et~al.}(2019)\citenamefont {Cai},
  \citenamefont {Fogarty}, \citenamefont {Schaal}, \citenamefont
  {Patom{\"a}ki}, \citenamefont {Benjamin},\ and\ \citenamefont
  {Morton}}]{cai2019silicon}%
  \BibitemOpen
  \bibfield  {author} {\bibinfo {author} {\bibfnamefont {Z.}~\bibnamefont
  {Cai}}, \bibinfo {author} {\bibfnamefont {M.~A.}\ \bibnamefont {Fogarty}},
  \bibinfo {author} {\bibfnamefont {S.}~\bibnamefont {Schaal}}, \bibinfo
  {author} {\bibfnamefont {S.}~\bibnamefont {Patom{\"a}ki}}, \bibinfo {author}
  {\bibfnamefont {S.~C.}\ \bibnamefont {Benjamin}}, \ and\ \bibinfo {author}
  {\bibfnamefont {J.~J.~L.}\ \bibnamefont {Morton}},\ }\bibfield  {title}
  {\enquote {\bibinfo {title} {A silicon surface code architecture resilient
  against leakage errors},}\ }\href@noop {} {\bibfield  {journal} {\bibinfo
  {journal} {Quantum}\ }\textbf {\bibinfo {volume} {3}},\ \bibinfo {pages}
  {212} (\bibinfo {year} {2019})}\BibitemShut {NoStop}%
\bibitem [{\citenamefont {Srinivasa}\ \emph {et~al.}(2015)\citenamefont
  {Srinivasa}, \citenamefont {Xu},\ and\ \citenamefont
  {Taylor}}]{srinivasa2015tunable}%
  \BibitemOpen
  \bibfield  {author} {\bibinfo {author} {\bibfnamefont {V.}~\bibnamefont
  {Srinivasa}}, \bibinfo {author} {\bibfnamefont {H.}~\bibnamefont {Xu}}, \
  and\ \bibinfo {author} {\bibfnamefont {J.~M.}\ \bibnamefont {Taylor}},\
  }\bibfield  {title} {\enquote {\bibinfo {title} {Tunable spin-qubit coupling
  mediated by a multielectron quantum dot},}\ }\href@noop {} {\bibfield
  {journal} {\bibinfo  {journal} {Physical review letters}\ }\textbf {\bibinfo
  {volume} {114}},\ \bibinfo {pages} {226803} (\bibinfo {year}
  {2015})}\BibitemShut {NoStop}%
\bibitem [{\citenamefont {Malinowski}\ \emph {et~al.}(2019)\citenamefont
  {Malinowski}, \citenamefont {Martins}, \citenamefont {Smith}, \citenamefont
  {Bartlett}, \citenamefont {Doherty}, \citenamefont {Nissen}, \citenamefont
  {Fallahi}, \citenamefont {Gardner}, \citenamefont {Manfra}, \citenamefont
  {Marcus},\ and\ \citenamefont {Kuemmeth}}]{malinowski2019fast}%
  \BibitemOpen
  \bibfield  {author} {\bibinfo {author} {\bibfnamefont {F.~K.}\ \bibnamefont
  {Malinowski}}, \bibinfo {author} {\bibfnamefont {F.}~\bibnamefont {Martins}},
  \bibinfo {author} {\bibfnamefont {T.~B.}\ \bibnamefont {Smith}}, \bibinfo
  {author} {\bibfnamefont {S.~D.}\ \bibnamefont {Bartlett}}, \bibinfo {author}
  {\bibfnamefont {A.~C.}\ \bibnamefont {Doherty}}, \bibinfo {author}
  {\bibfnamefont {P.~D.}\ \bibnamefont {Nissen}}, \bibinfo {author}
  {\bibfnamefont {S.}~\bibnamefont {Fallahi}}, \bibinfo {author} {\bibfnamefont
  {G.~C.}\ \bibnamefont {Gardner}}, \bibinfo {author} {\bibfnamefont {M.~J.}\
  \bibnamefont {Manfra}}, \bibinfo {author} {\bibfnamefont {C.~M.}\
  \bibnamefont {Marcus}}, \ and\ \bibinfo {author} {\bibfnamefont
  {F.}~\bibnamefont {Kuemmeth}},\ }\bibfield  {title} {\enquote {\bibinfo
  {title} {Fast spin exchange across a multielectron mediator},}\ }\href@noop
  {} {\bibfield  {journal} {\bibinfo  {journal} {Nature communications}\
  }\textbf {\bibinfo {volume} {10}},\ \bibinfo {pages} {1--6} (\bibinfo {year}
  {2019})}\BibitemShut {NoStop}%
\bibitem [{\citenamefont {Rochette}\ \emph {et~al.}(2019)\citenamefont
  {Rochette}, \citenamefont {Rudolph}, \citenamefont {Roy}, \citenamefont
  {Curry}, \citenamefont {Ten~Eyck}, \citenamefont {Manginell}, \citenamefont
  {Wendt}, \citenamefont {Pluym}, \citenamefont {Carr}, \citenamefont {Ward},
  \citenamefont {Lilly}, \citenamefont {Carroll},\ and\ \citenamefont
  {Pioro-Ladrière}}]{rochette2019quantum}%
  \BibitemOpen
  \bibfield  {author} {\bibinfo {author} {\bibfnamefont {S.}~\bibnamefont
  {Rochette}}, \bibinfo {author} {\bibfnamefont {M.}~\bibnamefont {Rudolph}},
  \bibinfo {author} {\bibfnamefont {A.-M.}\ \bibnamefont {Roy}}, \bibinfo
  {author} {\bibfnamefont {M.~J.}\ \bibnamefont {Curry}}, \bibinfo {author}
  {\bibfnamefont {G.}~\bibnamefont {Ten~Eyck}}, \bibinfo {author}
  {\bibfnamefont {R.~P.}\ \bibnamefont {Manginell}}, \bibinfo {author}
  {\bibfnamefont {J.~R.}\ \bibnamefont {Wendt}}, \bibinfo {author}
  {\bibfnamefont {T.}~\bibnamefont {Pluym}}, \bibinfo {author} {\bibfnamefont
  {S.~M.}\ \bibnamefont {Carr}}, \bibinfo {author} {\bibfnamefont {D.~R.}\
  \bibnamefont {Ward}}, \bibinfo {author} {\bibfnamefont {M.~P.}\ \bibnamefont
  {Lilly}}, \bibinfo {author} {\bibfnamefont {M.~S.}\ \bibnamefont {Carroll}},
  \ and\ \bibinfo {author} {\bibfnamefont {M.}~\bibnamefont
  {Pioro-Ladrière}},\ }\bibfield  {title} {\enquote {\bibinfo {title} {Quantum
  dots with split enhancement gate tunnel barrier control},}\ }\href@noop {}
  {\bibfield  {journal} {\bibinfo  {journal} {Applied Physics Letters}\
  }\textbf {\bibinfo {volume} {114}},\ \bibinfo {pages} {083101} (\bibinfo
  {year} {2019})}\BibitemShut {NoStop}%
\bibitem [{\citenamefont {Yang}\ \emph {et~al.}(2013)\citenamefont {Yang},
  \citenamefont {Rossi}, \citenamefont {Ruskov}, \citenamefont {Lai},
  \citenamefont {Mohiyaddin}, \citenamefont {Lee}, \citenamefont {Tahan},
  \citenamefont {Klimeck}, \citenamefont {Morello},\ and\ \citenamefont
  {Dzurak}}]{yang2013spin}%
  \BibitemOpen
  \bibfield  {author} {\bibinfo {author} {\bibfnamefont {C.~H.}\ \bibnamefont
  {Yang}}, \bibinfo {author} {\bibfnamefont {A.}~\bibnamefont {Rossi}},
  \bibinfo {author} {\bibfnamefont {R.}~\bibnamefont {Ruskov}}, \bibinfo
  {author} {\bibfnamefont {N.~S.}\ \bibnamefont {Lai}}, \bibinfo {author}
  {\bibfnamefont {F.~A.}\ \bibnamefont {Mohiyaddin}}, \bibinfo {author}
  {\bibfnamefont {S.}~\bibnamefont {Lee}}, \bibinfo {author} {\bibfnamefont
  {C.}~\bibnamefont {Tahan}}, \bibinfo {author} {\bibfnamefont
  {G.}~\bibnamefont {Klimeck}}, \bibinfo {author} {\bibfnamefont
  {A.}~\bibnamefont {Morello}}, \ and\ \bibinfo {author} {\bibfnamefont
  {A.~S.}\ \bibnamefont {Dzurak}},\ }\bibfield  {title} {\enquote {\bibinfo
  {title} {Spin-valley lifetimes in a silicon quantum dot with tunable valley
  splitting},}\ }\href@noop {} {\bibfield  {journal} {\bibinfo  {journal}
  {Nature communications}\ }\textbf {\bibinfo {volume} {4}},\ \bibinfo {pages}
  {1--8} (\bibinfo {year} {2013})}\BibitemShut {NoStop}%
\bibitem [{\citenamefont {Zajac}\ \emph {et~al.}(2018)\citenamefont {Zajac},
  \citenamefont {Sigillito}, \citenamefont {Russ}, \citenamefont {Borjans},
  \citenamefont {Taylor}, \citenamefont {Burkard},\ and\ \citenamefont
  {Petta}}]{zajac2018resonantly}%
  \BibitemOpen
  \bibfield  {author} {\bibinfo {author} {\bibfnamefont {D.~M.}\ \bibnamefont
  {Zajac}}, \bibinfo {author} {\bibfnamefont {A.~J.}\ \bibnamefont
  {Sigillito}}, \bibinfo {author} {\bibfnamefont {M.}~\bibnamefont {Russ}},
  \bibinfo {author} {\bibfnamefont {F.}~\bibnamefont {Borjans}}, \bibinfo
  {author} {\bibfnamefont {J.~M.}\ \bibnamefont {Taylor}}, \bibinfo {author}
  {\bibfnamefont {G.}~\bibnamefont {Burkard}}, \ and\ \bibinfo {author}
  {\bibfnamefont {J.~R.}\ \bibnamefont {Petta}},\ }\bibfield  {title} {\enquote
  {\bibinfo {title} {Resonantly driven cnot gate for electron spins},}\
  }\href@noop {} {\bibfield  {journal} {\bibinfo  {journal} {Science}\ }\textbf
  {\bibinfo {volume} {359}},\ \bibinfo {pages} {439--442} (\bibinfo {year}
  {2018})}\BibitemShut {NoStop}%
\bibitem [{\citenamefont {Jones}\ \emph {et~al.}(2018)\citenamefont {Jones},
  \citenamefont {Fogarty}, \citenamefont {Morello}, \citenamefont {Gyure},
  \citenamefont {Dzurak},\ and\ \citenamefont {Ladd}}]{jones2018logical}%
  \BibitemOpen
  \bibfield  {author} {\bibinfo {author} {\bibfnamefont {C.}~\bibnamefont
  {Jones}}, \bibinfo {author} {\bibfnamefont {M.~A.}\ \bibnamefont {Fogarty}},
  \bibinfo {author} {\bibfnamefont {A.}~\bibnamefont {Morello}}, \bibinfo
  {author} {\bibfnamefont {M.~F.}\ \bibnamefont {Gyure}}, \bibinfo {author}
  {\bibfnamefont {A.~S.}\ \bibnamefont {Dzurak}}, \ and\ \bibinfo {author}
  {\bibfnamefont {T.~D.}\ \bibnamefont {Ladd}},\ }\bibfield  {title} {\enquote
  {\bibinfo {title} {Logical qubit in a linear array of semiconductor quantum
  dots},}\ }\href@noop {} {\bibfield  {journal} {\bibinfo  {journal} {Physical
  Review X}\ }\textbf {\bibinfo {volume} {8}},\ \bibinfo {pages} {021058}
  (\bibinfo {year} {2018})}\BibitemShut {NoStop}%
\bibitem [{\citenamefont {Yang}\ \emph {et~al.}(2019)\citenamefont {Yang},
  \citenamefont {Chan}, \citenamefont {Harper}, \citenamefont {Huang},
  \citenamefont {Evans}, \citenamefont {Hwang}, \citenamefont {Hensen},
  \citenamefont {Laucht}, \citenamefont {Tanttu}, \citenamefont {Hudson},
  \citenamefont {Flammia}, \citenamefont {Itoh}, \citenamefont {Morello},
  \citenamefont {Bartlett},\ and\ \citenamefont {Dzurak}}]{yang2019silicon}%
  \BibitemOpen
  \bibfield  {author} {\bibinfo {author} {\bibfnamefont {C.~H.}\ \bibnamefont
  {Yang}}, \bibinfo {author} {\bibfnamefont {K.~W.}\ \bibnamefont {Chan}},
  \bibinfo {author} {\bibfnamefont {R.}~\bibnamefont {Harper}}, \bibinfo
  {author} {\bibfnamefont {W.}~\bibnamefont {Huang}}, \bibinfo {author}
  {\bibfnamefont {T.}~\bibnamefont {Evans}}, \bibinfo {author} {\bibfnamefont
  {J.~C.~C.}\ \bibnamefont {Hwang}}, \bibinfo {author} {\bibfnamefont
  {B.}~\bibnamefont {Hensen}}, \bibinfo {author} {\bibfnamefont
  {A.}~\bibnamefont {Laucht}}, \bibinfo {author} {\bibfnamefont
  {T.}~\bibnamefont {Tanttu}}, \bibinfo {author} {\bibfnamefont {F.~E.}\
  \bibnamefont {Hudson}}, \bibinfo {author} {\bibfnamefont {S.~T.}\
  \bibnamefont {Flammia}}, \bibinfo {author} {\bibfnamefont {K.~M.}\
  \bibnamefont {Itoh}}, \bibinfo {author} {\bibfnamefont {A.}~\bibnamefont
  {Morello}}, \bibinfo {author} {\bibfnamefont {S.~D.}\ \bibnamefont
  {Bartlett}}, \ and\ \bibinfo {author} {\bibfnamefont {A.~S.}\ \bibnamefont
  {Dzurak}},\ }\bibfield  {title} {\enquote {\bibinfo {title} {Silicon qubit
  fidelities approaching incoherent noise limits via pulse engineering},}\
  }\href@noop {} {\bibfield  {journal} {\bibinfo  {journal} {Nature
  Electronics}\ }\textbf {\bibinfo {volume} {2}},\ \bibinfo {pages} {151--158}
  (\bibinfo {year} {2019})}\BibitemShut {NoStop}%
\bibitem [{\citenamefont {Ono}\ \emph {et~al.}(2002)\citenamefont {Ono},
  \citenamefont {Austing}, \citenamefont {Tokura},\ and\ \citenamefont
  {Tarucha}}]{ono2002current}%
  \BibitemOpen
  \bibfield  {author} {\bibinfo {author} {\bibfnamefont {K.}~\bibnamefont
  {Ono}}, \bibinfo {author} {\bibfnamefont {D.~G.}\ \bibnamefont {Austing}},
  \bibinfo {author} {\bibfnamefont {Y.}~\bibnamefont {Tokura}}, \ and\ \bibinfo
  {author} {\bibfnamefont {S.}~\bibnamefont {Tarucha}},\ }\bibfield  {title}
  {\enquote {\bibinfo {title} {Current rectification by pauli exclusion in a
  weakly coupled double quantum dot system},}\ }\href@noop {} {\bibfield
  {journal} {\bibinfo  {journal} {Science}\ }\textbf {\bibinfo {volume}
  {297}},\ \bibinfo {pages} {1313--1317} (\bibinfo {year} {2002})}\BibitemShut
  {NoStop}%
\bibitem [{\citenamefont {Petta}\ \emph {et~al.}(2005)\citenamefont {Petta},
  \citenamefont {Johnson}, \citenamefont {Taylor}, \citenamefont {Laird},
  \citenamefont {Yacoby}, \citenamefont {Lukin}, \citenamefont {Marcus},
  \citenamefont {Hanson},\ and\ \citenamefont {Gossard}}]{petta2005coherent}%
  \BibitemOpen
  \bibfield  {author} {\bibinfo {author} {\bibfnamefont {J.~R.}\ \bibnamefont
  {Petta}}, \bibinfo {author} {\bibfnamefont {A.~C.}\ \bibnamefont {Johnson}},
  \bibinfo {author} {\bibfnamefont {J.~M.}\ \bibnamefont {Taylor}}, \bibinfo
  {author} {\bibfnamefont {E.~A.}\ \bibnamefont {Laird}}, \bibinfo {author}
  {\bibfnamefont {A.}~\bibnamefont {Yacoby}}, \bibinfo {author} {\bibfnamefont
  {M.~D.}\ \bibnamefont {Lukin}}, \bibinfo {author} {\bibfnamefont {C.~M.}\
  \bibnamefont {Marcus}}, \bibinfo {author} {\bibfnamefont {M.~P.}\
  \bibnamefont {Hanson}}, \ and\ \bibinfo {author} {\bibfnamefont {A.~C.}\
  \bibnamefont {Gossard}},\ }\bibfield  {title} {\enquote {\bibinfo {title}
  {Coherent manipulation of coupled electron spins in semiconductor quantum
  dots},}\ }\href@noop {} {\bibfield  {journal} {\bibinfo  {journal} {Science}\
  }\textbf {\bibinfo {volume} {309}},\ \bibinfo {pages} {2180--2184} (\bibinfo
  {year} {2005})}\BibitemShut {NoStop}%
\bibitem [{\citenamefont {Harvey-Collard}\ \emph {et~al.}(2018)\citenamefont
  {Harvey-Collard}, \citenamefont {D’Anjou}, \citenamefont {Rudolph},
  \citenamefont {Jacobson}, \citenamefont {Dominguez}, \citenamefont
  {Ten~Eyck}, \citenamefont {Wendt}, \citenamefont {Pluym}, \citenamefont
  {Lilly}, \citenamefont {Coish}, \citenamefont {Pioro-Ladrière},\ and\
  \citenamefont {Carroll}}]{harvey2018high}%
  \BibitemOpen
  \bibfield  {author} {\bibinfo {author} {\bibfnamefont {P.}~\bibnamefont
  {Harvey-Collard}}, \bibinfo {author} {\bibfnamefont {B.}~\bibnamefont
  {D’Anjou}}, \bibinfo {author} {\bibfnamefont {M.}~\bibnamefont {Rudolph}},
  \bibinfo {author} {\bibfnamefont {N.~T.}\ \bibnamefont {Jacobson}}, \bibinfo
  {author} {\bibfnamefont {J.}~\bibnamefont {Dominguez}}, \bibinfo {author}
  {\bibfnamefont {G.~A.}\ \bibnamefont {Ten~Eyck}}, \bibinfo {author}
  {\bibfnamefont {J.~R.}\ \bibnamefont {Wendt}}, \bibinfo {author}
  {\bibfnamefont {T.}~\bibnamefont {Pluym}}, \bibinfo {author} {\bibfnamefont
  {M.~P.}\ \bibnamefont {Lilly}}, \bibinfo {author} {\bibfnamefont {W.~A.}\
  \bibnamefont {Coish}}, \bibinfo {author} {\bibfnamefont {M.}~\bibnamefont
  {Pioro-Ladrière}}, \ and\ \bibinfo {author} {\bibfnamefont {M.~S.}\
  \bibnamefont {Carroll}},\ }\bibfield  {title} {\enquote {\bibinfo {title}
  {High-fidelity single-shot readout for a spin qubit via an enhanced latching
  mechanism},}\ }\href@noop {} {\bibfield  {journal} {\bibinfo  {journal}
  {Physical Review X}\ }\textbf {\bibinfo {volume} {8}},\ \bibinfo {pages}
  {021046} (\bibinfo {year} {2018})}\BibitemShut {NoStop}%
\bibitem [{\citenamefont {Urdampilleta}\ \emph {et~al.}(2019)\citenamefont
  {Urdampilleta}, \citenamefont {Niegemann}, \citenamefont {Chanrion},
  \citenamefont {Jadot}, \citenamefont {Spence}, \citenamefont {Mortemousque},
  \citenamefont {B{\"a}uerle}, \citenamefont {Hutin}, \citenamefont {Bertrand},
  \citenamefont {Barraud}, \citenamefont {Maurand}, \citenamefont {Sanquer},
  \citenamefont {Jehl}, \citenamefont {De~Franceschi}, \citenamefont {Vinet},\
  and\ \citenamefont {Meunier}}]{urdampilleta2019gate}%
  \BibitemOpen
  \bibfield  {author} {\bibinfo {author} {\bibfnamefont {M.}~\bibnamefont
  {Urdampilleta}}, \bibinfo {author} {\bibfnamefont {D.~J.}\ \bibnamefont
  {Niegemann}}, \bibinfo {author} {\bibfnamefont {E.}~\bibnamefont {Chanrion}},
  \bibinfo {author} {\bibfnamefont {B.}~\bibnamefont {Jadot}}, \bibinfo
  {author} {\bibfnamefont {C.}~\bibnamefont {Spence}}, \bibinfo {author}
  {\bibfnamefont {P.-A.}\ \bibnamefont {Mortemousque}}, \bibinfo {author}
  {\bibfnamefont {C.}~\bibnamefont {B{\"a}uerle}}, \bibinfo {author}
  {\bibfnamefont {L.}~\bibnamefont {Hutin}}, \bibinfo {author} {\bibfnamefont
  {B.}~\bibnamefont {Bertrand}}, \bibinfo {author} {\bibfnamefont
  {S.}~\bibnamefont {Barraud}}, \bibinfo {author} {\bibfnamefont
  {R.}~\bibnamefont {Maurand}}, \bibinfo {author} {\bibfnamefont
  {M.}~\bibnamefont {Sanquer}}, \bibinfo {author} {\bibfnamefont
  {X.}~\bibnamefont {Jehl}}, \bibinfo {author} {\bibfnamefont {S.}~\bibnamefont
  {De~Franceschi}}, \bibinfo {author} {\bibfnamefont {M.}~\bibnamefont
  {Vinet}}, \ and\ \bibinfo {author} {\bibfnamefont {T.}~\bibnamefont
  {Meunier}},\ }\bibfield  {title} {\enquote {\bibinfo {title} {Gate-based high
  fidelity spin readout in a cmos device},}\ }\href@noop {} {\bibfield
  {journal} {\bibinfo  {journal} {Nature nanotechnology}\ }\textbf {\bibinfo
  {volume} {14}},\ \bibinfo {pages} {737--741} (\bibinfo {year}
  {2019})}\BibitemShut {NoStop}%
\bibitem [{\citenamefont {Zhao}\ \emph {et~al.}(2019)\citenamefont {Zhao},
  \citenamefont {Tanttu}, \citenamefont {Tan}, \citenamefont {Hensen},
  \citenamefont {Chan}, \citenamefont {Hwang}, \citenamefont {Leon},
  \citenamefont {Yang}, \citenamefont {Gilbert}, \citenamefont {Hudson},
  \citenamefont {Itoh}, \citenamefont {Kiselev}, \citenamefont {Ladd},
  \citenamefont {Morello}, \citenamefont {Dzurak},\ and\ \citenamefont
  {Laucht}}]{zhao2019single}%
  \BibitemOpen
  \bibfield  {author} {\bibinfo {author} {\bibfnamefont {R.}~\bibnamefont
  {Zhao}}, \bibinfo {author} {\bibfnamefont {T.}~\bibnamefont {Tanttu}},
  \bibinfo {author} {\bibfnamefont {K.~Y.}\ \bibnamefont {Tan}}, \bibinfo
  {author} {\bibfnamefont {B.}~\bibnamefont {Hensen}}, \bibinfo {author}
  {\bibfnamefont {K.~W.}\ \bibnamefont {Chan}}, \bibinfo {author}
  {\bibfnamefont {J.~C.~C.}\ \bibnamefont {Hwang}}, \bibinfo {author}
  {\bibfnamefont {R.~C.~C.}\ \bibnamefont {Leon}}, \bibinfo {author}
  {\bibfnamefont {C.~H.}\ \bibnamefont {Yang}}, \bibinfo {author}
  {\bibfnamefont {W.}~\bibnamefont {Gilbert}}, \bibinfo {author} {\bibfnamefont
  {F.~E.}\ \bibnamefont {Hudson}}, \bibinfo {author} {\bibfnamefont {K.~M.}\
  \bibnamefont {Itoh}}, \bibinfo {author} {\bibfnamefont {A.~A.}\ \bibnamefont
  {Kiselev}}, \bibinfo {author} {\bibfnamefont {T.~D.}\ \bibnamefont {Ladd}},
  \bibinfo {author} {\bibfnamefont {A.}~\bibnamefont {Morello}}, \bibinfo
  {author} {\bibfnamefont {A.~S.}\ \bibnamefont {Dzurak}}, \ and\ \bibinfo
  {author} {\bibfnamefont {A.}~\bibnamefont {Laucht}},\ }\bibfield  {title}
  {\enquote {\bibinfo {title} {Single-spin qubits in isotopically enriched
  silicon at low magnetic field},}\ }\href@noop {} {\bibfield  {journal}
  {\bibinfo  {journal} {Nature communications}\ }\textbf {\bibinfo {volume}
  {10}},\ \bibinfo {pages} {1--9} (\bibinfo {year} {2019})}\BibitemShut
  {NoStop}%
\bibitem [{\citenamefont {West}\ \emph {et~al.}(2019)\citenamefont {West},
  \citenamefont {Hensen}, \citenamefont {Jouan}, \citenamefont {Tanttu},
  \citenamefont {Yang}, \citenamefont {Rossi}, \citenamefont {Gonzalez-Zalba},
  \citenamefont {Hudson}, \citenamefont {Morello}, \citenamefont {Reilly},\
  and\ \citenamefont {Dzurak}}]{west2019gate}%
  \BibitemOpen
  \bibfield  {author} {\bibinfo {author} {\bibfnamefont {A.}~\bibnamefont
  {West}}, \bibinfo {author} {\bibfnamefont {B.}~\bibnamefont {Hensen}},
  \bibinfo {author} {\bibfnamefont {A.}~\bibnamefont {Jouan}}, \bibinfo
  {author} {\bibfnamefont {T.}~\bibnamefont {Tanttu}}, \bibinfo {author}
  {\bibfnamefont {C.-H.}\ \bibnamefont {Yang}}, \bibinfo {author}
  {\bibfnamefont {A.}~\bibnamefont {Rossi}}, \bibinfo {author} {\bibfnamefont
  {M.~F.}\ \bibnamefont {Gonzalez-Zalba}}, \bibinfo {author} {\bibfnamefont
  {F.~E.}\ \bibnamefont {Hudson}}, \bibinfo {author} {\bibfnamefont
  {A.}~\bibnamefont {Morello}}, \bibinfo {author} {\bibfnamefont {D.~J.}\
  \bibnamefont {Reilly}}, \ and\ \bibinfo {author} {\bibfnamefont {A.~S.}\
  \bibnamefont {Dzurak}},\ }\bibfield  {title} {\enquote {\bibinfo {title}
  {Gate-based single-shot readout of spins in silicon},}\ }\href@noop {}
  {\bibfield  {journal} {\bibinfo  {journal} {Nature nanotechnology}\ }\textbf
  {\bibinfo {volume} {14}},\ \bibinfo {pages} {437--441} (\bibinfo {year}
  {2019})}\BibitemShut {NoStop}%
\bibitem [{\citenamefont {Ibberson}\ \emph {et~al.}(2021)\citenamefont
  {Ibberson}, \citenamefont {Lundberg}, \citenamefont {Haigh}, \citenamefont
  {Hutin}, \citenamefont {Bertrand}, \citenamefont {Barraud}, \citenamefont
  {Lee}, \citenamefont {Stelmashenko}, \citenamefont {Oakes}, \citenamefont
  {Cochrane}, \citenamefont {Robinson}, \citenamefont {Vinet}, \citenamefont
  {Gonzalez-Zalba},\ and\ \citenamefont {Ibberson}}]{ibberson2021large}%
  \BibitemOpen
  \bibfield  {author} {\bibinfo {author} {\bibfnamefont {D.~J.}\ \bibnamefont
  {Ibberson}}, \bibinfo {author} {\bibfnamefont {T.}~\bibnamefont {Lundberg}},
  \bibinfo {author} {\bibfnamefont {J.~A.}\ \bibnamefont {Haigh}}, \bibinfo
  {author} {\bibfnamefont {L.}~\bibnamefont {Hutin}}, \bibinfo {author}
  {\bibfnamefont {B.}~\bibnamefont {Bertrand}}, \bibinfo {author}
  {\bibfnamefont {S.}~\bibnamefont {Barraud}}, \bibinfo {author} {\bibfnamefont
  {C.-M.}\ \bibnamefont {Lee}}, \bibinfo {author} {\bibfnamefont {N.~A.}\
  \bibnamefont {Stelmashenko}}, \bibinfo {author} {\bibfnamefont {G.~A.}\
  \bibnamefont {Oakes}}, \bibinfo {author} {\bibfnamefont {L.}~\bibnamefont
  {Cochrane}}, \bibinfo {author} {\bibfnamefont {J.~W.~A.}\ \bibnamefont
  {Robinson}}, \bibinfo {author} {\bibfnamefont {M.}~\bibnamefont {Vinet}},
  \bibinfo {author} {\bibfnamefont {M.~F.}\ \bibnamefont {Gonzalez-Zalba}}, \
  and\ \bibinfo {author} {\bibfnamefont {L.~A.}\ \bibnamefont {Ibberson}},\
  }\bibfield  {title} {\enquote {\bibinfo {title} {Large dispersive interaction
  between a cmos double quantum dot and microwave photons},}\ }\href@noop {}
  {\bibfield  {journal} {\bibinfo  {journal} {PRX Quantum}\ }\textbf {\bibinfo
  {volume} {2}},\ \bibinfo {pages} {020315} (\bibinfo {year}
  {2021})}\BibitemShut {NoStop}%
\bibitem [{\citenamefont {Zheng}\ \emph {et~al.}(2019)\citenamefont {Zheng},
  \citenamefont {Samkharadze}, \citenamefont {Noordam}, \citenamefont {Kalhor},
  \citenamefont {Brousse}, \citenamefont {Sammak}, \citenamefont {Scappucci},\
  and\ \citenamefont {Vandersypen}}]{zheng2019rapid}%
  \BibitemOpen
  \bibfield  {author} {\bibinfo {author} {\bibfnamefont {G.}~\bibnamefont
  {Zheng}}, \bibinfo {author} {\bibfnamefont {N.}~\bibnamefont {Samkharadze}},
  \bibinfo {author} {\bibfnamefont {M.~L.}\ \bibnamefont {Noordam}}, \bibinfo
  {author} {\bibfnamefont {N.}~\bibnamefont {Kalhor}}, \bibinfo {author}
  {\bibfnamefont {D.}~\bibnamefont {Brousse}}, \bibinfo {author} {\bibfnamefont
  {A.}~\bibnamefont {Sammak}}, \bibinfo {author} {\bibfnamefont
  {G.}~\bibnamefont {Scappucci}}, \ and\ \bibinfo {author} {\bibfnamefont
  {L.~M.~K.}\ \bibnamefont {Vandersypen}},\ }\bibfield  {title} {\enquote
  {\bibinfo {title} {Rapid gate-based spin read-out in silicon using an on-chip
  resonator},}\ }\href@noop {} {\bibfield  {journal} {\bibinfo  {journal}
  {Nature nanotechnology}\ }\textbf {\bibinfo {volume} {14}},\ \bibinfo {pages}
  {742--746} (\bibinfo {year} {2019})}\BibitemShut {NoStop}%
\bibitem [{\citenamefont {Seedhouse}\ \emph {et~al.}(2021)\citenamefont
  {Seedhouse}, \citenamefont {Tanttu}, \citenamefont {Leon}, \citenamefont
  {Zhao}, \citenamefont {Tan}, \citenamefont {Hensen}, \citenamefont {Hudson},
  \citenamefont {Itoh}, \citenamefont {Yoneda}, \citenamefont {Yang},
  \citenamefont {Morello}, \citenamefont {Laucht}, \citenamefont {Coppersmith},
  \citenamefont {Saraiva},\ and\ \citenamefont {Dzurak}}]{seedhouse2021pauli}%
  \BibitemOpen
  \bibfield  {author} {\bibinfo {author} {\bibfnamefont {A.~E.}\ \bibnamefont
  {Seedhouse}}, \bibinfo {author} {\bibfnamefont {T.}~\bibnamefont {Tanttu}},
  \bibinfo {author} {\bibfnamefont {R.~C.~C.}\ \bibnamefont {Leon}}, \bibinfo
  {author} {\bibfnamefont {R.}~\bibnamefont {Zhao}}, \bibinfo {author}
  {\bibfnamefont {K.~Y.}\ \bibnamefont {Tan}}, \bibinfo {author} {\bibfnamefont
  {B.}~\bibnamefont {Hensen}}, \bibinfo {author} {\bibfnamefont {F.~E.}\
  \bibnamefont {Hudson}}, \bibinfo {author} {\bibfnamefont {K.~M.}\
  \bibnamefont {Itoh}}, \bibinfo {author} {\bibfnamefont {J.}~\bibnamefont
  {Yoneda}}, \bibinfo {author} {\bibfnamefont {C.~H.}\ \bibnamefont {Yang}},
  \bibinfo {author} {\bibfnamefont {A.}~\bibnamefont {Morello}}, \bibinfo
  {author} {\bibfnamefont {A.}~\bibnamefont {Laucht}}, \bibinfo {author}
  {\bibfnamefont {S.~N.}\ \bibnamefont {Coppersmith}}, \bibinfo {author}
  {\bibfnamefont {A.}~\bibnamefont {Saraiva}}, \ and\ \bibinfo {author}
  {\bibfnamefont {A.~S.}\ \bibnamefont {Dzurak}},\ }\bibfield  {title}
  {\enquote {\bibinfo {title} {Pauli blockade in silicon quantum dots with
  spin-orbit control},}\ }\href@noop {} {\bibfield  {journal} {\bibinfo
  {journal} {PRX Quantum}\ }\textbf {\bibinfo {volume} {2}},\ \bibinfo {pages}
  {010303} (\bibinfo {year} {2021})}\BibitemShut {NoStop}%
\bibitem [{\citenamefont {Srinivasa}\ \emph {et~al.}(2013)\citenamefont
  {Srinivasa}, \citenamefont {Nowack}, \citenamefont {Shafiei}, \citenamefont
  {Vandersypen},\ and\ \citenamefont {Taylor}}]{srinivasa2013simultaneous}%
  \BibitemOpen
  \bibfield  {author} {\bibinfo {author} {\bibfnamefont {V.}~\bibnamefont
  {Srinivasa}}, \bibinfo {author} {\bibfnamefont {K.~C.}\ \bibnamefont
  {Nowack}}, \bibinfo {author} {\bibfnamefont {M.}~\bibnamefont {Shafiei}},
  \bibinfo {author} {\bibfnamefont {L.~M.~K.}\ \bibnamefont {Vandersypen}}, \
  and\ \bibinfo {author} {\bibfnamefont {J.~M.}\ \bibnamefont {Taylor}},\
  }\bibfield  {title} {\enquote {\bibinfo {title} {Simultaneous spin-charge
  relaxation in double quantum dots},}\ }\href@noop {} {\bibfield  {journal}
  {\bibinfo  {journal} {Physical review letters}\ }\textbf {\bibinfo {volume}
  {110}},\ \bibinfo {pages} {196803} (\bibinfo {year} {2013})}\BibitemShut
  {NoStop}%
\bibitem [{\citenamefont {Huang}\ \emph {et~al.}(2019)\citenamefont {Huang},
  \citenamefont {Yang}, \citenamefont {Chan}, \citenamefont {Tanttu},
  \citenamefont {Hensen}, \citenamefont {Leon}, \citenamefont {Fogarty},
  \citenamefont {Hwang}, \citenamefont {Hudson}, \citenamefont {Itoh},
  \citenamefont {Morello}, \citenamefont {Laucht},\ and\ \citenamefont
  {Dzurak}}]{huang2019fidelity}%
  \BibitemOpen
  \bibfield  {author} {\bibinfo {author} {\bibfnamefont {W.}~\bibnamefont
  {Huang}}, \bibinfo {author} {\bibfnamefont {C.~H.}\ \bibnamefont {Yang}},
  \bibinfo {author} {\bibfnamefont {K.~W.}\ \bibnamefont {Chan}}, \bibinfo
  {author} {\bibfnamefont {T.}~\bibnamefont {Tanttu}}, \bibinfo {author}
  {\bibfnamefont {B.}~\bibnamefont {Hensen}}, \bibinfo {author} {\bibfnamefont
  {R.~C.~C.}\ \bibnamefont {Leon}}, \bibinfo {author} {\bibfnamefont {M.~A.}\
  \bibnamefont {Fogarty}}, \bibinfo {author} {\bibfnamefont {J.~C.~C.}\
  \bibnamefont {Hwang}}, \bibinfo {author} {\bibfnamefont {F.~E.}\ \bibnamefont
  {Hudson}}, \bibinfo {author} {\bibfnamefont {K.~M.}\ \bibnamefont {Itoh}},
  \bibinfo {author} {\bibfnamefont {A.}~\bibnamefont {Morello}}, \bibinfo
  {author} {\bibfnamefont {A.}~\bibnamefont {Laucht}}, \ and\ \bibinfo {author}
  {\bibfnamefont {A.~S.}\ \bibnamefont {Dzurak}},\ }\bibfield  {title}
  {\enquote {\bibinfo {title} {Fidelity benchmarks for two-qubit gates in
  silicon},}\ }\href@noop {} {\bibfield  {journal} {\bibinfo  {journal}
  {Nature}\ }\textbf {\bibinfo {volume} {569}},\ \bibinfo {pages} {532--536}
  (\bibinfo {year} {2019})}\BibitemShut {NoStop}%
\bibitem [{\citenamefont {Maune}\ \emph {et~al.}(2012)\citenamefont {Maune},
  \citenamefont {Borselli}, \citenamefont {Huang}, \citenamefont {Ladd},
  \citenamefont {Deelman}, \citenamefont {Holabird}, \citenamefont {Kiselev},
  \citenamefont {Alvarado-Rodriguez}, \citenamefont {Ross}, \citenamefont
  {Schmitz}, \citenamefont {Sokolich}, \citenamefont {Watson}, \citenamefont
  {Gyure},\ and\ \citenamefont {Hunter}}]{maune2012coherent}%
  \BibitemOpen
  \bibfield  {author} {\bibinfo {author} {\bibfnamefont {B.~M.}\ \bibnamefont
  {Maune}}, \bibinfo {author} {\bibfnamefont {M.~G.}\ \bibnamefont {Borselli}},
  \bibinfo {author} {\bibfnamefont {B.}~\bibnamefont {Huang}}, \bibinfo
  {author} {\bibfnamefont {T.~D.}\ \bibnamefont {Ladd}}, \bibinfo {author}
  {\bibfnamefont {P.~W.}\ \bibnamefont {Deelman}}, \bibinfo {author}
  {\bibfnamefont {K.~S.}\ \bibnamefont {Holabird}}, \bibinfo {author}
  {\bibfnamefont {A.~A.}\ \bibnamefont {Kiselev}}, \bibinfo {author}
  {\bibfnamefont {I.}~\bibnamefont {Alvarado-Rodriguez}}, \bibinfo {author}
  {\bibfnamefont {R.~S.}\ \bibnamefont {Ross}}, \bibinfo {author}
  {\bibfnamefont {A.~E.}\ \bibnamefont {Schmitz}}, \bibinfo {author}
  {\bibfnamefont {M.}~\bibnamefont {Sokolich}}, \bibinfo {author}
  {\bibfnamefont {C.~A.}\ \bibnamefont {Watson}}, \bibinfo {author}
  {\bibfnamefont {M.~F.}\ \bibnamefont {Gyure}}, \ and\ \bibinfo {author}
  {\bibfnamefont {A.~T.}\ \bibnamefont {Hunter}},\ }\bibfield  {title}
  {\enquote {\bibinfo {title} {Coherent singlet-triplet oscillations in a
  silicon-based double quantum dot},}\ }\href@noop {} {\bibfield  {journal}
  {\bibinfo  {journal} {Nature}\ }\textbf {\bibinfo {volume} {481}},\ \bibinfo
  {pages} {344--347} (\bibinfo {year} {2012})}\BibitemShut {NoStop}%
\bibitem [{\citenamefont {Jock}\ \emph {et~al.}(2022)\citenamefont {Jock},
  \citenamefont {Jacobson}, \citenamefont {Rudolph}, \citenamefont {Ward},
  \citenamefont {Carroll},\ and\ \citenamefont {Luhman}}]{jock2021silicon}%
  \BibitemOpen
  \bibfield  {author} {\bibinfo {author} {\bibfnamefont {R.~M.}\ \bibnamefont
  {Jock}}, \bibinfo {author} {\bibfnamefont {N.~T.}\ \bibnamefont {Jacobson}},
  \bibinfo {author} {\bibfnamefont {M.}~\bibnamefont {Rudolph}}, \bibinfo
  {author} {\bibfnamefont {D.~R.}\ \bibnamefont {Ward}}, \bibinfo {author}
  {\bibfnamefont {M.~S.}\ \bibnamefont {Carroll}}, \ and\ \bibinfo {author}
  {\bibfnamefont {D.~R.}\ \bibnamefont {Luhman}},\ }\bibfield  {title}
  {\enquote {\bibinfo {title} {A silicon singlet--triplet qubit driven by
  spin-valley coupling},}\ }\href@noop {} {\bibfield  {journal} {\bibinfo
  {journal} {Nature communications}\ }\textbf {\bibinfo {volume} {13}},\
  \bibinfo {pages} {1--9} (\bibinfo {year} {2022})}\BibitemShut {NoStop}%
\bibitem [{\citenamefont {Watson}\ \emph {et~al.}(2018)\citenamefont {Watson},
  \citenamefont {Philips}, \citenamefont {Kawakami}, \citenamefont {Ward},
  \citenamefont {Scarlino}, \citenamefont {Veldhorst}, \citenamefont {Savage},
  \citenamefont {Lagally}, \citenamefont {Friesen}, \citenamefont
  {Coppersmith}, \citenamefont {Eriksson},\ and\ \citenamefont
  {Vandersypen}}]{watson2018programmable}%
  \BibitemOpen
  \bibfield  {author} {\bibinfo {author} {\bibfnamefont {T.~F.}\ \bibnamefont
  {Watson}}, \bibinfo {author} {\bibfnamefont {S.~G.~J.}\ \bibnamefont
  {Philips}}, \bibinfo {author} {\bibfnamefont {E.}~\bibnamefont {Kawakami}},
  \bibinfo {author} {\bibfnamefont {D.~R.}\ \bibnamefont {Ward}}, \bibinfo
  {author} {\bibfnamefont {P.}~\bibnamefont {Scarlino}}, \bibinfo {author}
  {\bibfnamefont {M.}~\bibnamefont {Veldhorst}}, \bibinfo {author}
  {\bibfnamefont {D.~E.}\ \bibnamefont {Savage}}, \bibinfo {author}
  {\bibfnamefont {M.~G.}\ \bibnamefont {Lagally}}, \bibinfo {author}
  {\bibfnamefont {M.}~\bibnamefont {Friesen}}, \bibinfo {author} {\bibfnamefont
  {S.~N.}\ \bibnamefont {Coppersmith}}, \bibinfo {author} {\bibfnamefont
  {M.~A.}\ \bibnamefont {Eriksson}}, \ and\ \bibinfo {author} {\bibfnamefont
  {L.~M.~K.}\ \bibnamefont {Vandersypen}},\ }\bibfield  {title} {\enquote
  {\bibinfo {title} {A programmable two-qubit quantum processor in silicon},}\
  }\href@noop {} {\bibfield  {journal} {\bibinfo  {journal} {nature}\ }\textbf
  {\bibinfo {volume} {555}},\ \bibinfo {pages} {633--637} (\bibinfo {year}
  {2018})}\BibitemShut {NoStop}%
\bibitem [{\citenamefont {He}\ \emph {et~al.}(2019)\citenamefont {He},
  \citenamefont {Gorman}, \citenamefont {Keith}, \citenamefont {Kranz},
  \citenamefont {Keizer},\ and\ \citenamefont {Simmons}}]{he2019two}%
  \BibitemOpen
  \bibfield  {author} {\bibinfo {author} {\bibfnamefont {Y.}~\bibnamefont
  {He}}, \bibinfo {author} {\bibfnamefont {S.~K.}\ \bibnamefont {Gorman}},
  \bibinfo {author} {\bibfnamefont {D.}~\bibnamefont {Keith}}, \bibinfo
  {author} {\bibfnamefont {L.}~\bibnamefont {Kranz}}, \bibinfo {author}
  {\bibfnamefont {J.~G.}\ \bibnamefont {Keizer}}, \ and\ \bibinfo {author}
  {\bibfnamefont {M.~Y.}\ \bibnamefont {Simmons}},\ }\bibfield  {title}
  {\enquote {\bibinfo {title} {A two-qubit gate between phosphorus donor
  electrons in silicon},}\ }\href@noop {} {\bibfield  {journal} {\bibinfo
  {journal} {Nature}\ }\textbf {\bibinfo {volume} {571}},\ \bibinfo {pages}
  {371--375} (\bibinfo {year} {2019})}\BibitemShut {NoStop}%
\bibitem [{\citenamefont {Sigillito}\ \emph {et~al.}(2019)\citenamefont
  {Sigillito}, \citenamefont {Gullans}, \citenamefont {Edge}, \citenamefont
  {Borselli},\ and\ \citenamefont {Petta}}]{sigillito2019coherent}%
  \BibitemOpen
  \bibfield  {author} {\bibinfo {author} {\bibfnamefont {A.~J.}\ \bibnamefont
  {Sigillito}}, \bibinfo {author} {\bibfnamefont {M.~J.}\ \bibnamefont
  {Gullans}}, \bibinfo {author} {\bibfnamefont {L.~F.}\ \bibnamefont {Edge}},
  \bibinfo {author} {\bibfnamefont {M.}~\bibnamefont {Borselli}}, \ and\
  \bibinfo {author} {\bibfnamefont {J.~R.}\ \bibnamefont {Petta}},\ }\bibfield
  {title} {\enquote {\bibinfo {title} {Coherent transfer of quantum information
  in a silicon double quantum dot using resonant swap gates},}\ }\href@noop {}
  {\bibfield  {journal} {\bibinfo  {journal} {npj Quantum Information}\
  }\textbf {\bibinfo {volume} {5}},\ \bibinfo {pages} {1--7} (\bibinfo {year}
  {2019})}\BibitemShut {NoStop}%
\bibitem [{\citenamefont {Boter}\ \emph {et~al.}(2019)\citenamefont {Boter},
  \citenamefont {Dehollain}, \citenamefont {van Dijk}, \citenamefont
  {Hensgens}, \citenamefont {Versluis}, \citenamefont {Clarke}, \citenamefont
  {Veldhorst}, \citenamefont {Sebastiano},\ and\ \citenamefont
  {Vandersypen}}]{boter2019sparse}%
  \BibitemOpen
  \bibfield  {author} {\bibinfo {author} {\bibfnamefont {J.~M.}\ \bibnamefont
  {Boter}}, \bibinfo {author} {\bibfnamefont {J.~P.}\ \bibnamefont
  {Dehollain}}, \bibinfo {author} {\bibfnamefont {J.~P.~G.}\ \bibnamefont {van
  Dijk}}, \bibinfo {author} {\bibfnamefont {T.}~\bibnamefont {Hensgens}},
  \bibinfo {author} {\bibfnamefont {R.}~\bibnamefont {Versluis}}, \bibinfo
  {author} {\bibfnamefont {J.~S.}\ \bibnamefont {Clarke}}, \bibinfo {author}
  {\bibfnamefont {M.}~\bibnamefont {Veldhorst}}, \bibinfo {author}
  {\bibfnamefont {F.}~\bibnamefont {Sebastiano}}, \ and\ \bibinfo {author}
  {\bibfnamefont {L.~M.~K.}\ \bibnamefont {Vandersypen}},\ }\bibfield  {title}
  {\enquote {\bibinfo {title} {A sparse spin qubit array with integrated
  control electronics},}\ \ }(\bibinfo {organization} {IEEE},\ \bibinfo {year}
  {2019})\ pp.\ \bibinfo {pages} {31--4}\BibitemShut {NoStop}%
\bibitem [{\citenamefont {Borjans}\ \emph {et~al.}(2020)\citenamefont
  {Borjans}, \citenamefont {Croot}, \citenamefont {Mi}, \citenamefont
  {Gullans},\ and\ \citenamefont {Petta}}]{borjans2020resonant}%
  \BibitemOpen
  \bibfield  {author} {\bibinfo {author} {\bibfnamefont {F.}~\bibnamefont
  {Borjans}}, \bibinfo {author} {\bibfnamefont {X.~G.}\ \bibnamefont {Croot}},
  \bibinfo {author} {\bibfnamefont {X.}~\bibnamefont {Mi}}, \bibinfo {author}
  {\bibfnamefont {M.~J.}\ \bibnamefont {Gullans}}, \ and\ \bibinfo {author}
  {\bibfnamefont {J.~R.}\ \bibnamefont {Petta}},\ }\bibfield  {title} {\enquote
  {\bibinfo {title} {Resonant microwave-mediated interactions between distant
  electron spins},}\ }\href@noop {} {\bibfield  {journal} {\bibinfo  {journal}
  {Nature}\ }\textbf {\bibinfo {volume} {577}},\ \bibinfo {pages} {195--198}
  (\bibinfo {year} {2020})}\BibitemShut {NoStop}%
\bibitem [{\citenamefont {Clerk}\ \emph {et~al.}(2020)\citenamefont {Clerk},
  \citenamefont {Lehnert}, \citenamefont {Bertet}, \citenamefont {Petta},\ and\
  \citenamefont {Nakamura}}]{clerk2020hybrid}%
  \BibitemOpen
  \bibfield  {author} {\bibinfo {author} {\bibfnamefont {A.~A.}\ \bibnamefont
  {Clerk}}, \bibinfo {author} {\bibfnamefont {K.~W.}\ \bibnamefont {Lehnert}},
  \bibinfo {author} {\bibfnamefont {P.}~\bibnamefont {Bertet}}, \bibinfo
  {author} {\bibfnamefont {J.~R.}\ \bibnamefont {Petta}}, \ and\ \bibinfo
  {author} {\bibfnamefont {Y.}~\bibnamefont {Nakamura}},\ }\bibfield  {title}
  {\enquote {\bibinfo {title} {Hybrid quantum systems with circuit quantum
  electrodynamics},}\ }\href@noop {} {\bibfield  {journal} {\bibinfo  {journal}
  {Nature Physics}\ }\textbf {\bibinfo {volume} {16}},\ \bibinfo {pages}
  {257--267} (\bibinfo {year} {2020})}\BibitemShut {NoStop}%
\bibitem [{\citenamefont {Baart}\ \emph {et~al.}(2017)\citenamefont {Baart},
  \citenamefont {Fujita}, \citenamefont {Reichl}, \citenamefont {Wegscheider},\
  and\ \citenamefont {Vandersypen}}]{baart2017coherent}%
  \BibitemOpen
  \bibfield  {author} {\bibinfo {author} {\bibfnamefont {T.~A.}\ \bibnamefont
  {Baart}}, \bibinfo {author} {\bibfnamefont {T.}~\bibnamefont {Fujita}},
  \bibinfo {author} {\bibfnamefont {C.}~\bibnamefont {Reichl}}, \bibinfo
  {author} {\bibfnamefont {W.}~\bibnamefont {Wegscheider}}, \ and\ \bibinfo
  {author} {\bibfnamefont {L.~M.~K.}\ \bibnamefont {Vandersypen}},\ }\bibfield
  {title} {\enquote {\bibinfo {title} {Coherent spin-exchange via a quantum
  mediator},}\ }\href@noop {} {\bibfield  {journal} {\bibinfo  {journal}
  {Nature nanotechnology}\ }\textbf {\bibinfo {volume} {12}},\ \bibinfo {pages}
  {26--30} (\bibinfo {year} {2017})}\BibitemShut {NoStop}%
\bibitem [{\citenamefont {Yoneda}\ \emph {et~al.}(2021)\citenamefont {Yoneda},
  \citenamefont {Huang}, \citenamefont {Feng}, \citenamefont {Yang},
  \citenamefont {Chan}, \citenamefont {Tanttu}, \citenamefont {Gilbert},
  \citenamefont {Leon}, \citenamefont {Hudson}, \citenamefont {Itoh},
  \citenamefont {Morello}, \citenamefont {Bartlett}, \citenamefont {Laucht},
  \citenamefont {Saraiva},\ and\ \citenamefont {Dzurak}}]{yoneda2021coherent}%
  \BibitemOpen
  \bibfield  {author} {\bibinfo {author} {\bibfnamefont {J.}~\bibnamefont
  {Yoneda}}, \bibinfo {author} {\bibfnamefont {W.}~\bibnamefont {Huang}},
  \bibinfo {author} {\bibfnamefont {M.}~\bibnamefont {Feng}}, \bibinfo {author}
  {\bibfnamefont {C.~H.}\ \bibnamefont {Yang}}, \bibinfo {author}
  {\bibfnamefont {K.~W.}\ \bibnamefont {Chan}}, \bibinfo {author}
  {\bibfnamefont {T.}~\bibnamefont {Tanttu}}, \bibinfo {author} {\bibfnamefont
  {W.}~\bibnamefont {Gilbert}}, \bibinfo {author} {\bibfnamefont {R.~C.~C.}\
  \bibnamefont {Leon}}, \bibinfo {author} {\bibfnamefont {F.~E.}\ \bibnamefont
  {Hudson}}, \bibinfo {author} {\bibfnamefont {K.~M.}\ \bibnamefont {Itoh}},
  \bibinfo {author} {\bibfnamefont {A.}~\bibnamefont {Morello}}, \bibinfo
  {author} {\bibfnamefont {S.~D.}\ \bibnamefont {Bartlett}}, \bibinfo {author}
  {\bibfnamefont {A.}~\bibnamefont {Laucht}}, \bibinfo {author} {\bibfnamefont
  {A.}~\bibnamefont {Saraiva}}, \ and\ \bibinfo {author} {\bibfnamefont
  {A.~S.}\ \bibnamefont {Dzurak}},\ }\bibfield  {title} {\enquote {\bibinfo
  {title} {Coherent spin qubit transport in silicon},}\ }\href@noop {}
  {\bibfield  {journal} {\bibinfo  {journal} {Nature Communications}\ }\textbf
  {\bibinfo {volume} {12}},\ \bibinfo {pages} {1--9} (\bibinfo {year}
  {2021})}\BibitemShut {NoStop}%
\bibitem [{\citenamefont {Mills}\ \emph {et~al.}(2019)\citenamefont {Mills},
  \citenamefont {Zajac}, \citenamefont {Gullans}, \citenamefont {Schupp},
  \citenamefont {Hazard},\ and\ \citenamefont {Petta}}]{mills2019shuttling}%
  \BibitemOpen
  \bibfield  {author} {\bibinfo {author} {\bibfnamefont {A.~R.}\ \bibnamefont
  {Mills}}, \bibinfo {author} {\bibfnamefont {D.~M.}\ \bibnamefont {Zajac}},
  \bibinfo {author} {\bibfnamefont {M.~J.}\ \bibnamefont {Gullans}}, \bibinfo
  {author} {\bibfnamefont {F.~J.}\ \bibnamefont {Schupp}}, \bibinfo {author}
  {\bibfnamefont {T.~M.}\ \bibnamefont {Hazard}}, \ and\ \bibinfo {author}
  {\bibfnamefont {J.~R.}\ \bibnamefont {Petta}},\ }\bibfield  {title} {\enquote
  {\bibinfo {title} {Shuttling a single charge across a one-dimensional array
  of silicon quantum dots},}\ }\href@noop {} {\bibfield  {journal} {\bibinfo
  {journal} {Nature communications}\ }\textbf {\bibinfo {volume} {10}},\
  \bibinfo {pages} {1--6} (\bibinfo {year} {2019})}\BibitemShut {NoStop}%
\bibitem [{\citenamefont {Rossi}\ \emph {et~al.}(2014)\citenamefont {Rossi},
  \citenamefont {Tanttu}, \citenamefont {Tan}, \citenamefont {Iisakka},
  \citenamefont {Zhao}, \citenamefont {Chan}, \citenamefont {Tettamanzi},
  \citenamefont {Rogge}, \citenamefont {Dzurak},\ and\ \citenamefont
  {M\"ott\"onen}}]{rossi2014accurate}%
  \BibitemOpen
  \bibfield  {author} {\bibinfo {author} {\bibfnamefont {A.}~\bibnamefont
  {Rossi}}, \bibinfo {author} {\bibfnamefont {T.}~\bibnamefont {Tanttu}},
  \bibinfo {author} {\bibfnamefont {K.~Y.}\ \bibnamefont {Tan}}, \bibinfo
  {author} {\bibfnamefont {I.}~\bibnamefont {Iisakka}}, \bibinfo {author}
  {\bibfnamefont {R.}~\bibnamefont {Zhao}}, \bibinfo {author} {\bibfnamefont
  {K.~W.}\ \bibnamefont {Chan}}, \bibinfo {author} {\bibfnamefont {G.~C.}\
  \bibnamefont {Tettamanzi}}, \bibinfo {author} {\bibfnamefont
  {S.}~\bibnamefont {Rogge}}, \bibinfo {author} {\bibfnamefont {Andrew~S.}\
  \bibnamefont {Dzurak}}, \ and\ \bibinfo {author} {\bibfnamefont
  {M.}~\bibnamefont {M\"ott\"onen}},\ }\bibfield  {title} {\enquote {\bibinfo
  {title} {An accurate single-electron pump based on a highly tunable silicon
  quantum dot},}\ }\href@noop {} {\bibfield  {journal} {\bibinfo  {journal}
  {Nano letters}\ }\textbf {\bibinfo {volume} {14}},\ \bibinfo {pages}
  {3405--3411} (\bibinfo {year} {2014})}\BibitemShut {NoStop}%
\bibitem [{\citenamefont {Vahapoglu}\ \emph {et~al.}(2021)\citenamefont
  {Vahapoglu}, \citenamefont {Slack-Smith}, \citenamefont {Leon}, \citenamefont
  {Lim}, \citenamefont {Hudson}, \citenamefont {Day}, \citenamefont {Tanttu},
  \citenamefont {Yang}, \citenamefont {Laucht}, \citenamefont {Dzurak},\ and\
  \citenamefont {Pla}}]{vahapoglu2020single}%
  \BibitemOpen
  \bibfield  {author} {\bibinfo {author} {\bibfnamefont {E.}~\bibnamefont
  {Vahapoglu}}, \bibinfo {author} {\bibfnamefont {J.~P.}\ \bibnamefont
  {Slack-Smith}}, \bibinfo {author} {\bibfnamefont {R.~C.~C.}\ \bibnamefont
  {Leon}}, \bibinfo {author} {\bibfnamefont {W.~H.}\ \bibnamefont {Lim}},
  \bibinfo {author} {\bibfnamefont {F.~E.}\ \bibnamefont {Hudson}}, \bibinfo
  {author} {\bibfnamefont {T.}~\bibnamefont {Day}}, \bibinfo {author}
  {\bibfnamefont {T.}~\bibnamefont {Tanttu}}, \bibinfo {author} {\bibfnamefont
  {C.~H.}\ \bibnamefont {Yang}}, \bibinfo {author} {\bibfnamefont
  {A.}~\bibnamefont {Laucht}}, \bibinfo {author} {\bibfnamefont {A.~S.}\
  \bibnamefont {Dzurak}}, \ and\ \bibinfo {author} {\bibfnamefont {J.~J.}\
  \bibnamefont {Pla}},\ }\bibfield  {title} {\enquote {\bibinfo {title}
  {Single-electron spin resonance in a nanoelectronic device using a global
  field},}\ }\href@noop {} {\bibfield  {journal} {\bibinfo  {journal} {Science
  Advances}\ }\textbf {\bibinfo {volume} {7}},\ \bibinfo {pages} {eabg9158}
  (\bibinfo {year} {2021})}\BibitemShut {NoStop}%
\bibitem [{\citenamefont {Ferdous}\ \emph {et~al.}(2018)\citenamefont
  {Ferdous}, \citenamefont {Chan}, \citenamefont {Veldhorst}, \citenamefont
  {Hwang}, \citenamefont {Yang}, \citenamefont {Sahasrabudhe}, \citenamefont
  {Klimeck}, \citenamefont {Morello}, \citenamefont {Dzurak},\ and\
  \citenamefont {Rahman}}]{ferdous2018interface}%
  \BibitemOpen
  \bibfield  {author} {\bibinfo {author} {\bibfnamefont {R.}~\bibnamefont
  {Ferdous}}, \bibinfo {author} {\bibfnamefont {K.~W.}\ \bibnamefont {Chan}},
  \bibinfo {author} {\bibfnamefont {M.}~\bibnamefont {Veldhorst}}, \bibinfo
  {author} {\bibfnamefont {J.~C.~C.}\ \bibnamefont {Hwang}}, \bibinfo {author}
  {\bibfnamefont {C.~H.}\ \bibnamefont {Yang}}, \bibinfo {author}
  {\bibfnamefont {H.}~\bibnamefont {Sahasrabudhe}}, \bibinfo {author}
  {\bibfnamefont {G.}~\bibnamefont {Klimeck}}, \bibinfo {author} {\bibfnamefont
  {A.}~\bibnamefont {Morello}}, \bibinfo {author} {\bibfnamefont {A.~S.}\
  \bibnamefont {Dzurak}}, \ and\ \bibinfo {author} {\bibfnamefont
  {R.}~\bibnamefont {Rahman}},\ }\bibfield  {title} {\enquote {\bibinfo {title}
  {Interface-induced spin-orbit interaction in silicon quantum dots and
  prospects for scalability},}\ }\href@noop {} {\bibfield  {journal} {\bibinfo
  {journal} {Physical Review B}\ }\textbf {\bibinfo {volume} {97}},\ \bibinfo
  {pages} {241401} (\bibinfo {year} {2018})}\BibitemShut {NoStop}%
\bibitem [{\citenamefont {Hwang}\ \emph {et~al.}(2017)\citenamefont {Hwang},
  \citenamefont {Yang}, \citenamefont {Veldhorst}, \citenamefont {Hendrickx},
  \citenamefont {Fogarty}, \citenamefont {Huang}, \citenamefont {Hudson},
  \citenamefont {Morello},\ and\ \citenamefont {Dzurak}}]{hwang2017impact}%
  \BibitemOpen
  \bibfield  {author} {\bibinfo {author} {\bibfnamefont {J.~C.~C.}\
  \bibnamefont {Hwang}}, \bibinfo {author} {\bibfnamefont {C.~H.}\ \bibnamefont
  {Yang}}, \bibinfo {author} {\bibfnamefont {M.}~\bibnamefont {Veldhorst}},
  \bibinfo {author} {\bibfnamefont {N.}~\bibnamefont {Hendrickx}}, \bibinfo
  {author} {\bibfnamefont {M.~A.}\ \bibnamefont {Fogarty}}, \bibinfo {author}
  {\bibfnamefont {W.}~\bibnamefont {Huang}}, \bibinfo {author} {\bibfnamefont
  {F.~E.}\ \bibnamefont {Hudson}}, \bibinfo {author} {\bibfnamefont
  {A.}~\bibnamefont {Morello}}, \ and\ \bibinfo {author} {\bibfnamefont
  {A.~S.}\ \bibnamefont {Dzurak}},\ }\bibfield  {title} {\enquote {\bibinfo
  {title} {Impact of g-factors and valleys on spin qubits in a silicon double
  quantum dot},}\ }\href@noop {} {\bibfield  {journal} {\bibinfo  {journal}
  {Physical Review B}\ }\textbf {\bibinfo {volume} {96}},\ \bibinfo {pages}
  {045302} (\bibinfo {year} {2017})}\BibitemShut {NoStop}%
\bibitem [{\citenamefont {Jock}\ \emph {et~al.}(2018)\citenamefont {Jock},
  \citenamefont {Jacobson}, \citenamefont {Harvey-collard}, \citenamefont
  {Mounce}, \citenamefont {Srinivasa}, \citenamefont {Ward}, \citenamefont
  {Anderson}, \citenamefont {Manginell}, \citenamefont {Wendt}, \citenamefont
  {Rudolph}, \citenamefont {Pluym}, \citenamefont {Gamble}, \citenamefont
  {Baczewski}, \citenamefont {Witzel},\ and\ \citenamefont
  {Carroll}}]{jock2017probing}%
  \BibitemOpen
  \bibfield  {author} {\bibinfo {author} {\bibfnamefont {R.~M.}\ \bibnamefont
  {Jock}}, \bibinfo {author} {\bibfnamefont {N.~T.}\ \bibnamefont {Jacobson}},
  \bibinfo {author} {\bibfnamefont {P.}~\bibnamefont {Harvey-collard}},
  \bibinfo {author} {\bibfnamefont {A.~M.}\ \bibnamefont {Mounce}}, \bibinfo
  {author} {\bibfnamefont {V.}~\bibnamefont {Srinivasa}}, \bibinfo {author}
  {\bibfnamefont {D.~R.}\ \bibnamefont {Ward}}, \bibinfo {author}
  {\bibfnamefont {J.}~\bibnamefont {Anderson}}, \bibinfo {author}
  {\bibfnamefont {R.}~\bibnamefont {Manginell}}, \bibinfo {author}
  {\bibfnamefont {J.~R.}\ \bibnamefont {Wendt}}, \bibinfo {author}
  {\bibfnamefont {M.}~\bibnamefont {Rudolph}}, \bibinfo {author} {\bibfnamefont
  {T.}~\bibnamefont {Pluym}}, \bibinfo {author} {\bibfnamefont {J.~K.}\
  \bibnamefont {Gamble}}, \bibinfo {author} {\bibfnamefont {A.~D.}\
  \bibnamefont {Baczewski}}, \bibinfo {author} {\bibfnamefont {W.~M.}\
  \bibnamefont {Witzel}}, \ and\ \bibinfo {author} {\bibfnamefont {M.~S.}\
  \bibnamefont {Carroll}},\ }\bibfield  {title} {\enquote {\bibinfo {title} {A
  silicon metal-oxide-semiconductor electron spin-orbit qubit},}\ }\href@noop
  {} {\bibfield  {journal} {\bibinfo  {journal} {Nature communications}\
  }\textbf {\bibinfo {volume} {9}},\ \bibinfo {pages} {1--8} (\bibinfo {year}
  {2018})}\BibitemShut {NoStop}%
\bibitem [{\citenamefont {Vandersypen}\ and\ \citenamefont
  {Chuang}(2005)}]{vandersypen2005nmr}%
  \BibitemOpen
  \bibfield  {author} {\bibinfo {author} {\bibfnamefont {L.~M.~K.}\
  \bibnamefont {Vandersypen}}\ and\ \bibinfo {author} {\bibfnamefont {I.~L.}\
  \bibnamefont {Chuang}},\ }\bibfield  {title} {\enquote {\bibinfo {title} {Nmr
  techniques for quantum control and computation},}\ }\href@noop {} {\bibfield
  {journal} {\bibinfo  {journal} {Reviews of modern physics}\ }\textbf
  {\bibinfo {volume} {76}},\ \bibinfo {pages} {1037} (\bibinfo {year}
  {2005})}\BibitemShut {NoStop}%
\bibitem [{\citenamefont {Khaneja}\ \emph {et~al.}(2005)\citenamefont
  {Khaneja}, \citenamefont {Reiss}, \citenamefont {Kehlet}, \citenamefont
  {Schulte-Herbr{\"u}ggen},\ and\ \citenamefont {Glaser}}]{khaneja2005optimal}%
  \BibitemOpen
  \bibfield  {author} {\bibinfo {author} {\bibfnamefont {N.}~\bibnamefont
  {Khaneja}}, \bibinfo {author} {\bibfnamefont {T.}~\bibnamefont {Reiss}},
  \bibinfo {author} {\bibfnamefont {C.}~\bibnamefont {Kehlet}}, \bibinfo
  {author} {\bibfnamefont {T.}~\bibnamefont {Schulte-Herbr{\"u}ggen}}, \ and\
  \bibinfo {author} {\bibfnamefont {S.~J.}\ \bibnamefont {Glaser}},\ }\bibfield
   {title} {\enquote {\bibinfo {title} {Optimal control of coupled spin
  dynamics: design of nmr pulse sequences by gradient ascent algorithms},}\
  }\href@noop {} {\bibfield  {journal} {\bibinfo  {journal} {Journal of
  magnetic resonance}\ }\textbf {\bibinfo {volume} {172}},\ \bibinfo {pages}
  {296--305} (\bibinfo {year} {2005})}\BibitemShut {NoStop}%
\bibitem [{\citenamefont {De~Fouquieres}\ \emph {et~al.}(2011)\citenamefont
  {De~Fouquieres}, \citenamefont {Schirmer}, \citenamefont {Glaser},\ and\
  \citenamefont {Kuprov}}]{de2011second}%
  \BibitemOpen
  \bibfield  {author} {\bibinfo {author} {\bibfnamefont {P.}~\bibnamefont
  {De~Fouquieres}}, \bibinfo {author} {\bibfnamefont {S.~G.}\ \bibnamefont
  {Schirmer}}, \bibinfo {author} {\bibfnamefont {S.~J.}\ \bibnamefont
  {Glaser}}, \ and\ \bibinfo {author} {\bibfnamefont {I.}~\bibnamefont
  {Kuprov}},\ }\bibfield  {title} {\enquote {\bibinfo {title} {Second order
  gradient ascent pulse engineering},}\ }\href@noop {} {\bibfield  {journal}
  {\bibinfo  {journal} {Journal of Magnetic Resonance}\ }\textbf {\bibinfo
  {volume} {212}},\ \bibinfo {pages} {412--417} (\bibinfo {year}
  {2011})}\BibitemShut {NoStop}%
\bibitem [{\citenamefont {Green}\ \emph {et~al.}(2013)\citenamefont {Green},
  \citenamefont {Sastrawan}, \citenamefont {Uys},\ and\ \citenamefont
  {Biercuk}}]{green2013arbitrary}%
  \BibitemOpen
  \bibfield  {author} {\bibinfo {author} {\bibfnamefont {T.~J.}\ \bibnamefont
  {Green}}, \bibinfo {author} {\bibfnamefont {J.}~\bibnamefont {Sastrawan}},
  \bibinfo {author} {\bibfnamefont {H.}~\bibnamefont {Uys}}, \ and\ \bibinfo
  {author} {\bibfnamefont {M.~J.}\ \bibnamefont {Biercuk}},\ }\bibfield
  {title} {\enquote {\bibinfo {title} {Arbitrary quantum control of qubits in
  the presence of universal noise},}\ }\href@noop {} {\bibfield  {journal}
  {\bibinfo  {journal} {New Journal of Physics}\ }\textbf {\bibinfo {volume}
  {15}},\ \bibinfo {pages} {095004} (\bibinfo {year} {2013})}\BibitemShut
  {NoStop}%
\bibitem [{\citenamefont {Chan}\ \emph {et~al.}(2018)\citenamefont {Chan},
  \citenamefont {Huang}, \citenamefont {Yang}, \citenamefont {Hwang},
  \citenamefont {Hensen}, \citenamefont {Tanttu}, \citenamefont {Hudson},
  \citenamefont {Itoh}, \citenamefont {Laucht}, \citenamefont {Morello},\ and\
  \citenamefont {Dzurak}}]{chan2018assessment}%
  \BibitemOpen
  \bibfield  {author} {\bibinfo {author} {\bibfnamefont {K.~W.}\ \bibnamefont
  {Chan}}, \bibinfo {author} {\bibfnamefont {W.}~\bibnamefont {Huang}},
  \bibinfo {author} {\bibfnamefont {C.~H.}\ \bibnamefont {Yang}}, \bibinfo
  {author} {\bibfnamefont {J.~C.~C.}\ \bibnamefont {Hwang}}, \bibinfo {author}
  {\bibfnamefont {B.}~\bibnamefont {Hensen}}, \bibinfo {author} {\bibfnamefont
  {T.}~\bibnamefont {Tanttu}}, \bibinfo {author} {\bibfnamefont {F.~E.}\
  \bibnamefont {Hudson}}, \bibinfo {author} {\bibfnamefont {K.~M.}\
  \bibnamefont {Itoh}}, \bibinfo {author} {\bibfnamefont {A.}~\bibnamefont
  {Laucht}}, \bibinfo {author} {\bibfnamefont {A.}~\bibnamefont {Morello}}, \
  and\ \bibinfo {author} {\bibfnamefont {A.~S.}\ \bibnamefont {Dzurak}},\
  }\bibfield  {title} {\enquote {\bibinfo {title} {Assessment of a silicon
  quantum dot spin qubit environment via noise spectroscopy},}\ }\href@noop {}
  {\bibfield  {journal} {\bibinfo  {journal} {Physical Review Applied}\
  }\textbf {\bibinfo {volume} {10}},\ \bibinfo {pages} {044017} (\bibinfo
  {year} {2018})}\BibitemShut {NoStop}%
\bibitem [{\citenamefont {Ball}\ \emph {et~al.}(2021)\citenamefont {Ball},
  \citenamefont {Biercuk}, \citenamefont {Carvalho}, \citenamefont {Chen},
  \citenamefont {Hush}, \citenamefont {De~Castro}, \citenamefont {Li},
  \citenamefont {Liebermann}, \citenamefont {Slatyer}, \citenamefont {Edmunds},
  \citenamefont {Frey}, \citenamefont {Hempel},\ and\ \citenamefont
  {Milne}}]{ball2021software}%
  \BibitemOpen
  \bibfield  {author} {\bibinfo {author} {\bibfnamefont {H.}~\bibnamefont
  {Ball}}, \bibinfo {author} {\bibfnamefont {M.~J.}\ \bibnamefont {Biercuk}},
  \bibinfo {author} {\bibfnamefont {A.~R.~R.}\ \bibnamefont {Carvalho}},
  \bibinfo {author} {\bibfnamefont {J.}~\bibnamefont {Chen}}, \bibinfo {author}
  {\bibfnamefont {M.}~\bibnamefont {Hush}}, \bibinfo {author} {\bibfnamefont
  {L.~A.}\ \bibnamefont {De~Castro}}, \bibinfo {author} {\bibfnamefont
  {L.}~\bibnamefont {Li}}, \bibinfo {author} {\bibfnamefont {P.~J.}\
  \bibnamefont {Liebermann}}, \bibinfo {author} {\bibfnamefont {H.~J.}\
  \bibnamefont {Slatyer}}, \bibinfo {author} {\bibfnamefont {C.}~\bibnamefont
  {Edmunds}}, \bibinfo {author} {\bibfnamefont {V.}~\bibnamefont {Frey}},
  \bibinfo {author} {\bibfnamefont {C.}~\bibnamefont {Hempel}}, \ and\ \bibinfo
  {author} {\bibfnamefont {A.}~\bibnamefont {Milne}},\ }\bibfield  {title}
  {\enquote {\bibinfo {title} {Software tools for quantum control: Improving
  quantum computer performance through noise and error suppression},}\
  }\href@noop {} {\bibfield  {journal} {\bibinfo  {journal} {Quantum Science
  and Technology}\ }\textbf {\bibinfo {volume} {6}},\ \bibinfo {pages} {044011}
  (\bibinfo {year} {2021})}\BibitemShut {NoStop}%
\bibitem [{\citenamefont {Meunier}\ \emph {et~al.}(2011)\citenamefont
  {Meunier}, \citenamefont {Calado},\ and\ \citenamefont
  {Vandersypen}}]{meunier2011efficient}%
  \BibitemOpen
  \bibfield  {author} {\bibinfo {author} {\bibfnamefont {T.}~\bibnamefont
  {Meunier}}, \bibinfo {author} {\bibfnamefont {V.~E.}\ \bibnamefont {Calado}},
  \ and\ \bibinfo {author} {\bibfnamefont {L.~M.~K.}\ \bibnamefont
  {Vandersypen}},\ }\bibfield  {title} {\enquote {\bibinfo {title} {Efficient
  controlled-phase gate for single-spin qubits in quantum dots},}\ }\href@noop
  {} {\bibfield  {journal} {\bibinfo  {journal} {Physical Review B}\ }\textbf
  {\bibinfo {volume} {83}},\ \bibinfo {pages} {121403} (\bibinfo {year}
  {2011})}\BibitemShut {NoStop}%
\bibitem [{\citenamefont {Hasler}\ \emph {et~al.}(2021)\citenamefont {Hasler},
  \citenamefont {Dick}, \citenamefont {Das}, \citenamefont {Degnans},
  \citenamefont {Moini},\ and\ \citenamefont {Reilly}}]{hasler2021cryogenic}%
  \BibitemOpen
  \bibfield  {author} {\bibinfo {author} {\bibfnamefont {J.}~\bibnamefont
  {Hasler}}, \bibinfo {author} {\bibfnamefont {N.}~\bibnamefont {Dick}},
  \bibinfo {author} {\bibfnamefont {K.}~\bibnamefont {Das}}, \bibinfo {author}
  {\bibfnamefont {B.}~\bibnamefont {Degnans}}, \bibinfo {author} {\bibfnamefont
  {A.}~\bibnamefont {Moini}}, \ and\ \bibinfo {author} {\bibfnamefont
  {D.}~\bibnamefont {Reilly}},\ }\bibfield  {title} {\enquote {\bibinfo {title}
  {Cryogenic floating-gate cmos circuits for quantum control},}\ }\href@noop {}
  {\bibfield  {journal} {\bibinfo  {journal} {IEEE Transactions on Quantum
  Engineering}\ } (\bibinfo {year} {2021})}\BibitemShut {NoStop}%
\bibitem [{\citenamefont {Schaal}\ \emph {et~al.}(2018)\citenamefont {Schaal},
  \citenamefont {Barraud}, \citenamefont {Morton},\ and\ \citenamefont
  {Gonzalez-Zalba}}]{schaal2018conditional}%
  \BibitemOpen
  \bibfield  {author} {\bibinfo {author} {\bibfnamefont {S.}~\bibnamefont
  {Schaal}}, \bibinfo {author} {\bibfnamefont {S.}~\bibnamefont {Barraud}},
  \bibinfo {author} {\bibfnamefont {J.~J.~L.}\ \bibnamefont {Morton}}, \ and\
  \bibinfo {author} {\bibfnamefont {M.~F.}\ \bibnamefont {Gonzalez-Zalba}},\
  }\bibfield  {title} {\enquote {\bibinfo {title} {Conditional dispersive
  readout of a cmos single-electron memory cell},}\ }\href@noop {} {\bibfield
  {journal} {\bibinfo  {journal} {Physical Review Applied}\ }\textbf {\bibinfo
  {volume} {9}},\ \bibinfo {pages} {054016} (\bibinfo {year}
  {2018})}\BibitemShut {NoStop}%
\bibitem [{\citenamefont {Laucht}\ \emph {et~al.}(2021)\citenamefont {Laucht},
  \citenamefont {Hohls}, \citenamefont {Ubbelohde}, \citenamefont
  {Gonzalez-Zalba}, \citenamefont {Reilly}, \citenamefont {Stobbe},
  \citenamefont {Schr{\"o}der}, \citenamefont {Scarlino}, \citenamefont
  {Koski}, \citenamefont {Dzurak}, \citenamefont {Yang}, \citenamefont
  {Yoneda}, \citenamefont {Kuemmeth}, , \citenamefont {Bluhm}, \citenamefont
  {Pla}, \citenamefont {Hill}, \citenamefont {Salfi}, \citenamefont {Oiwa},
  \citenamefont {Muhonen}, \citenamefont {Verhagen}, \citenamefont {LaHaye},
  \citenamefont {Kim}, \citenamefont {Tsen}, \citenamefont {Culcer},
  \citenamefont {Geresdi}, \citenamefont {Mol}, \citenamefont {Mohan},
  \citenamefont {Jain},\ and\ \citenamefont {Baugh}}]{laucht2021roadmap}%
  \BibitemOpen
  \bibfield  {author} {\bibinfo {author} {\bibfnamefont {A.}~\bibnamefont
  {Laucht}}, \bibinfo {author} {\bibfnamefont {F.}~\bibnamefont {Hohls}},
  \bibinfo {author} {\bibfnamefont {N.}~\bibnamefont {Ubbelohde}}, \bibinfo
  {author} {\bibfnamefont {M.~F.}\ \bibnamefont {Gonzalez-Zalba}}, \bibinfo
  {author} {\bibfnamefont {D.~J.}\ \bibnamefont {Reilly}}, \bibinfo {author}
  {\bibfnamefont {S.}~\bibnamefont {Stobbe}}, \bibinfo {author} {\bibfnamefont
  {T.}~\bibnamefont {Schr{\"o}der}}, \bibinfo {author} {\bibfnamefont
  {P.}~\bibnamefont {Scarlino}}, \bibinfo {author} {\bibfnamefont {J.~V.}\
  \bibnamefont {Koski}}, \bibinfo {author} {\bibfnamefont {A.}~\bibnamefont
  {Dzurak}}, \bibinfo {author} {\bibfnamefont {C.-H.}\ \bibnamefont {Yang}},
  \bibinfo {author} {\bibfnamefont {J.}~\bibnamefont {Yoneda}}, \bibinfo
  {author} {\bibfnamefont {F.}~\bibnamefont {Kuemmeth}}, , \bibinfo {author}
  {\bibfnamefont {H.}~\bibnamefont {Bluhm}}, \bibinfo {author} {\bibfnamefont
  {J.}~\bibnamefont {Pla}}, \bibinfo {author} {\bibfnamefont {C.}~\bibnamefont
  {Hill}}, \bibinfo {author} {\bibfnamefont {J.}~\bibnamefont {Salfi}},
  \bibinfo {author} {\bibfnamefont {A.}~\bibnamefont {Oiwa}}, \bibinfo {author}
  {\bibfnamefont {J.~T.}\ \bibnamefont {Muhonen}}, \bibinfo {author}
  {\bibfnamefont {E.}~\bibnamefont {Verhagen}}, \bibinfo {author}
  {\bibfnamefont {M.~D.}\ \bibnamefont {LaHaye}}, \bibinfo {author}
  {\bibfnamefont {H.~H.}\ \bibnamefont {Kim}}, \bibinfo {author} {\bibfnamefont
  {A.~W.}\ \bibnamefont {Tsen}}, \bibinfo {author} {\bibfnamefont
  {D.}~\bibnamefont {Culcer}}, \bibinfo {author} {\bibfnamefont
  {A.}~\bibnamefont {Geresdi}}, \bibinfo {author} {\bibfnamefont {J.~A.}\
  \bibnamefont {Mol}}, \bibinfo {author} {\bibfnamefont {V.}~\bibnamefont
  {Mohan}}, \bibinfo {author} {\bibfnamefont {P.~K.}\ \bibnamefont {Jain}}, \
  and\ \bibinfo {author} {\bibfnamefont {J.}~\bibnamefont {Baugh}},\ }\bibfield
   {title} {\enquote {\bibinfo {title} {Roadmap on quantum nanotechnologies},}\
  }\href@noop {} {\bibfield  {journal} {\bibinfo  {journal} {Nanotechnology}\
  }\textbf {\bibinfo {volume} {32}},\ \bibinfo {pages} {162003} (\bibinfo
  {year} {2021})}\BibitemShut {NoStop}%
\bibitem [{\citenamefont {Singh}\ \emph {et~al.}(2020)\citenamefont {Singh},
  \citenamefont {Clarke}, \citenamefont {Veldhorst},\ and\ \citenamefont
  {Vandersypen}}]{singh2020quantum}%
  \BibitemOpen
  \bibfield  {author} {\bibinfo {author} {\bibfnamefont {K.}~\bibnamefont
  {Singh}}, \bibinfo {author} {\bibfnamefont {J.~S.}\ \bibnamefont {Clarke}},
  \bibinfo {author} {\bibfnamefont {M.}~\bibnamefont {Veldhorst}}, \ and\
  \bibinfo {author} {\bibfnamefont {L.~M.~K.}\ \bibnamefont {Vandersypen}},\
  }\href@noop {} {\enquote {\bibinfo {title} {Quantum dot devices},}\ }
  (\bibinfo {year} {2020}),\ \bibinfo {note} {\ US Patent App.
  16/616,427}\BibitemShut {NoStop}%
\bibitem [{\citenamefont {Pioro-Ladriere}\ \emph {et~al.}(2008)\citenamefont
  {Pioro-Ladriere}, \citenamefont {Obata}, \citenamefont {Tokura},
  \citenamefont {Shin}, \citenamefont {Kubo}, \citenamefont {Yoshida},
  \citenamefont {Taniyama},\ and\ \citenamefont
  {Tarucha}}]{pioro2008electrically}%
  \BibitemOpen
  \bibfield  {author} {\bibinfo {author} {\bibfnamefont {M.}~\bibnamefont
  {Pioro-Ladriere}}, \bibinfo {author} {\bibfnamefont {T.}~\bibnamefont
  {Obata}}, \bibinfo {author} {\bibfnamefont {Y.}~\bibnamefont {Tokura}},
  \bibinfo {author} {\bibfnamefont {Y.-S.}\ \bibnamefont {Shin}}, \bibinfo
  {author} {\bibfnamefont {T.}~\bibnamefont {Kubo}}, \bibinfo {author}
  {\bibfnamefont {K.}~\bibnamefont {Yoshida}}, \bibinfo {author} {\bibfnamefont
  {T.}~\bibnamefont {Taniyama}}, \ and\ \bibinfo {author} {\bibfnamefont
  {S.}~\bibnamefont {Tarucha}},\ }\bibfield  {title} {\enquote {\bibinfo
  {title} {Electrically driven single-electron spin resonance in a slanting
  zeeman field},}\ }\href@noop {} {\bibfield  {journal} {\bibinfo  {journal}
  {Nature Physics}\ }\textbf {\bibinfo {volume} {4}},\ \bibinfo {pages}
  {776--779} (\bibinfo {year} {2008})}\BibitemShut {NoStop}%
\bibitem [{\citenamefont {Wu}\ \emph {et~al.}(2014)\citenamefont {Wu},
  \citenamefont {Ward}, \citenamefont {Prance}, \citenamefont {Kim},
  \citenamefont {Gamble}, \citenamefont {Mohr}, \citenamefont {Shi},
  \citenamefont {Savage}, \citenamefont {Lagally}, \citenamefont {Friesen},
  \citenamefont {Coppersmith},\ and\ \citenamefont {Eriksson}}]{wu2014two}%
  \BibitemOpen
  \bibfield  {author} {\bibinfo {author} {\bibfnamefont {X.}~\bibnamefont
  {Wu}}, \bibinfo {author} {\bibfnamefont {D.~R.}\ \bibnamefont {Ward}},
  \bibinfo {author} {\bibfnamefont {J.~R.}\ \bibnamefont {Prance}}, \bibinfo
  {author} {\bibfnamefont {D.}~\bibnamefont {Kim}}, \bibinfo {author}
  {\bibfnamefont {J.~K.}\ \bibnamefont {Gamble}}, \bibinfo {author}
  {\bibfnamefont {R.~T.}\ \bibnamefont {Mohr}}, \bibinfo {author}
  {\bibfnamefont {Z.}~\bibnamefont {Shi}}, \bibinfo {author} {\bibfnamefont
  {D.~E.}\ \bibnamefont {Savage}}, \bibinfo {author} {\bibfnamefont {M.~G.}\
  \bibnamefont {Lagally}}, \bibinfo {author} {\bibfnamefont {M.}~\bibnamefont
  {Friesen}}, \bibinfo {author} {\bibfnamefont {S.~N.}\ \bibnamefont
  {Coppersmith}}, \ and\ \bibinfo {author} {\bibfnamefont {M.~A.}\ \bibnamefont
  {Eriksson}},\ }\bibfield  {title} {\enquote {\bibinfo {title} {Two-axis
  control of a singlet--triplet qubit with an integrated micromagnet},}\
  }\href@noop {} {\bibfield  {journal} {\bibinfo  {journal} {Proceedings of the
  National Academy of Sciences}\ }\textbf {\bibinfo {volume} {111}},\ \bibinfo
  {pages} {11938--11942} (\bibinfo {year} {2014})}\BibitemShut {NoStop}%
\bibitem [{\citenamefont {Maurand}\ \emph {et~al.}(2016)\citenamefont
  {Maurand}, \citenamefont {Jehl}, \citenamefont {Kotekar-Patil}, \citenamefont
  {Corna}, \citenamefont {Bohuslavskyi}, \citenamefont {Lavi{\'e}ville},
  \citenamefont {Hutin}, \citenamefont {Barraud}, \citenamefont {Vinet},
  \citenamefont {Sanquer},\ and\ \citenamefont
  {De~Franceschi}}]{maurand2016cmos}%
  \BibitemOpen
  \bibfield  {author} {\bibinfo {author} {\bibfnamefont {R.}~\bibnamefont
  {Maurand}}, \bibinfo {author} {\bibfnamefont {X.}~\bibnamefont {Jehl}},
  \bibinfo {author} {\bibfnamefont {D.}~\bibnamefont {Kotekar-Patil}}, \bibinfo
  {author} {\bibfnamefont {A.}~\bibnamefont {Corna}}, \bibinfo {author}
  {\bibfnamefont {H.}~\bibnamefont {Bohuslavskyi}}, \bibinfo {author}
  {\bibfnamefont {R.}~\bibnamefont {Lavi{\'e}ville}}, \bibinfo {author}
  {\bibfnamefont {L.}~\bibnamefont {Hutin}}, \bibinfo {author} {\bibfnamefont
  {S.}~\bibnamefont {Barraud}}, \bibinfo {author} {\bibfnamefont
  {M.}~\bibnamefont {Vinet}}, \bibinfo {author} {\bibfnamefont
  {M.}~\bibnamefont {Sanquer}}, \ and\ \bibinfo {author} {\bibfnamefont
  {S.}~\bibnamefont {De~Franceschi}},\ }\bibfield  {title} {\enquote {\bibinfo
  {title} {A cmos silicon spin qubit},}\ }\href@noop {} {\bibfield  {journal}
  {\bibinfo  {journal} {Nature communications}\ }\textbf {\bibinfo {volume}
  {7}},\ \bibinfo {pages} {1--6} (\bibinfo {year} {2016})}\BibitemShut
  {NoStop}%
\bibitem [{\citenamefont {Shi}\ \emph {et~al.}(2012)\citenamefont {Shi},
  \citenamefont {Simmons}, \citenamefont {Prance}, \citenamefont {Gamble},
  \citenamefont {Koh}, \citenamefont {Shim}, \citenamefont {Hu}, \citenamefont
  {Savage}, \citenamefont {Lagally}, \citenamefont {Eriksson}, \citenamefont
  {Friesen},\ and\ \citenamefont {Coppersmith}}]{shi2012fast}%
  \BibitemOpen
  \bibfield  {author} {\bibinfo {author} {\bibfnamefont {Z.}~\bibnamefont
  {Shi}}, \bibinfo {author} {\bibfnamefont {C.~B.}\ \bibnamefont {Simmons}},
  \bibinfo {author} {\bibfnamefont {J.~R.}\ \bibnamefont {Prance}}, \bibinfo
  {author} {\bibfnamefont {J.~K.}\ \bibnamefont {Gamble}}, \bibinfo {author}
  {\bibfnamefont {T.~S.}\ \bibnamefont {Koh}}, \bibinfo {author} {\bibfnamefont
  {Y.-P.}\ \bibnamefont {Shim}}, \bibinfo {author} {\bibfnamefont
  {X.}~\bibnamefont {Hu}}, \bibinfo {author} {\bibfnamefont {D.~E.}\
  \bibnamefont {Savage}}, \bibinfo {author} {\bibfnamefont {M.~G.}\
  \bibnamefont {Lagally}}, \bibinfo {author} {\bibfnamefont {M.~A.}\
  \bibnamefont {Eriksson}}, \bibinfo {author} {\bibfnamefont {M.}~\bibnamefont
  {Friesen}}, \ and\ \bibinfo {author} {\bibfnamefont {S.~N.}\ \bibnamefont
  {Coppersmith}},\ }\bibfield  {title} {\enquote {\bibinfo {title} {Fast hybrid
  silicon double-quantum-dot qubit},}\ }\href@noop {} {\bibfield  {journal}
  {\bibinfo  {journal} {Physical review letters}\ }\textbf {\bibinfo {volume}
  {108}},\ \bibinfo {pages} {140503} (\bibinfo {year} {2012})}\BibitemShut
  {NoStop}%
\bibitem [{\citenamefont {Hayashi}\ \emph {et~al.}(2003)\citenamefont
  {Hayashi}, \citenamefont {Fujisawa}, \citenamefont {Cheong}, \citenamefont
  {Jeong},\ and\ \citenamefont {Hirayama}}]{hayashi2003coherent}%
  \BibitemOpen
  \bibfield  {author} {\bibinfo {author} {\bibfnamefont {T.}~\bibnamefont
  {Hayashi}}, \bibinfo {author} {\bibfnamefont {T.}~\bibnamefont {Fujisawa}},
  \bibinfo {author} {\bibfnamefont {H.-D.}\ \bibnamefont {Cheong}}, \bibinfo
  {author} {\bibfnamefont {Y.~H.}\ \bibnamefont {Jeong}}, \ and\ \bibinfo
  {author} {\bibfnamefont {Y.}~\bibnamefont {Hirayama}},\ }\bibfield  {title}
  {\enquote {\bibinfo {title} {Coherent manipulation of electronic states in a
  double quantum dot},}\ }\href@noop {} {\bibfield  {journal} {\bibinfo
  {journal} {Physical review letters}\ }\textbf {\bibinfo {volume} {91}},\
  \bibinfo {pages} {226804} (\bibinfo {year} {2003})}\BibitemShut {NoStop}%
\bibitem [{\citenamefont {Oakes}\ \emph {et~al.}(2022)\citenamefont {Oakes},
  \citenamefont {Ciriano-Tejel}, \citenamefont {Wise}, \citenamefont {Fogarty},
  \citenamefont {Lundberg}, \citenamefont {Lain{\'e}}, \citenamefont {Schaal},
  \citenamefont {Martins}, \citenamefont {Ibberson}, \citenamefont {Hutin},
  \citenamefont {Bertrand}, \citenamefont {Stelmashenko}, \citenamefont
  {Robinson}, \citenamefont {Ibberson}, \citenamefont {Hashim}, \citenamefont
  {Siddiqi}, \citenamefont {Lee}, \citenamefont {Vinet}, \citenamefont {Smith},
  \citenamefont {Morton},\ and\ \citenamefont
  {Gonzalez-Zalba}}]{oakes2022fast}%
  \BibitemOpen
  \bibfield  {author} {\bibinfo {author} {\bibfnamefont {G.~A.}\ \bibnamefont
  {Oakes}}, \bibinfo {author} {\bibfnamefont {V.~N.}\ \bibnamefont
  {Ciriano-Tejel}}, \bibinfo {author} {\bibfnamefont {D.}~\bibnamefont {Wise}},
  \bibinfo {author} {\bibfnamefont {M.~A.}\ \bibnamefont {Fogarty}}, \bibinfo
  {author} {\bibfnamefont {T.}~\bibnamefont {Lundberg}}, \bibinfo {author}
  {\bibfnamefont {C.}~\bibnamefont {Lain{\'e}}}, \bibinfo {author}
  {\bibfnamefont {S.}~\bibnamefont {Schaal}}, \bibinfo {author} {\bibfnamefont
  {F.}~\bibnamefont {Martins}}, \bibinfo {author} {\bibfnamefont {D.~J.}\
  \bibnamefont {Ibberson}}, \bibinfo {author} {\bibfnamefont {L.}~\bibnamefont
  {Hutin}}, \bibinfo {author} {\bibfnamefont {B.}~\bibnamefont {Bertrand}},
  \bibinfo {author} {\bibfnamefont {N.}~\bibnamefont {Stelmashenko}}, \bibinfo
  {author} {\bibfnamefont {J.~A.~W.}\ \bibnamefont {Robinson}}, \bibinfo
  {author} {\bibfnamefont {L.}~\bibnamefont {Ibberson}}, \bibinfo {author}
  {\bibfnamefont {A.}~\bibnamefont {Hashim}}, \bibinfo {author} {\bibfnamefont
  {I.}~\bibnamefont {Siddiqi}}, \bibinfo {author} {\bibfnamefont
  {A.}~\bibnamefont {Lee}}, \bibinfo {author} {\bibfnamefont {M.}~\bibnamefont
  {Vinet}}, \bibinfo {author} {\bibfnamefont {C.~G.}\ \bibnamefont {Smith}},
  \bibinfo {author} {\bibfnamefont {J.~L.~L.}\ \bibnamefont {Morton}}, \ and\
  \bibinfo {author} {\bibfnamefont {M.~F.}\ \bibnamefont {Gonzalez-Zalba}},\
  }\bibfield  {title} {\enquote {\bibinfo {title} {Fast high-fidelity
  single-shot readout of spins in silicon using a single-electron box},}\
  }\href@noop {} {\bibfield  {journal} {\bibinfo  {journal} {arXiv:2203.06608}\
  } (\bibinfo {year} {2022})}\BibitemShut {NoStop}%
\bibitem [{\citenamefont {Schaal}\ \emph {et~al.}(2019)\citenamefont {Schaal},
  \citenamefont {Rossi}, \citenamefont {Ciriano-Tejel}, \citenamefont
  {Barraud}, \citenamefont {Morton},\ and\ \citenamefont
  {Gonzalez-Zalba}}]{schaal2019cmos}%
  \BibitemOpen
  \bibfield  {author} {\bibinfo {author} {\bibfnamefont {S.}~\bibnamefont
  {Schaal}}, \bibinfo {author} {\bibfnamefont {A.}~\bibnamefont {Rossi}},
  \bibinfo {author} {\bibfnamefont {T.-Y.}\ \bibnamefont {Ciriano-Tejel},
  \bibfnamefont {V.~N .and~Yang}}, \bibinfo {author} {\bibfnamefont
  {S.}~\bibnamefont {Barraud}}, \bibinfo {author} {\bibfnamefont {J.~J.~L.}\
  \bibnamefont {Morton}}, \ and\ \bibinfo {author} {\bibfnamefont {M.~F.}\
  \bibnamefont {Gonzalez-Zalba}},\ }\bibfield  {title} {\enquote {\bibinfo
  {title} {A cmos dynamic random access architecture for radio-frequency
  readout of quantum devices},}\ }\href@noop {} {\bibfield  {journal} {\bibinfo
   {journal} {Nature Electronics}\ }\textbf {\bibinfo {volume} {2}},\ \bibinfo
  {pages} {236--242} (\bibinfo {year} {2019})}\BibitemShut {NoStop}%
\bibitem [{\citenamefont {Horsman}\ \emph {et~al.}(2012)\citenamefont
  {Horsman}, \citenamefont {Fowler}, \citenamefont {Devitt},\ and\
  \citenamefont {Van~Meter}}]{horsman2012surface}%
  \BibitemOpen
  \bibfield  {author} {\bibinfo {author} {\bibfnamefont {C.}~\bibnamefont
  {Horsman}}, \bibinfo {author} {\bibfnamefont {A.~G.}\ \bibnamefont {Fowler}},
  \bibinfo {author} {\bibfnamefont {S.}~\bibnamefont {Devitt}}, \ and\ \bibinfo
  {author} {\bibfnamefont {R.}~\bibnamefont {Van~Meter}},\ }\bibfield  {title}
  {\enquote {\bibinfo {title} {Surface code quantum computing by lattice
  surgery},}\ }\href@noop {} {\bibfield  {journal} {\bibinfo  {journal} {New
  Journal of Physics}\ }\textbf {\bibinfo {volume} {14}},\ \bibinfo {pages}
  {123011} (\bibinfo {year} {2012})}\BibitemShut {NoStop}%
\bibitem [{\citenamefont {Strikis}\ \emph {et~al.}(2021)\citenamefont
  {Strikis}, \citenamefont {Benjamin},\ and\ \citenamefont
  {Brown}}]{strikis2021quantum}%
  \BibitemOpen
  \bibfield  {author} {\bibinfo {author} {\bibfnamefont {A.}~\bibnamefont
  {Strikis}}, \bibinfo {author} {\bibfnamefont {S.~C.}\ \bibnamefont
  {Benjamin}}, \ and\ \bibinfo {author} {\bibfnamefont {B.~J.}\ \bibnamefont
  {Brown}},\ }\bibfield  {title} {\enquote {\bibinfo {title} {Quantum computing
  is scalable on a planar array of qubits with fabrication defects},}\
  }\href@noop {} {\bibfield  {journal} {\bibinfo  {journal} {arXiv:2111.06432}\
  } (\bibinfo {year} {2021})}\BibitemShut {NoStop}%
\bibitem [{\citenamefont {Fujita}\ and\ \citenamefont
  {Amemiya}(1993)}]{fujita1993floating}%
  \BibitemOpen
  \bibfield  {author} {\bibinfo {author} {\bibfnamefont {O.}~\bibnamefont
  {Fujita}}\ and\ \bibinfo {author} {\bibfnamefont {Y.}~\bibnamefont
  {Amemiya}},\ }\bibfield  {title} {\enquote {\bibinfo {title} {A floating-gate
  analog memory device for neural networks},}\ }\href@noop {} {\bibfield
  {journal} {\bibinfo  {journal} {IEEE transactions on electron devices}\
  }\textbf {\bibinfo {volume} {40}},\ \bibinfo {pages} {2029--2035} (\bibinfo
  {year} {1993})}\BibitemShut {NoStop}%
\end{thebibliography}%

\newpage
\linenumbers

\renewcommand{\thefigure}{S\arabic{figure}}
\setcounter{figure}{0}

\onecolumngrid{
\newpage
\section*{Supplementary Figures}
\vspace*{-5mm}
}

\begin{figure*}[h]
 \centering
  \begin{center}
  \includegraphics[width=8cm]{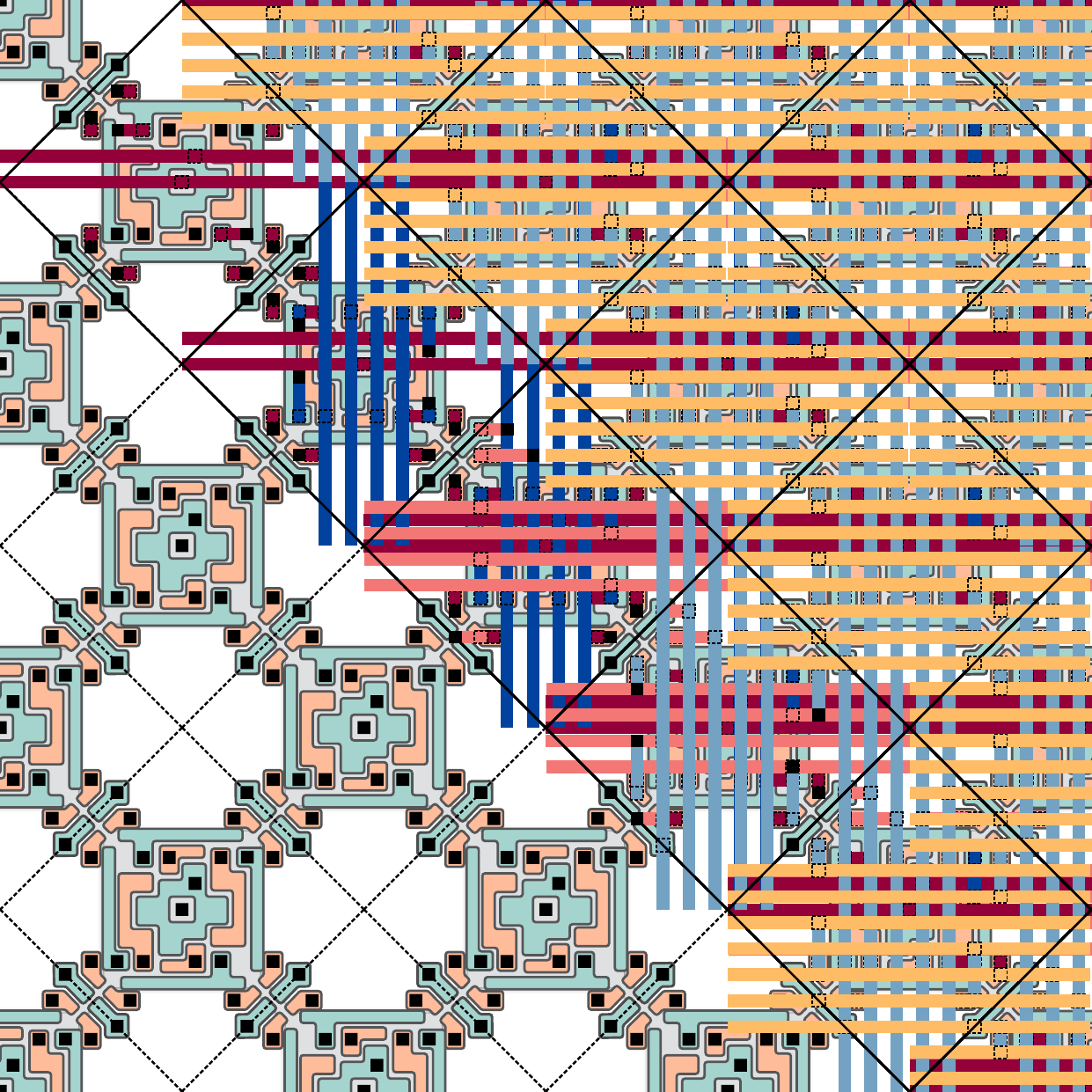}
  \end{center}
  \vspace*{-6mm}
  \caption{\textbf{$\vert$Integration of unit cells.} The complete tiling of the unit cell FEOL combined with staggered layering of BEOL routing layers is shown. This solution illustrates direct connection of cells, however the space between cells amounts to $\sim(5\times p)^2$, where $p$ represents the routing pitch, and could be utilised for the layout of hardware in the FEOL such as elements in \S~\ref{SSec:Trim}. A solution with integrated supporting hardware would require an alternative routing scheme compared to the one depicted here.} \label{fig:UCFab_Int}
\end{figure*}

\begin{figure*}[h]
 \centering
  \begin{center}
  \includegraphics[width=18cm]{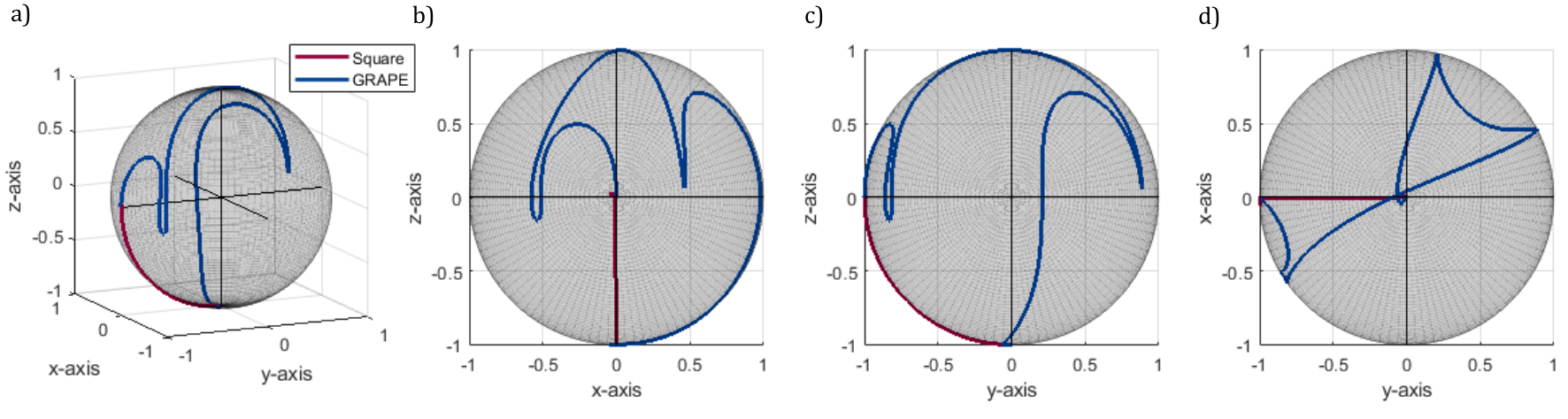}
  \end{center}
  \vspace*{-6mm}
  \caption{\textbf{$\vert$Projections of the  GRAPE pulse. a)} Oblique view on the Bloch sphere showing the time evolution of the Hadamard pulse optimised through the GRAPE method (blue), with the initial state vector $\ket{\psi} =\ket{\downarrow}$ (reproduced from the main text). The spin trajectory is made clearer through the additional projections of this pulse onto a plane normal to the \textbf{b)} y-axis, \textbf{c)} x-axis and \textbf{d)} z-axis. The Hadamard gate executed by square pulses is also illustrated (red) with a small detuning offset added to accentuate the component of rotation around the x-axis.} \label{fig:GrapeSpheres}
\end{figure*}

\begin{figure*}[h]
 \centering
  \begin{center}
  \includegraphics[width=7cm]{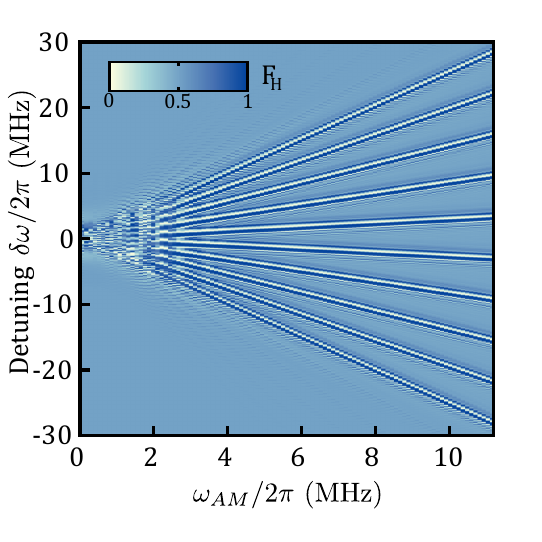}
  \end{center}
  \vspace*{-6mm}
  \caption{\textbf{$\vert$Amplitude Modulation cross-talk} Fidelity surface for the Amplitude Modulation method described in the main text and Supplementary Note~\ref{SupNote:AM_Scheme}. Each band is superimposed with the GRAPE solution for the Hadamard as described in the main text. Here N=10 side-bands are shown for increasing side-band separation $\omega_{AM}$.}\label{fig:Cavity_AM10Env}
\end{figure*}

\begin{figure*}[h]
 \centering
  \begin{center}
  \includegraphics[width=18cm]{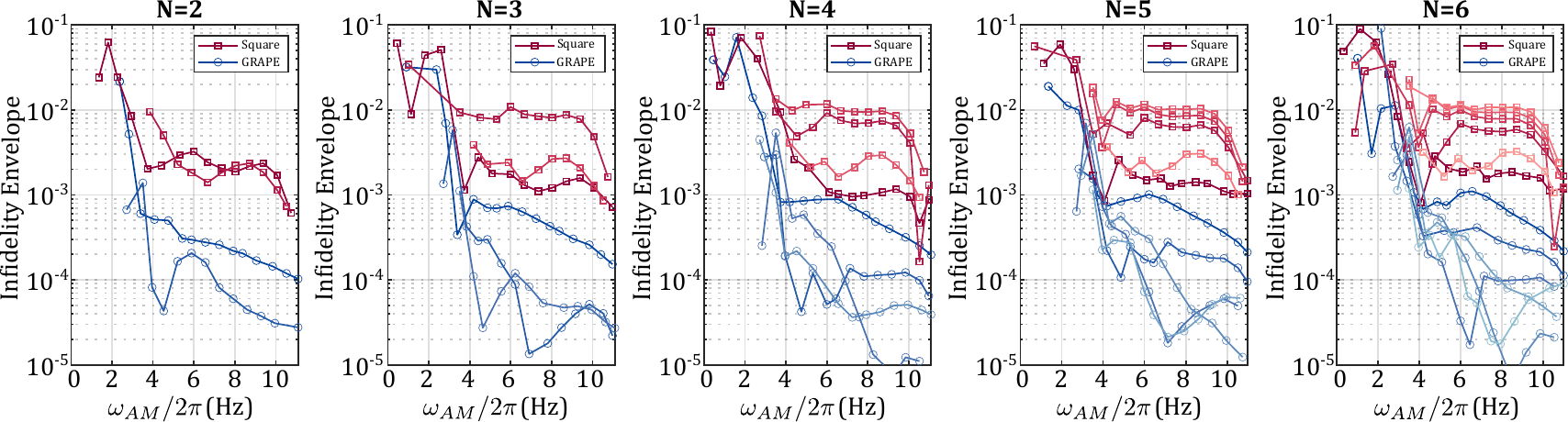}
  \end{center}
  \vspace*{-6mm}
  \caption{\textbf{$\vert$Amplitude Modulation cross-talk} Comparison of the Hadamard gate as executed via a set of square pulses (red squares) against the GRAPE solution (blue circles) for increasing number of sidebands N under the Amplitude Modulation scheme discussed int he main text. It is seen that the degree of cross-talk between sidebands is much higher for the square pulse solution compared to the GRAPE, allowing the GRAPE pulses to be spaced much closer together, catering for a smaller range in the stark-shift of the electron $g$-factor. It is noted that the lowest frequency band for the GRAPE solution has the worst performance, compared to the outermost bands for the square solution. This is attributed to the asymmetry of the GRAPE pulse trace as seen in Fig.~\ref{fig:CavityControl}b in the main text.}\label{fig:AM_squareVSgrape}
\end{figure*}

\newpage
\onecolumngrid{
\newpage
\renewcommand*{\thesection}{\Alph{section}.}
\setcounter{section}{19}
\renewcommand*{\thesubsection}{S\arabic{subsection}}

\nolinenumbers

\section*{Supplementary Notes}
\vspace*{-5mm}

\subsection{Time-Steps for the Surface Code}\label{SupNote:SC_TimeSteps}
A breakdown of the individual time steps within the surface code cycle is as follows:
\begin{itemize}
    \item[1.] Initialisation of the $X$ ancilla as a singlet $\ket{S}$. Data qubits located in D1$_{A}$ and D2$_{A}$ dots are idle. This time step is also coincident with the measurement of the $Z$ ancilla from the previous cycle (grey in Fig.~\ref{fig:SC_Unit}a).
    \item[2.] Hadamard gates applied to all active spin qubits (D1$_{A}$,D2$_{A}$ and individual elements of $X$) coincident with initialisation of the $Z$ ancilla as a singlet $\ket{S}$.
    \item[3.] Concurrently applied CZ operations, internal to the hardware unit cell, between each element of the $X$ ancilla and D1$_{A}$ or D2${_A}$. These CZ operations are applied via mediator quantum dots~\cite{cai2019silicon,malinowski2019fast}.
    \item[4.] Hadamard gate globally applied to all individual spins. 
    \item[5.] Shuttling of the data qubit form within dot location $A$ to $B$. This constitutes a charge shuttling from one side of a nanowire to another.
    \item[6.] Concurrently applied CZ operations between each element of the $Z$ ancilla and D1$_{B}$ or D2${_B}$, internal to the hardware unit cell.
    \item[7.] Hadamard gate applied to all spins. For the $X$ ancilla, these Hadamard operations cancel with the operations form time step 4.
    \item[8.] Concurrently applied CZ operations between each element of the $X$ ancilla and D1${_B'}$ or D2${_B'}$. CZ gates are external to the hardware unit cell, shown as grey in Fig.~\ref{fig:SC_Unit}a).
    \item[9.] Hadamard gate applied to all spin qubits.
    \item[10.] Shuttling of the data qubit form within dot location $B$ to $A$.
    \item[11.] Concurrently applied CZ operations between each element of the $Z$ ancilla and D1${_A'}$ or D2${_A'}$. Gates are external to the hardware unit cell.
    \item[12.] Hadamard gate applied to all active spin qubits coincident with the measurement of the $X$ ancilla. The unassigned Hadamard gate applied to D1$_{A}$ and D2${_A}$ effectively swaps $Z$ and $X$ ancillas for the next cycle unless a global Hadamard is applied in step 1 of the next cycle. 
\end{itemize}

\subsection{Amplitude Modulation Scheme}\label{SupNote:AM_Scheme}
An amplitude modulation scheme which can be utilised to increase the number of resonant peaks output by the cavity from a single peak, to N peaks centred at $\omega_0$, is given by: 
\begin{align*}
    I_{N}(t) &= A_{AM}(\omega_{AM},N,t)\cdot I_0(t)\\
    Q_{N}(t) &= A_{AM}(\omega_{AM},N,t)\cdot Q_0(t)\\
    A_{AM}(\omega_{AM},N,t) &=  \begin{cases}
    \Sigma_{n=1}^{(N-1)/2} [2\cos{(\frac{n\omega_{AM}t}{2} - n\phi)}]+1, & \text{odd }n\\
    \Sigma_{n=1}^{N/2} [2\cos{(\frac{(2n-1)\omega_{AM}t}{4} - n\phi)}], & \text{even }n
    \end{cases}\\
    \phi &= \frac{\omega_{AM}\tau}{2}.\tag{S1}\label{Eq.AM}
\end{align*}
Here, $A_{AM}$ is the modulation envelope which results in the replication of the pulse defined by $I_0(t)$ and $Q_0(t)$ at frequency detunings separated by $\omega_{AM}$. For each sideband, a phase correction term $\phi_f$ is included, which is dependent upon the detuning frequency and pulse duration $\tau$. 

\subsection{Resistive Trimmer Feasibility}\label{SupNote:Trimmer_resistive}
The trimmer configuration in Fig.~\ref{fig:Trimmer}a) of the main text works on the ratio $R_D : R_{T,ch}$, the system has substantially more tolerance to variations in devices due to fabrication in both the trimming circuitry and quantum dots, as the value of $R_{T,ch}$ is quasi-continuous~\cite{fujita1993floating}. Regarding the necessary operating range for $V_{QD}$ compared to $V_{DD}$, and thus the $R_{T,ch} : R_D$ ratio, by setting $V_{DD}$ over the nominal value of the second electron $V_{2e}$ for all quantum dots, when $R_{T,ch} \ll R_D$, $V_{QD} \simeq V_{DD}$. Conversely, as the addition voltage $V_{1e} $for the first electron on the dot is  much larger than the potential equivalent of the addition energy (i.e. $V_{2e}-V_{1e}$), thus for $V_{DD}\simeq V_{2e}$, achieving $R_{T,ch} 
\simeq R_D$ results in the condition where $V_{QD} \simeq V_{DD}/2 < V_{1e}$. Therefore, $V_{QD}$ can be tuned over the entire voltage range where the dot holds exactly one electron. The key limitation for the operation range of this circuit is the deep sub-threshold (maximum) channel resistance, which is primarily set by the device dimensions.

}


\end{document}